\documentclass[11pt]{article}
\usepackage{calc}
\usepackage{rotating}
\usepackage[english]{babel}
\usepackage{graphicx}
\usepackage{subfig}
\usepackage{float}
\usepackage{amsmath}
\usepackage{amssymb}
\usepackage{amsthm}
\usepackage{latexsym}
\usepackage{dcolumn}
\usepackage{geometry}
\usepackage{color}
\usepackage{hyperref}
\usepackage{mciteplus}

\def\gtwid{\mathrel{\raise.3ex\hbox{$>$\kern-.75em\lower1ex\hbox{$\sim
$}}}}
\def\vio{\mathrel{\hbox{$E$\kern-.60em\hbox{$/
$}}}}
\newcommand{\newc}{\newcommand*}

\long\def\begincomment#1\endcomment{%
        \begingroup\sf\baselineskip12pt#1\endgroup}
\newc{\etal}{\textrm{et al.}} 
\newc{\eg}{\textrm{e.g.}} 
\newc{\ie}{\textrm{i.e.}}
\newc{\etc}{\textrm{etc}}
\newc\vs{\textrm{vs.}}
\newc{\cl}{\rm {CL}}

\newc{\ev}{\ensuremath{\,\mathrm{eV}}}
\newc{\kev}{\ensuremath{\,\mathrm{keV}}}
\newc{\mev}{\ensuremath{\,\mathrm{MeV}}}
\newc{\gev}{\ensuremath{\,\mathrm{GeV}}}
\newc{\tev}{\ensuremath{\,\mathrm{TeV}}}

\newc{\MeV}{\mev} 
\newc{\TeV}{\tev}
\newc{\invpb}{\ensuremath{/\text{pb}}}
\newc{\invfb}{\ensuremath{/\text{fb}}}

\newc\nb{\ensuremath{\,\mathrm{nb}}} \newc\pb{\ensuremath{\,\mathrm{pb}}} \newc\fb{\ensuremath{\,\mathrm{fb}}}

\newc\pc{\ensuremath{\,\mathrm{pc}}}
\newc\kpc{\ensuremath{\,\mathrm{kpc}}}
\newc\mpc{\ensuremath{\,\mathrm{Mpc}}}

\newc\ps{\ensuremath{\,\mathrm{ps}}} 

\newc\cmeter{\ensuremath{\,\mathrm{cm}}} 
\newc\meter{\ensuremath{\,\mathrm{m}}} 
\newc\kmeter{\ensuremath{\,\mathrm{km}}}

\newc\second{\ensuremath{\,\mathrm{s}}}
\newc\msecond{\ensuremath{\,\mathrm{ms}}}
\newc\nsecond{\ensuremath{\,\mathrm{ns}}}
\newc\psecond{\ensuremath{\,\mathrm{ps}}}

\newc{\chisqmin}{\ensuremath{\chi^2_{\mathrm{min}}}}
\newc{\Delchisq}{\ensuremath{\Delta\chi^2}}
\newc{\chisq}{\ensuremath{\chi^2}}

\newc{\like}{\ensuremath{\mathcal{L}}}
\newc\lsim{\ensuremath{\mathrel{\rlap{\lower4pt\hbox{\hskip1pt$\sim$}}\raise1pt\hbox{$<$}}}}
\newc\gsim{\ensuremath{\mathrel{\rlap{\lower4pt\hbox{\hskip1pt$\sim$}}\raise1pt\hbox{$>$}}}}

\newc{\VEV}[1]{\ensuremath{\langle #1 \rangle}}

\newc{\dl}{\ensuremath{\stackrel{\leftarrow}{D}}}
\newc{\dr}{\ensuremath{\stackrel{\rightarrow}{D}}}

\newc{\bcenter}{\begin{center}}    \newc{\ecenter}{\end{center}}
\newc{\bfl}{\begin{flushleft}}    \newc{\efl}{\end{flushleft}}
\newc{\bfr}{\begin{flushright}}    \newc{\efr}{\end{flushright}}

\newc{\bi}{\begin{itemize}}
\newc{\ei}{\end{itemize}}
\newc{\bed}{\begin{description}}
\newc{\eed}{\end{description}}
\newc{\ben}{\begin{enumerate}}
\newc{\een}{\end{enumerate}}

\newc{\be}{\begin{equation}}
\newc{\ee}{\end{equation}}
\newc{\bea}{\begin{eqnarray}}
\newc{\eea}{\end{eqnarray}}
\newc{\ra}{\rightarrow}

\newc{\alphas}{\ensuremath{\alpha_s}}
\newc{\alphatwo}{\ensuremath{\alpha_2}}
\newc{\alphaone}{\ensuremath{\alpha_1}}
\newc{\alphai}[1]{\ensuremath{\alpha_{#1}}}
\newc{\alphaem}{\ensuremath{\alpha_{\mathrm{em}}}}

\newc{\alphaeff}{\ensuremath{\alpha_{\mathrm{eff}}}}

\newc{\sineff}{\ensuremath{\sin \theta_{\mathrm{eff}}}}
\newc{\sinsqeff}{\ensuremath{\sin^2 \theta_{\mathrm{eff}}}}
\newc{\dalphahad}{\ensuremath{\Delta \alpha_{\mathrm{had}}}}

\newc{\yt}{\ensuremath{h_t}} \newc{\yb}{\ensuremath{h_b}} \newc{\ytau}{\ensuremath{h_{\tau}}}

\newc\mz{\ensuremath{M_Z}} 
\newc\mw{\ensuremath{m_W}}
\newc\mZ{\mz}        \newc\mW{\mw}

\newc\mhsm{\ensuremath{ m_{H_{\mathrm{SM}}}}}

\newc{\mtop}{\ensuremath{ m_t}}               \newc{\mtpole}{\ensuremath{ M_t}}
\newc{\mbottom}{\ensuremath{ m_b}} 
\newc{\mtau}{\ensuremath{ m_{\tau}}}
\newc{\mt}{\mtpole}
\newc{\mb}{\mbottom} 

\newc{\rtwogg}{\ensuremath{R_{h_2}(\gamma\gamma)}}
\newc{\rtwozz}{\ensuremath{R_{h_2}(ZZ)}}
\newc{\ronegg}{\ensuremath{R_{h_1}(\gamma\gamma)}}
\newc{\ronezz}{\ensuremath{R_{h_1}(ZZ)}}
\newc{\rsiggg}{\ensuremath{R_{h_\textrm{sig}}(\gamma\gamma)}}
\newc{\rsigzz}{\ensuremath{R_{h_\textrm{sig}}(ZZ)}}

\newc{\llbar}{\ensuremath{\ell\bar{\ell}}}
\newc{\tauptaum}{\ensuremath{ \tau^+\tau^-}}

\newc{\qqbar}{\ensuremath{ q\bar{q}}} \newc{\ppbar}{\ensuremath{ p\bar{p}}}
\newc{\bbbar}{\ensuremath{ b\bar{b}}} \newc{\ttbar}{\ensuremath{ t\bar{t}}}
\newc{\ffbar}{\ensuremath{ f\bar{f}}} \newc{\tautaubar}{\ensuremath{ \tau\bar{\tau}}}

\newc{\mchi}{\ensuremath{m_\neutone}}
\newc{\squark}{\ensuremath{\tilde{q}}}
\newc{\slepton}{\ensuremath{\tilde{l}}}
\newc{\gluino}{\ensuremath{\tilde{g}}} 
\newc{\mgluino}{\ensuremath{{m_{\gluino}}}}
\newc{\sthw}{\ensuremath{ \sin\theta_W}}              \newc{\cthw}{\ensuremath{\cos\theta_W}}
\newc{\tanthw}{\ensuremath{ \tan\theta_W}}              \newc{\cotthw}{\ensuremath{\cot\theta_W}}

\newc{\ssqthw}{\ensuremath{\sin^2 \theta_W}}

\newc{\msbar}{\ensuremath{\overline{MS}}} \newc{\drbar}{\ensuremath{\overline{DR}}}

\newc{\mtmtsmmsbar}{\ensuremath{ m_t(m_t)^{\msbar}_{{\mathrm{SM}}}}}
\newc{\mtmtsmdrbar}{\ensuremath{ m_t(m_t)^{\drbar}_{{\mathrm{SM}}}}}
\newc{\mtmtmssmdrbar}{\ensuremath{ m_t(m_t)^{\drbar}_{{\mathrm{SUSY}}}}}

\newc{\mbmbmsbar}{\ensuremath{ m_b(m_b)^{\msbar} }}

\newc{\mbmbsmmsbar}{\ensuremath{ m_b(m_b)^{\msbar}_{{\mathrm{SM}}}}}
\newc{\mbmzsmmsbar}{\ensuremath{ m_b(\mz)^{\msbar}_{{\mathrm{SM}}}}}
\newc{\mbmzsmdrbar}{\ensuremath{ m_b(\mz)^{\drbar}_{{\mathrm{SM}}}}}
\newc{\mbmzmssmdrbar}{\ensuremath{ m_b(\mz)^{\drbar}_{{\mathrm{SUSY}}}}}

\newc{\mtaumzsmmsbar}{\ensuremath{ m_{\tau}(\mz)^{\msbar}_{{\mathrm{SM}}}}}
\newc{\mtaumzsmdrbar}{\ensuremath{ m_{\tau}(\mz)^{\drbar}_{{\mathrm{SM}}}}}
\newc{\mtaumzmssmdrbar}{\ensuremath{ m_{\tau}(\mz)^{\drbar}_{{\mathrm{SUSY}}}}}

\newc{\alphasmzms}{\ensuremath{\alpha_s(M_Z)^{\overline{MS}}}}
\newc{\alphaimzms}[1]{\ensuremath{\alpha_{#1}(M_Z)^{\overline{MS}}}}

\newc{\alphaemmz}{\ensuremath{\alpha_{\mathrm{em}}(M_Z)^{\overline{MS}}}}

\newc{\mzero}{\ensuremath{{m_0}}}
\newc{\mhalf}{\ensuremath{ m_{1/2}}}
\newc{\tanb}{\ensuremath{\tan\beta}}
\newc{\azero}{\ensuremath{ A_0}}
\newc{\signmu}{\ensuremath{\rm{sgn}\,\mu}}
\newc{\atau}{\ensuremath{{A_{\tau}}}}

\newc{\mueff}{\ensuremath{\mu_{\rm{eff}}}}

\newc{\lam}{\ensuremath{{\lambda}}}
\newc{\kap}{\ensuremath{{\kappa}}}

\newc{\alam}{\ensuremath{{A_{\lambda}}}}
\newc{\akap}{\ensuremath{{A_{\kappa}}}}

 \newc{\hs}{\ensuremath{ H_s}}      
\newc{\mhs}{\ensuremath{ m_{H_s}}} 

\newc{\mgut}{\ensuremath{ M_{\rm GUT}}}
\newc{\mplanck}{\ensuremath{ M_{\rm P}}}      \newc{\mpl}{\ensuremath{ M_{\rm Pl}}}
\newc{\msusy}{\ensuremath{ M_{\rm SUSY}}}      \newc{\ms}{\ensuremath{ M_{\rm S}}}

 \newc{\hu}{\ensuremath{ H_u}}       \newc{\hd}{\ensuremath{ H_d}}
 \newc{\mhu}{\ensuremath{ m_{H_u}}}       \newc{\mhd}{\ensuremath{ m_{H_d}}}
 \newc{\mhuew}{\ensuremath{ m^{\ast}_{H_u}}}       \newc{\mhdew}{\ensuremath{ m^{\ast}_{H_d}}}
 \newc{\mhuewsq}{\ensuremath{ m^{\ast\, 2}_{H_u}}}       \newc{\mhdewsq}{\ensuremath{ m^{\ast\, 2}_{H_d}}}
 \newc{\mhl}{\ensuremath{m_\hl}} 
 \newc{\mhone}{\ensuremath{m_{h_1}}} 
 \newc{\mhtwo}{\ensuremath{m_{h_2}}} 
 \newc{\mglu}{\ensuremath{m_{\tilde g}}} 
 \newc{\mul}{\ensuremath{m_{\tilde{u}_L}}} 
 \newc{\mtone}{\ensuremath{m_{\tilde{t}_1}}} 
 \newc{\ma}{\ensuremath{m_A}} 
 \newc{\maone}{\ensuremath{m_{a_1}}} 
 \newc{\matwo}{\ensuremath{m_{a_2}}}
 \newc{\hone}{\ensuremath{h_1}}
 \newc{\htwo}{\ensuremath{h_2}}
 \newc{\aone}{\ensuremath{a_1}}
 \newc{\atwo}{\ensuremath{a_2}}

\newc{\sigsip}{\ensuremath{\sigma^{\rm SI}_{p}}}	\newc{\sigsin}{\ensuremath{\sigma^{\rm SI}_{n}}}
\newc{\sigsdp}{\ensuremath{\sigma^{\rm SD}_{p}}}	\newc{\sigsdn}{\ensuremath{\sigma^{\rm SD}_{n}}}
\newc{\sigsi}{\ensuremath{\sigma^{\rm SI}}}	\newc{\sigsd}{\ensuremath{\sigma^{\rm SD}}}

\newc{\abund}{\ensuremath{ \Omega h^2}}
\newc{\omegadm}{\ensuremath{ \Omega_{{\rm DM}}}}     \newc{\abunddm}{\ensuremath{ \Omega_{{\rm DM}} h^2}} 
\newc{\omegam}{\ensuremath{ \Omega_{{\rm m}}}}       \newc{\abundm}{\ensuremath{ \Omega_{{\rm m}} h^2}}
\newc{\omegab}{\ensuremath{ \Omega_{{\rm b}}}}	\newc{\abundb}{\ensuremath{ \Omega_{{\rm b}} h^2}}
\newc{\omegatot}{\ensuremath{ \Omega_{{\rm TOT}}}}
\newc{\omegacdm}{\ensuremath{ \Omega_{{\rm CDM}}}}   \newc{\abundcdm}{\ensuremath{ \Omega_{{\rm CDM}} h^2}}
\newc{\omegalambda}{\ensuremath{ \Omega_{\Lambda}}} \newc{\abundlambda}{\ensuremath{ \Omega_{\Lambda} h^2}}
\newc{\omegarad}{\ensuremath{ \Omega_{{\rm rad}}}}  \newc{\abundrad}{\ensuremath{ \Omega_{{\rm rad}} h^2}}

\newc{\rhocrit}{\ensuremath{ \rho_{\rm crit}}}
\newc{\rhochi}{\ensuremath{ \rho_{\chi}}}

\newc{\abunchi}{\ensuremath{\Omega_\chi h^2}}
\newc{\abundlsp}{\ensuremath{\Omega_{\rm LSP}h^2}}

\newcommand*{\abundchi}{\ensuremath{\Omega_\chi h^2}}% For multiple citations with one key
\newc{\amu}{\ensuremath{ a_{\mu}}}        \newc{\amususy}{\ensuremath{ a_{\mu}^{\mathrm{SUSY}}}}
\newc{\amuexpt}{\ensuremath{ a_{\mu}^{\mathrm{expt}}}}        \newc{\amusm}{\ensuremath{ a_{\mu}^{\mathrm{SM}}}}
\newc\deltaamu{\ensuremath{\Delta a_{\mu}}} \newc{\deltaamususy}{\ensuremath{\delta a_{\mu}^{\mathrm{SUSY}}}}
\newc\gmtwo{\ensuremath{ (g-2)_{\mu}}} 
\newc{\deltagmtwomususy}{\ensuremath{\delta\left(g-2\right)_{\mu}^{\mathrm{SUSY}}}}
\newc{\deltagmtwomu}{\ensuremath{\delta\left(g-2\right)_{\mu}}}

\newc\BR{\ensuremath{\rm BR}}

\newc\bsgamma{\ensuremath{ b\rightarrow s \gamma }}
\newc\bxsgamma{\ensuremath{\overline{B}\rightarrow X_{s}\gamma}}

\newc\brbsgamma{\ensuremath{\BR\left(\bsgamma\right)}}
\newc\brbxsgamma{\ensuremath{\BR\left(\bxsgamma\right)}}

\newc\bsmumu{\ensuremath{B_s\to\mu^+\mu^-}}
\newc\brbsmumu{\ensuremath{\BR\left(B_s\to\mu^+\mu^-\right)}}

\newc\bdmmumu{\ensuremath{\overline{B}_d\to\mu^+\mu^-}}

\newc\bbbarmix{\ensuremath{\overline{B}_s\mbox{-}B_s}}      % B_s mixing
\newc\delmbs{\ensuremath{\Delta M_{B_s}}}

\newc{\butaunu}{\ensuremath{B_u \rightarrow \tau \nu}}
\newc{\brbutaunu}{\ensuremath{\BR\left(B_u \rightarrow \tau \nu\right)}}

     \newcommand*{\refsec}[1]{Sec.~\ref{#1}}

\newcommand*{\neutone}{\ensuremath{\chi}}

\newcommand*{\razor}{\textrm{razor}}
\newcommand*{\razorfourfb}{\ensuremath{\cms\ \razor\ 4.4\invfb} }
\newcommand*{\razorexp}{\ensuremath{\cms\ \razor\ 4.4\invfb} analysis}

\newcommand*{\softsusy}{SOFTSUSY}

\newcommand*{\micromegas}{MicrOMEGAs}

\newcommand*{\cms}{\text{CMS}}

\newcommand*{\superiso}{\text{SuperIso}}

\newcommand*{\nmssmtools}{\text{NMSSMTools}}

\let\oldcite\cite
\renewcommand*{\cite}{~\oldcite}

\newcommand*{\hl}{\ensuremath{h}}

\newcommand*{\mh}{\ensuremath{m_h}}

\restylefloat{figure}

\begin{document}

\title{Constrained NMSSM with a 126\gev\ Higgs boson:\\ A global
  analysis}

\author{Kamila Kowalska,~Shoaib Munir,~Leszek Roszkowski\footnote{On leave of absence from the University of Sheffield, UK.},~Enrico Maria Sessolo,\\
~Sebastian Trojanowski,~and Yue-Lin Sming Tsai \\[2ex]
\small\it BayesFITS Group \\ National Centre for Nuclear Research,
  Ho{\. z}a 69, 00-681 Warsaw, Poland \\
\bigskip \\
\url{Kamila.Kowalska@fuw.edu.pl,Shoaib.Munir@fuw.edu.pl},\\ \url{L.Roszkowski@sheffield.ac.uk,Enrico-Maria.Sessolo@fuw.edu.pl},\\ \url{Sebastian.Trojanowski@fuw.edu.pl,Sming.Tsai@fuw.edu.pl}}

% Appears in brackets after all authors
%\collaboration{\bayesfits}

% Date
%\date{}

\maketitle

%%%%%%%%%%%%%%%%%%%%%%%%%%%%%%%%%%%%%%%%%%%%%%%%%%%%%%%%%%%%%%%%%%%%%%%%%%%%%%%%
\begin{abstract} We present the first global analysis of the
  Constrained NMSSM that investigates the impact of the recent
  discovery of a 126\gev\ Higgs-like boson, of the observation of a signal for branching 
  ratio \brbsmumu, and of constraints on
  supersymmetry from $\sim5$/fb of data accumulated at the LHC, as
  well as of other relevant constraints from colliders, flavor physics
  and dark matter.  We consider three possible cases, assuming in turn
  that the discovered Higgs boson is (i) the lightest Higgs boson of
  the model; (ii) the next-to-lightest Higgs boson; and (iii) a
  combination of both roughly degenerate in mass. The likelihood
  function for the Higgs signal uses signal rates in the
  $\gamma\gamma$ and $ZZ\rightarrow 4l$ channels, while that for the
  Higgs exclusion limits assumes decay through the $\gamma\gamma$,
  $\tau\tau$, $ZZ$ and $W^+ W^-$ channels.  In all cases considered we
  identify the $68$\% and $95$\% credible posterior probability
  regions in a Bayesian approach.  We find that, when the constraints
  are applied with their respective uncertainties, the first case
  shows strong CMSSM-like behavior, with the stau coannihilation
  region featuring highest posterior probability, the best-fit point,
  a correct mass of the lightest Higgs boson and the lighter stop mass
  in the ballpark of 1\tev. We also expose in this region a linear
  relationship between the trilinear couplings of the stau and the
  stop, with both of them being strongly negative as enforced by the
  Higgs mass and the relic density, which outside of the stau
  coannihilation region show some tension. The second and the
  third case, on the other hand, while allowed are disfavored by the
  constraints from direct detection of dark matter and from \brbsmumu.
  Without the anomalous magnetic moment of the muon the fit improves
  considerably, especially for negative effective $\mu$ parameter.  We
  discuss how the considered scenarios could be tested further at
  the LHC and in dark matter searches.
\end{abstract}

%%%%%%%%%%%%%%%%%%%%%%%%%%%%%%%%%%%%%%%%%%%%%%%%%%%%%%%%%%%

\newpage
%%%%%%%%%%%%%%%%%%%%%%%%%%%%%%%%%%%%%%%%%%%%%%%%%%%%%%%%
\section{\label{intro}Introduction}

%wherein the naturalness
%problem of the Standard Model (SM) is addressed by invoking supersymmetry
 %(SUSY) and introducing a ${\it superpartner}$ for every SM
 %particle. Since such superpartners remain unobserved at colliders
 %hitherto, SUSY is assumed to be broken not far above the electroweak (EW)
 %scale. Without any specification of the underlying mechanism
 %responsible for it, SUSY breaking is accounted for by introducing {\it soft} 
 %dimensionful coupling parametrs for the superpartners in the MSSM
 %lagrangian. However, this leads to a large number of new arbitrary parameters in
 %the theory and, for phenomenological studies of the model, it is
 %customary to postulate unification of some of these parameters
 %at the GUT scale, resulting in only five free parameters. This
 %version of the model is referred to as the constrained MSSM,
 %(CMSSM). Other versions of the model with some relaxations in
%the unification condition, such as the non-universal Higgs model
%(NUHM) and a more general phenomenological MSSM
%such as the pMSSM, are also frequently exploited. 

In July 2012 the CMS and ATLAS collaborations at the LHC made the
announcement of a $\sim5\sigma$ discovery of a boson with mass
$125.3\pm0.6\gev$\cite{CMS:2012gu} and
$126.0\pm0.6\gev$\cite{ATLAS:2012gk}, respectively, consistent with
the Higgs boson predicted by the electroweak (EW) Standard Model
(SM). The CMS Collaboration has recently updated its results\cite{CMS-PAS-HIG-12-045}
to $125.8\pm0.6\gev$.  The updated result is based on the 
analysis of the data corresponding
to integrated luminosities of 5.1/fb at $\sqrt{s}=7\tev$ and up to 12.2/fb at
$\sqrt{s}=8\tev$ in the $\gamma\gamma$, $ZZ$, $WW$, $\tau\tau$ and
$bb$ decay channels. The ATLAS analysis combined approximately 4.8/fb
of 2011 data at $\sqrt{s}=7\tev$ in the same five channels with 5.8/fb
of data at $\sqrt{s}=8\tev$ in the $ZZ$, $\gamma\gamma$ and $WW$
channels only. Evidently, the excess of events near the reported mass
is driven by the $\gamma\gamma$ and $ZZ$ channels, owing to the
highest mass resolution of these channels. Both collaborations also determined the
ratio $\mu(X)$ of the observed Higgs production cross section to the
one predicted by the SM, in each of the mentioned Higgs decay channels
$X$.  An enhancement in the $\gamma\gamma$ channel was reported by
ATLAS, with $\mu(\gamma\gamma) = 1.9\pm0.5$, as well as by CMS, with
$\mu(\gamma\gamma) = 1.6\pm0.4$. On the other hand, the updated value of
$\mu(ZZ)$ by CMS\cite{CMS-PAS-HIG-12-041} and ATLAS\cite{ATLAS-CONF-2012-162} is, within $1\sigma$ error, SM-like.

While its exact characteristics are still being carefully
examined, such a boson is not only consistent with the SM Higgs
particle but can also be easily accommodated into models of new
physics, particularly those based on softly broken low-energy
supersymmetry (SUSY) which actually predict a relatively light Higgs
boson. Most studies have been performed within the framework of the
two Higgs doublet Minimal Supersymmetric Standard Model (MSSM),
usually augmented by various unification boundary conditions. The
framework, however, suffers from the so-called
``$\mu$-problem''\cite{muproblem}: the SUSY-preserving parameter $\mu$
in the superpotential is expected to be of the order of the SUSY
breaking scale $\msusy\sim1\tev$ to ensure correct radiative EW
symmetry breaking.

Perhaps the most compelling and simplest solution to the $\mu$-problem
is to invoke an additional gauge singlet field coupled to the Higgs
doublets of the MSSM\cite{muproblem}. The $\mu$-term is then generated dynamically
through the vacuum expectation value (vev) of the singlet field which
is of the order of \msusy. The model is commonly referred to as the
Next-to-Minimal Supersymmetric Standard Model (NMSSM); for reviews
see, e.g.,\cite{Ellwanger:2009dp,Maniatis:2009re}.  After the
discovery of the Higgs-like boson, numerous studies have appeared in
the context of the
NMSSM\cite{Ellwanger:2011aa,Ellwanger:2012ke,King:2012is,*Cao:2012fz,*Vasquez:2012hn,*Rathsman:2012dp,*Das:2012ys,
Gunion:2012gc,*Gunion:2012he,Benbrik:2012rm,*Bae:2012am,*Kang:2012bv,*Cheng:2012pe,
  *Perelstein:2012qg,*Belanger:2012sd,*Agashe:2012zq,Cao:2012yn,Chalons:2012qe,*Belanger:2012tt,
  *Gogoladze:2012jp,*LopezFogliani:2012yq,Gunion:2012zd}, since the model presents
several interesting features.  In the NMSSM the particle content
remains the same as in the MSSM, except that the number of Higgs
bosons increases from three to five: three neutral scalars
$h_{1,2,3}$ and two neutral pseudoscalars $a_{1,2}$. The number of
neutralinos also increases from four to five, $\chi_{1,\ldots,5}$,
owing to the singlino partner of the singlet Higgs boson.

On the phenomenological front, a two-loop theoretical upper bound  on
the mass of the lightest $CP$-even Higgs boson, \mhone, can increase by a
few GeV compared to the MSSM for some combinations of the model's
parameters. Moreover, in the context of the recent LHC results, the
next-to-lightest $CP$-even Higgs boson of the model, $h_2$, could also
have mass around 126\gev\ with $h_1$ being even lighter than the LEP
bound on the SM-like Higgs boson mass, without violating it\cite{Ellwanger:2011aa,Ellwanger:2012ke}, owing to the singlet-doublet mixing
effects. Both features potentially allow us to achieve the correct
mass of the experimentally detected Higgs without excessive
fine-tuning\cite{BasteroGil:2000bw,Delgado:2010uj,Ellwanger:2011mu,Ross:2011xv},
making the NMSSM potentially a more ``natural'' model than the MSSM. 
Finally, the NMSSM also offers the additional possibility that the
observed excess in the $\gamma\gamma$ and $ZZ$ rates
could be due to the combination of $\hone$ and $\htwo$ decays (with
\mhone\ and \mhtwo\ being almost degenerate)\cite{Gunion:2012gc}. 
Besides masses, the decay rates of the Higgs boson(s) are also affected by
the modifications in the superpotential. For example, for large
doublet-singlet mixing, the $\hone b\bar{b}$ coupling becomes suppressed,
reducing the total decay width. As a result, the branching ratio (BR)
of $h_1\to\gamma\gamma$ becomes marginally enhanced compared to the SM or
the MSSM\cite{Ellwanger:2010nf}. An important consequence of this feature is that $h_1$
lighter than 115\gev\ could have escaped detection at LEP, and also at
the LHC, due to the possibility of its decay into a pair of the two
lighter pseudoscalars, $\hone\ra\aone\aone$~(see,
e.g.,\cite{Gunion:1996fb}), causing thus a suppression in all of the other BR's of
$\hone$. 

%Due to the incorporation of a new singlet field, certain new free
%parameters arise in the model besides the

The NMSSM in its most general form contains more than a hundred free
parameters: those appearing in the MSSM and some additional ones
relating to the extended Higgs sector.  Similarly to the MSSM, where
imposing universality conditions on the soft
SUSY-breaking parameters of the model at the scale of grand
unification (GUT scale) leads to the Constrained MSSM (CMSSM)\cite{hep-ph/9312272},
a GUT-constrained version of the NMSSM
(CNMSSM)\cite{PhysRevD.39.844,Djouadi:2008yj,Djouadi:2008uj} can be defined; see
\refsec{Model} for more details and discussion.

It is well known that in the CMSSM the lightest Higgs boson's mass, as
calculated at two-loop level with FeynHiggs\cite{feynhiggs:00} or
\softsusy\cite{softsusy}, is typically a few GeV below 126\gev, which
can be considered as somewhat unsatisfactory even if one takes into
account a residual error of some 2--3\gev\cite{Heinemeyer:2011aa} from
scheme dependence. (Values of \mh\ around 126\gev\ can still be obtained in the CMSSM for $\msusy\gg
1\tev$ at the expense of increased fine-tuning\cite{Fowlie:2012im}.) \textit{A priori} one could
expect that in the CNMSSM, with more freedom in the Higgs sector, the
tension will be reduced. However, recent studies\cite{Gunion:2012zd}
using random scans with fixed-window application of the constraints have shown that in the CNMSSM it is
extremely difficult to obtain a Higgs boson as heavy as 126\gev,
particularly one with an enhanced $\gamma\gamma$ decay rate, while also satisfying other phenomenological
constraints, thus nullifying the noted advantages of the
singlet-extension of the MSSM.

On the
other hand,
it has been demonstrated by our recent global analysis of the
CMSSM\cite{Fowlie:2012im} and by some earlier Bayesian
studies\cite{LopezFogliani:2009np,Gunion:2011hs}
%such studies relied on grid scans, over which the phenomenological
%constraints were imposed as a hard cut. It has been long
that a proper treatment of the experimental constraints through a
likelihood function can lead to significantly different results from
``top-hat'' scans where such constraints are typically implemented in a
more simplified boxlike fashion, with all points accepted when
satisfying experimental values within some fixed range (typically
within $2\sigma$), and otherwise rejected.  One of the
main advantages of the statistical approach is that scanned points are
instead ``weighted'' by the total \chisq, thus indicating how well they
fit all constraints. For example, in the top-hat approach a point
giving a value of even one constraint slightly beyond the allowed range
while reproducing central values of all the other constraints
would be rejected, while a point with values for all the constraints
barely within the allowed boxes would be completely allowed. In contrast, in a statistical approach
both points would be accepted but weighted with their respective
\chisq. Another advantage of the statistical approach based on the
likelihood function is that theoretical and experimental uncertainties
can be easily implemented in a consistent manner.  Thanks to some
recent developments in sampling algorithms~(see, e.g.,\cite{mackay03}
and\cite{skilling04}), multidimensional Bayesian scans can now be
carried out rather quickly and efficiently.

One should note two important changes in the data that have recently
taken place on the experimental front. In November 2012 the LHCb Collaboration published the 
most recent update of their search for the rare decay \bsmumu\cite{Aaij:2012ct}, reporting an excess of \bsmumu\ candidates over the background. The measured value, $\brbsmumu=3.2^{+1.5}_{-1.2}\times
10^{-9}$, is now very close to the time-averaged SM value,
$3.5\times10^{-9}$\cite{DeBruyn:2012wk,Buras:2012ru}. 
We include the
constraint in our likelihood function taking into account both
theoretical and experimental uncertainties, as will be described
below.

The other important update was the top pole mass by the Particle Data Group, obtained from
an average of data from Tevatron and the LHC at $\sqrt{s}=7\tev$,
$\mtpole=173.5\pm1.0\gev$\cite{Beringer:1900zz}.  As we shall see
below this is a  welcome increase relative to its previous value in
the context of the Higgs
sector of constrained SUSY models as it pushes the mass of $\hone$
up, closer to the experimentally observed Higgs-like resonance
mass.  

In this article, we present the first global Bayesian analysis of the
CNMSSM after the observation of the SM Higgs-like boson. We separately
consider the cases of this boson being $h_1$, or $h_2$, or a
combination of both.  We test the parameter space of the model against
the currently published, already stringent constraints from
SUSY searches at the LHC and other relevant constraints from
colliders, $b$-physics and dark matter~(DM) relic density.  Our goal
is to map out the regions of the parameter space of the CNMSSM that
are favored by these constraints.  As in our CMSSM
study\cite{Fowlie:2012im}, the \cms\ razor limit based on $4.4\invfb$
of data is implemented through an approximate but accurate likelihood
function. We also study the effects of relaxing the \gmtwo\ constraint.

The article is organized as follows. In \refsec{Model} we briefly
revisit the model, highlighting some of its salient features. In
\refsec{Method} we detail our methodology, including our statistical
approach and our construction of the likelihoods for the \brbsmumu\ signal, the \razorexp, and the CMS Higgs
searches. In \refsec{Results} we present the results from our scans
and discuss their novel features. We summarize our findings in
\refsec{Summary}.

%%%%%%%%%%%%%%%%%%%%%%%%%%%%%%%%%%%%%%%%%%%%%%%%%%%%%%%%%%%
%%%%%%%%%%%%%%%%%%%%%%%%%%%%%%%%%%%%%%%%%%%%%%%%%%%%%%%%%%%

\section{\label{Model}The NMSSM with GUT-scale universality}

The NMSSM is an economical extension of the MSSM, in which one adds a
gauge-singlet superfield $S$ whose scalar component couples only to
the two MSSM Higgs doublets $\hu$ and $\hd$ at the
tree level.\footnote{For simplicity we will be using the same notation
  for superfields and their bosonic components.} The scale-invariant
superpotential of the model has the form
%%%%
\be
W=\lam S \hu \hd+\frac{\kap}{3}S^3+(\textrm{MSSM Yukawa terms})\,,\label{NMSSM}
\ee
%%%%
where \lam\ and \kap\ are dimensionless couplings.  Upon spontaneous
symmetry breaking, the scalar Higgs field $S$ develops a vev, $s\equiv\langle
S\rangle$, and the first term in Eq.~(\ref{NMSSM}) assumes the role
of the effective $\mu$-term of the MSSM, $\mueff=\lam s$. The soft
SUSY-breaking terms in the Higgs sector are then given by 
%%%%
\be
V_{\rm{soft}}=m_{\hu}^2|\hu|^2+m_{\hd}^2|\hd|^2+m_{S}^2|S|^2+\left(\lam\alam
  S \hu \hd+\frac{1}{3}\kap\akap
  S^3+\textrm{h.c.}\right)\,,\label{soft} 
\ee
%%%%
where \alam\ and \akap\ are soft trilinear terms associated with the
\lam\ and \kap\ terms in the superpotential.
The vev $s$, determined by the minimization conditions of the Higgs potential, is effectively induced by the
SUSY-breaking terms in Eq.~(\ref{soft}), and is naturally set by \msusy, thus
solving the $\mu$-problem of the MSSM.

We define the CNMSSM in terms of
five continuous input parameters and one sign,
%%%%
\be
\mzero, \mhalf, \azero, \tanb, \lam, \textrm{sgn}(\mueff)\, ,\label{params}
\ee
%%%%
where unification conditions at a high scale require that all the
scalar soft SUSY-breaking masses in the superpotential (except $m_S$)
are unified to \mzero, the gaugino masses are unified to \mhalf, and
all trilinear couplings, including \alam\ and \akap, are unified to
\azero. This leaves us with two additional free parameters: 
\lam\ and the singlet soft-breaking mass $m_S^2$. 
The latter is not unified to $m_0^2$ for both theoretical and phenomenological reasons. 
From the theoretical point of view, it has been argued\cite{Hugonie:2007vd}
that the mechanism for SUSY breaking might treat the singlet field
differently from the other superfields. From the phenomenological
point of view, the freedom in $m_S$ allows for easier convergence when
the renormalization group equations (RGEs) are evolved from the GUT scale
down to \msusy. It also yields, in the limit $\lam\rightarrow 0$, and
with $\lam s$ fixed, effectively the CMSSM plus a singlet and singlino
fields that both decouple from the rest of the spectrum. 
Through the minimization equations of the Higgs potential, $m_S^2$ can then be traded for \tanb\ 
(the ratio of the vev's of the neutral components of the $\hu$ and $\hd$ fields) and either sgn(\mueff) or \kap. 
We choose sgn(\mueff) for conventional analogy with the CMSSM. Both \lam\ and \tanb\ are defined at \msusy.
Our choice of the parameter space is the same as the one used by one of us 
in a previous Bayesian analysis\cite{LopezFogliani:2009np}, 
of which this paper is, in some sense, an update.
Of course, there exist different possibilities that have been explored in the literature.
Some authors have studied the more constrained version of the CNMSSM, 
characterized by $m_S^2=m_0^2$\cite{Djouadi:2008uj}. But it is also true
that the underlying assumption employed here, of a different
treatment of the singlet field by the SUSY breaking mechanism, 
would allow for freedom in \akap\ at the GUT scale\cite{Hugonie:2007vd}.
We will give some comment in the Conclusions about 
the possible impact of relaxing the unification condition for \akap.

%%%%%%%%%%%%%%%%%%%%%%%%%%%%%%%%%%%%%%%%%%%%%%%%%%%%%%%%%%%
%%%%%%%%%%%%%%%%%%%%%%%%%%%%%%%%%%%%%%%%%%%%%%%%%%%%%%%%%%%

\section{\label{Method}Statistical treatment of experimental data}

We explore the parameter space of the model with the help of Bayesian
formalism. We follow the procedure outlined in detail in our previous
papers~\cite{Fowlie:2011mb,Roszkowski:2012uf,Fowlie:2012im}, of which
we give a short summary here. Our aim is to map out the 68\% and 95\%
credible regions of  $p(m|d)$, the posterior probability density function (pdf), given by
Bayes' theorem,
%%%%
\be
p(m|d)=\frac{p(d|\xi(m))\pi(m)}{p(d)}\,.
\label{eq:1}
\ee
%%%%
$p(d|\xi(m))\equiv\mathcal{L}$ is the likelihood function, which
describes the probability of obtaining the data $d$ given the computed
value of some observable $\xi(m)$, which is a function of the model's
parameters $m$. $\mathcal{L}$ also incorporates 
%lr the information on the measured values of the observables $\xi$,
%as it includes 
the experimental and theoretical uncertainties. Prior probability
$\pi(m)$ encodes assumed range and distribution of $m$.  Finally,
$p(d)$ is the evidence, which is a normalization constant as long as
only one model is considered, but serves as a comparative measure for
different models or scenarios.
%For several normally distributed uncorrelated observables the
%likelihood can be given in terms of the $\chi^2$ function as
%\bea
%\mathcal{L} = \exp[-\sum_i \frac{\chi_i^2}{2}]\,
%\label{eq:2}
%\eea
%with $\chi_i^2 = \frac{[\xi_i(m) - c_i]^2}{\sigma_i^2 + \tau_i^2}$,
%where $c_i$ is the measured central value of the observable $i$,
%$\sigma_i$ the experimental error in the measurement and $\tau_i$ the
%error in theoretical estimation of the observable. 

Bayes' theorem provides an efficient and natural procedure for drawing
inferences on a subset of $r$ specific model parameters (including
nuisance parameters), or observables, or a combination of both, which
we collectively denote by $\psi_i$.  They can be obtained through
marginalization of the full posterior pdf, carried out as
%%%%
\be
p(\psi_{i=1,..,r}|d)=\int p(m|d)d^{n-r}m\,,\label{marginalize}
\ee
%%%%
where $n$ is the total number of input parameters. An analogous
procedure can be performed with the observables and with a combination
of the model's parameters and observables.

%%%%%%%%%%%%%%%%%%%%%%%%%%%%   T   A   B   L   E   %%%%%%%%%%%%%%%%%%%%%%%%%%%%
% Table showing all experimental constraints in our scans
\begin{table}[t]\footnotesize
\begin{center}
%lr \input{./Tables/exp_constraints}
%%%%%%%%%%%%%%%%%%%%%%%%%%%%%%%%%%%%%%%%%%%%%%%%%%%%%%%%%%%%%%%%%%%%%%%%%%%%%%%%
\begin{tabular}{|l|l|l|l|l|l|}
\hline %\toprule
%%%%%%%%%%%%%%%%%%%%%%%%%%%%%%%%%%%%%%%%%%%%%%%%%%%%%%%%%%%%%%%%%%%%%%%%%%%%%%%%
Measurement & Mean or range & Error~(Exp.,~Th.) & Distribution & Ref.\\
%%%%%%%%%%%%%%%%%%%%%%%%%%%%%%%%%%%%%%%%%%%%%%%%%%%%%%%%%%%%%%%%%%%%%%%%%%%%%%%%
\hline %\cmidrule(r){1-6}
%lr \multicolumn{6}{|l|}{\razorexp} \\
%lr \hline %\cmidrule(r){1-6}
%%%%%%%%%%%%%%%%%%%%%%%%%%%%%%%%%%%%%%%%%%%%%%%%%%%%%%%%%%%%%%%%%%%%%%%%%%%%%%%%
\razorfourfb & See text. 	& See text. 	& Poisson &\cite{CMS-PAS-SUS-12-005}\\ \hline 
$m_{h_{\textrm{sig}}}$ & $125.8\gev$ & $0.6\gev, 3\gev$ & Gaussian &\cite{CMS-PAS-HIG-12-045} \\
%%%%%%%%%%%%%%%%%%%%%%%%%%%%%%%%%%%%%%%%%%%%%%%%%%%%%%%%%%%%%%%%%%%%%%%%%%%%%%%%
\hline
$R_{h_{\textrm{sig}}}(\gamma\gamma)$ & $1.6$ & $0.4$,~15\%	&
Gaussian & \cite{CMS:2012gu}\\

$R_{h_{\textrm{sig}}}(ZZ)$ & $0.80$ & $+0.35-0.28$,~15\% 	& Gaussian &
\cite{CMS-PAS-HIG-12-041}\\ \hline
%%%%%%%%%%%%%%%%%%%%%%%%%%%%%%%%%%%%%%%%%%%%%%%%%%%%%%%%%%%%%%%%%%%%%%%%%%%%%%%%
$m_{h_{\textrm{hid}}}$~(\gev) & $< 122.7\gev$, $> 128.9\gev$ & $0$,~$3\gev$ &
Error Fn& See text.\\ 
$R_{h_{\textrm{hid}}}(\gamma\gamma,ZZ,\tau\tau,WW)$ & See text. &
$0$,~$15\%$ & Error Fn& \cite{CMStwiki} \\ 
%%%%%%%%%%%%%%%%%%%%%%%%%%%%%%%%%%%%%%%%%%%%%%%%%%%%%%%%%%%%%%%%%%%%%%%%%%%%%%%%
\hline %\cmidrule(r){1-6}
%lr \multicolumn{6}{|l|}{Non-LHC} \\
%lr \hline %\cmidrule(r){1-6}
%%%%%%%%%%%%%%%%%%%%%%%%%%%%%%%%%%%%%%%%%%%%%%%%%%%%%%%%%%%%%%%%%%%%%%%%%%%%%%%%
\abundchi 			& $0.1120$ 	& $0.0056$,~$10\%$ 		& Gaussian &  \cite{Komatsu:2010fb}\\
\deltagmtwomususy $\times 10^{10}$ 	& $28.7 $  	& $8.0$,~$1.0$ 		& Gaussian &  \cite{Bennett:2006fi,Miller:2007kk} \\
\brbxsgamma $\times 10^{4}$ 		& $3.43$   	& $0.22$,~$0.21$ 		& Gaussian &  \cite{bsgamma}\\
\brbutaunu $\times 10^{4}$          & $1.66$  	& $0.66$,~$0.38$ 		& Gaussian &  \cite{Amhis:2012bh}\\
$\Delta M_{B_s}$ & $17.719\ps^{-1}$ & $0.043\ps^{-1},~2.400\ps^{-1}$ & Gaussian & \cite{Beringer:1900zz}\\
\brbsmumu			& $3.2 \times 10^{-9}$  	&  $+1.5-1.2$,~10\%  & Gaussian &  \cite{Aaij:2012ct}\\
%%%%%%%%%%%%%%%%%%%%%%%%%%%%%%%%%%%%%%%%%%%%%%%%%%%%%%%%%%%%%%%%%%%%%%%%%%%%%%%%
\hline %\bottomrule
\end{tabular}\caption{The experimental constraints that we apply to
  constrain model parameters. 
%lr Masses are in   GeV. 
$m_{h_{\textrm{sig}}}$, $m_{h_{\textrm{hid}}}$,
  $R_{h_{\textrm{sig}}}$ and
  $R_{h_{\textrm{hid}}}$ are defined in \refsec{Higgslike}.
} 
\label{tab:exp_constraints}
\end{center}
\end{table}
%%%%%%%%%%%%%%%%%%%%%%%%%%%%%%%%%%%%%%%%%%%%%%%%%%%%%%%%%%%%%%%%%%%%%%%%%%%%%%%%

In order to evaluate the posterior probability given by
Eq.~(\ref{eq:1}), one needs to first construct the likelihood
function. The constraints that we include in the current analysis are
listed in Table~\ref{tab:exp_constraints}. We shall be discussing them
in turn below.  As a rule, following the procedure developed
earlier\cite{deAustri:2006pe}, we implement positive measurements through
the usual Gaussian likelihood, while upper or lower limits through an
error function smeared with both theory and, when available, experimental error.
%In the case of the LHCb limit on
%\brbsmumu\cite{LHCb-CONF-2012-017} the experimental uncertainty is
%given through the $CL_s$ distribution function, our implementation of
%which is also described below.  
The construction of the likelihoods
for direct SUSY and Higgs searches is more involved, and will be
explained in detail later in this section.

%%%%%%%%%%%%%%%%%%%%%%%%%%%%%%%%%%%%%%%%%%%%%%%%%%%%%%%%%%%%%%%%%%%%%%

\subsection{Likelihood for \brbsmumu}
\label{sec:bsmlike}

In November 2012 the LHCb Collaboration published the most recent update of their search for the rare decay
\bsmumu\cite{Aaij:2012ct}, based on a combination of the 2012 data samples of 1.1/fb of proton-proton collisions 
at $\sqrt{s}=8\tev$ and the 2011, 1.0/fb data at $\sqrt{s}=7\tev$. The data superseded the combination of 2011 data 
from ATLAS, CMS and LHCb published in June 2012\cite{LHCb-CONF-2012-017}.

LHCb observed an excess of \bsmumu\ candidates over the background, consistent with the SM expectation. 
The measured value is $\brbsmumu=3.2^{+1.5}_{-1.2}\times
10^{-9}$, with a statistical significance of $3.5\,\sigma$. 
We used this information to construct an approximate
likelihood function for \brbsmumu\ which parametrizes the experimental
and theoretical uncertainties. 
The most important sources of theoretical uncertainty are the $B_s$
decay constant (main contribution) and its lifetime, the top quark mass
and the CKM matrix elements $V^{\ast}_{tb}V_{ts}$\cite{Buras:2012ru}
and their total amounts to approximately 11\% of the mean
value\cite{Mahmoudi:2012un}.  
Thus, we parametrized the likelihood function as a combination of two half-Gaussians, 
to take into account the asymmetry in the experimental uncertainty. A theoretical uncertainty equal 
to 10\% of the calculated value was added in quadrature. 
Notice that we neglected the
uncertainty due to the top pole mass ($\sim1\%$) since it is taken
care of by scanning over the SM nuisance parameters (see below). 

%%%%%%%%%%%%%%%
\subsection{Limits on SUSY from the LHC}
\label{sec:razor}

In Ref.\cite{Fowlie:2012im} we derived a methodology for constructing
an approximate likelihood map that reproduced the lower limit in the (\mzero, \mhalf) 
plane set by the CMS Collaboration with the razor method\cite{Chatrchyan:2011ek,CMS-PAS-SUS-12-005} based on 4.4\invfb\ of data.
We did so by
applying the razor method to simulated SUSY searches in all-hadronic
modes.   
We validated our map
against the experimental results by evaluating the resulting 95\%~C.L. contour in the (\mzero, \mhalf) plane and
comparing it with the corresponding 95\%~C.L. exclusion limit from the hadron box provided
by the CMS Collaboration. 
We obtained a very good agreement confirming
that our procedure for generating these likelihood maps was indeed
correct.  
Our methodology can be applied to produce the SUSY exclusion
limits in any $R$-parity conserving supersymmetric model, as long as
the supersymmetric spectra in that model present similar features to
the ones of the CMSSM, namely a sufficient mass difference between the
lightest SUSY particle (LSP) and gluinos or squarks of the first
two generations. 
When extending the procedure to other models, one
should also take into account possible changes in the production and
decay modes of the particles.  In this paper we follow the same
methodology for generating a SUSY likelihood map based on the
\razorexp\ for the NMSSM.

For each point in an $\mzero-\mhalf$ grid, with 50\gev\ step size in
both parameters, the supersymmetric mass spectrum and decay table are
generated using \nmssmtools\cite{NMSSMTools}, and fed into Herwig++\cite{Bahr:2008pv}
for parton shower
generation and calculation of the cross sections. 
Herwig++ allows one to work in the framework of the NMSSM,
and hence takes care of a possible contribution from the extended
neutralino and Higgs sectors. 
The production modes are not influenced
by these sectors, since the razor search is designed to detect squarks
and gluinos, which are produced through color interactions. 
%Moreover,
%the squarks of the first two generations are always significantly
%heavier than the LSP, so that neutralino composition should not affect
%the definition of missing energy or the cascade chain. 
The output of Herwig++ is passed in HepMC\cite{Dobbs:2001ck} format to
Delphes\cite{Delphes} for reconstruction of the physical objects by
simulating the CMS detector's response.  The output from Delphes is
distributed into 38 bins, defined in terms of the razor variables
$M_R$ and $R^2$, each containing the data (a number of background and
signal events) from the CMS-provided hadron box, to produce an efficiency map for
each bin, which is then translated into a likelihood map for the whole
grid.  We refer the reader to Ref.\cite{Fowlie:2012im} for more
details of the complete procedure adopted for the production of
efficiency maps.

%%%%%%%%%%%%%%%%%%%%%%%%%   F   I   G   U   R   E   %%%%%%%%%%%%%%%%%%%%%%%%%%%%
% 2 by 1: left: plot of 1d pdf of mhl,
% right: chi2 vs mhl
\begin{figure}[t]
\centering
%\subfloat[]{%
%\label{fig:-a}%
%\includegraphics[width=9.0cm, height=5.4cm]{figs_v5/BsmumuLike.pdf}
%}%
%\hspace{15pt}%
%\subfloat[]{%
%\label{fig:-b}%
\includegraphics[width=0.50\textwidth]{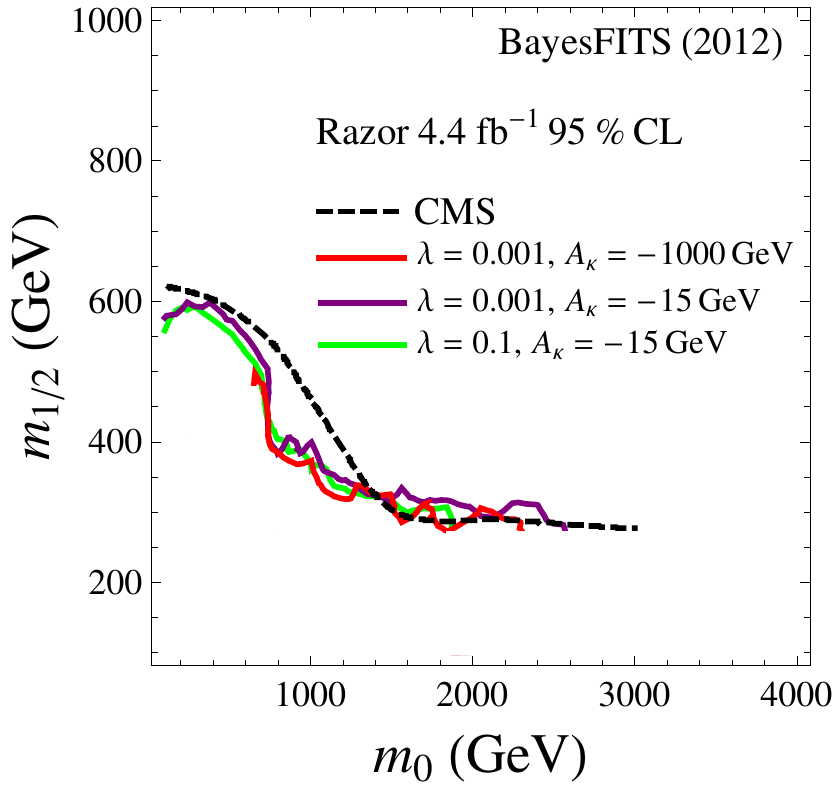}
%}%
\caption[]{The 95\%~C.L. lower bounds from our approximate razor likelihood, for different values of \lam\ and \akap, compared with the experimental line (in dashed black).}
\label{fig:bsmandrazor}
\end{figure} 
%%%%%%%%%%%%%%%%%%%%%%%%%%%%%%%%%%%%%%%%%%%%%%%%%%%%%%%%%%%%%%%%%%%%%%%%%%%%%%%%

The 95\% C.L. razor contour thus reproduced is shown in
Fig.~\ref{fig:bsmandrazor}.  The actual 95\% C.L. line
obtained by CMS is also shown as a dashed black line (we point out that the actual official line 
and the one obtained by CMS using the hadron box only are very close to each other). The relevant
CNMSSM parameters were fixed as $\tanb=5$, $\azero=0$,
$\akap=-15\gev$, $\lam=0.001$ and $\rm{sgn}(\mueff)=+1$.  The produced
line is indicated in purple. One can see that the CMS
limit is reproduced very well in the limits of $\mhalf\gg\mzero$ and
vice versa. A slight discrepancy in between those two extremes is
basically irrelevant, as we shall see below. The dependence of the limit on \lam\ and
\akap\ is negligible, since these parameters only affect the
boundaries of the physically allowed region but do not cause a shift
in the position of the contour. This is clear from
Fig.~\ref{fig:bsmandrazor}, where we show a contour for
$\lam=0.1$ (in green) and another for $\akap=-1000\gev$ (in red),
fixing the other parameters to the stated value in each case.  The
negligible dependence of the \razor\ limit on $\azero$, $\tanb$ and
\textrm{sgn}(\mueff)\, has already been verified in many analyses; see
\eg,\cite{Allanach:2011ut,Bechtle:2011dm,Fowlie:2012im}. Our choice of
these parameters thus only ensures the maximum allowed physical region in
the (\mzero, \mhalf) space, since the physicality aspect is later
taken into account during the scan when the unphysical points are
discarded.

%%%%%%%%%%%%%
\subsection{Likelihood for the Higgs bosons}
\label{Higgslike}

The measured mass of the Higgs-like boson has important
consequences for global properties of the CNMSSM. 

An absolute upper bound on the mass of a SM-like $h_1$ can be obtained in the limit
$\kap s\gg|\alam|,|\akap|$ (for which the singletlike scalar becomes very heavy) 
and is given by\cite{Drees:1988fc,Ellwanger:2009dp} 
%%%%
\be
\label{eq:mhiggs}
\mhone^2 \leq \mz^2\cos^2 2\beta 
+ \lam^2 v^2 \sin^2 2\beta -\frac{\lam^2}{\kap^2}v^2\left[\lam-\sin2\beta\left(\kap-\frac{\alam}{2s}\right)\right]^2
+\Delta m_h^2,  
\ee
%%%%
where $\Delta m_h^2$ denotes loop corrections, and $v=174\gev$ is the tree-level Higgs doublet vev.
The second, NMSSM-induced term on the right-hand side of Eq.~(\ref{eq:mhiggs}) can enhance the
$\hone$ mass by a few \gev\ beyond its
MSSM value, if \lam\ is large enough, $\lam\gsim
0.5$, and $\tanb\lesssim 2$. For larger values of \tanb, the second term on the right-hand side becomes suppressed, 
and the third, negative term becomes significant. Then, the value of the Higgs boson mass can only be enhanced by the loop corrections.  
As
we shall see, when
other constraints are taken into account, especially the ones coming
from the LEP Higgs searches, $b$-physics, the muon anomalous magnetic
moment, \gmtwo, and DM relic density, the aforementioned conditions on \lam\ and \tanb\ 
become hard to satisfy and drastically restrict the parameter space of the
model.

As noted in the Introduction, in the NMSSM either of the two lightest
$CP$-even Higgses could serve the role of the SM-Higgs-like boson
observed at the LHC.  There are thus three interesting possibilities
entailing such a discovery:\footnote{We assume that the third $CP$-even Higgs 
is always heavier than the current experimental reach.}\\

\noindent
Case 1: $\mhone \simeq 125.8$\gev,\\
Case 2: $\mhtwo \simeq 125.8$\gev,\\
Case 3: $\mhone \simeq \mhtwo \simeq 125.8$\gev. \\

\subsubsection{Mass likelihood}

In this paper we analyze all three cases mentioned above. We will
henceforth refer to our selected 125.8\gev\ signal Higgs for either of
the first two cases generically as $h_{\textrm{sig}}$.  We impose a
mass requirement on $h_{\textrm{sig}}$ through a Gaussian likelihood
which takes into account both the theoretical ($\tau$) and
experimental ($\sigma$) errors, following the procedure detailed in
Ref.\cite{Fowlie:2012im},
%%%%
\be
\mathcal{L}_{\textrm{Higgs mass}}(m_{h_{\textrm{sig}}})= \exp\left[-(125.8\gev -
  m_{h_{\textrm{sig}}})^2/2(\tau^2+\sigma^2)\right]\,.
\label{eq:like_sig_conv}
\ee 
%%%%
The theoretical error is due to residual differences between
calculations using different approaches and schemes, and it is
estimated in the literature to be of the order of 2--3\gev\cite{Djouadi:2005gj}. 
We assume $\tau=3\gev$. We use the CMS
determination of the Higgs mass rather than the one by ATLAS,
since it has been most recently updated, and also for consistency with our previous work\cite{Fowlie:2012im}.

%lr The corresponding value of Higgs mass provided by ATLAS is $126\pm0.6\gev$.
%lr The reason why we prefer the CMS value over the ATLAS one is two-fold. First,
%lr the Higgs mass determined by CMS, on account of being somewhat lower,
%lr is likely to therefore easier to reproduce in the CNMSSM. 
%lr Second, as noted in the Introduction, the CMS obtained a combined
%lr global significance of 5$\sigma$ for $\gamma\gamma$, $ZZ$, $WW$, $\tau\tau$ and
%lr $bb$ modes in the data collected at $\sqrt{s} = 8$\tev, while the
%lr combined significance of the discovery quoted by ATLAS is based on a 
%lr combination of the data for the first three of the given
%lr channels at $\sqrt{s}=8$\tev with the 2011 data for all the five channels at
%lr $\sqrt{s}=8$\tev. In our view, the figures presented by of the former are, therefore, more
%lr robust. 

\subsubsection{Cross section likelihood}

In order to be thoroughly consistent with the CMS measurement, we need
to calculate for both Higgs bosons $h_{1,2}$ the reduced cross sections, defined
in the literature as
%%%%
\be
R_{h_i}(X) =  \frac{\sigma(pp\rightarrow h_i)} {\sigma(pp\rightarrow
  h_{\textrm{SM}})}\times \frac{BR(h_i \rightarrow X)}{BR(h_{\textrm{SM}} \rightarrow
  X)}\,,
\label{eq:RX1}
\ee
%%%%
for a given Higgs decay channel, $X$. 

Equation~(\ref{eq:RX1}) can be approximated by  
% \bea
% R^{h_i}(X)&=&\sum_{Y\in\textrm{prod}}\frac{\sigma(pp\rightarrow Y\rightarrow h_{\textrm{SM}})\sigma(Y\rightarrow h_i)}{\sigma(Y\rightarrow h_{\textrm{SM}})\sigma(pp\rightarrow h_{\textrm{SM}})}\times\frac{\textrm{BR}(h_i\rightarrow X)}{\textrm{BR}(h_{\textrm{SM}}\rightarrow X)}\nonumber\\
% &=&\sum_{Y\in\textrm{prod}}\frac{\sigma(pp\rightarrow Y\rightarrow h_{\textrm{SM}})C^2(Y)}{\sigma(pp\rightarrow h_{\textrm{SM}})}\times\frac{\Gamma(h_i\rightarrow X)/\Gamma_{\textrm{tot}}}{\Gamma(h_{\textrm{SM}}\rightarrow X)/\Gamma_{\textrm{tot}}^{\textrm{SM}}}\nonumber\\
% &=&\frac{C^2(X)}{\sigma(pp\rightarrow h_{\textrm{SM}})}\sum_{Y\in\textrm{prod}}C^2(Y)\sigma(pp\rightarrow Y\rightarrow h_{\textrm{SM}})\times\frac{\Gamma_{\textrm{tot}}^{\textrm{SM}}}{\Gamma_{\textrm{tot}}}\nonumber\\
% &=&\frac{C^2(X)}{\sigma(pp\rightarrow
%   h_{\textrm{SM}})}\sum_{Z\in\textrm{decay}}\frac{\textrm{BR}(h_i\rightarrow
%   Z)}{C^2(Z)}\sum_{Y\in\textrm{prod}}C^2(Y)\sigma(pp\rightarrow
% Y\rightarrow h_{\textrm{SM}})\,.\label{Rcalc} \eea 

%%%%
\be
R_{h_i}(X)=\sum_{Y\in\textrm{ prod}}R_{h_i}^Y(X)\mathcal{R}_{\textrm{SM}}(Y)\,,\label{red_x_sec}
\ee
%%%%
where the sum runs over the Higgs production channels $Y$ ($Y=gg$ for
gluon-fusion, $VV$ for vector boson-fusion and Higgs-strahlung off a
$Z$ boson, $t\bar{t}$ and $b\bar{b}$ for associated Higgs production
with top and bottom quarks, respectively), and the ratios
$\mathcal{R}_{\textrm{SM}}(Y)\equiv\sigma(pp\rightarrow Y\rightarrow
h_{\textrm{SM}})/\sigma(pp\rightarrow h_{\textrm{SM}})$ are obtained
from the public tables provided by the LHC Higgs Cross Section Working
Group\cite{HXSWG:2011ti,Dittmaier:2012vm} for $\sqrt{s}=8\tev$.

The reduced cross sections $R^Y_{h_i}(X)$ for the individual
production channels are provided by the mass spectrum generator included in the
\nmssmtools\ package, which we use for our analysis. Alternatively, they can be
expressed in terms of the Higgs reduced couplings $C(X)$ (the
ratio of the couplings of the Higgs boson with a given mass to a pair
of $X$ particles in the NMSSM, to the ones calculated in the SM) and the decay branching ratios
$\textrm{BR}(h_i\rightarrow X)$, which are also provided by the mass
spectrum generator,
%%%%
\bea
R^Y_{h_i}(X)&\equiv&\frac{\sigma(Y\rightarrow
  h_i)}{\sigma(Y\rightarrow
  h_{\textrm{SM}})}\times\frac{\textrm{BR}(h_i\rightarrow
  X)}{\textrm{BR}(h_{\textrm{SM}}\rightarrow X)}\nonumber\\ 
 &=&C^2(Y)\times\frac{\Gamma(h_i\rightarrow X)/\Gamma_{\textrm{tot}}}{\Gamma(h_{\textrm{SM}}\rightarrow X)/\Gamma_{\textrm{tot}}^{\textrm{SM}}}\nonumber\\  
 &=&C^2(Y)C^2(X)\sum_{F\in\textrm{ SM
    decay}}\frac{\textrm{BR}(h_i\rightarrow
  F)}{C^2(F)}\,,\label{Rcalc} \eea
%%%%
where the sum runs over the decay channels $F$ open to the SM Higgs boson.

We require $R_{h_{\textrm{sig}}}(\gamma\gamma)$ and $R_{h_{\textrm{sig}}}(ZZ)$ to
comply with the measured rates, $R_{\textrm{obs}}(\gamma\gamma) =
1.6\pm 0.4$\cite{CMS:2012gu} and
$R_{\textrm{obs}}(ZZ)=0.8^{+0.35}_{-0.28}$\cite{CMS-PAS-HIG-12-041}, 
respectively.\footnote{We only use the dominant decay channels where a
  $\sim 5\sigma$ excess has been observed.} For the $\gamma\gamma$ channel
 the ``signal'' likelihood is taken to be a Gaussian around the measured central value,
%%%%
\be
\label{eq:like_R_sig}
\like_{\textrm{sig}}(\gamma\gamma) = \exp\left[-\left(R_{\textrm{obs}}(\gamma\gamma) -
 R_{h_{\textrm{sig}}}(\gamma\gamma)\right)^2/2 (\sigma_{\gamma\gamma}^2+\tau_{\gamma\gamma}^2)\right]\,,
\ee
%%%%
with $\sigma_{\gamma\gamma}$ being the experimental error. For the
$ZZ$ channel, since the experimental error is asymmetric, the signal
likelihood is defined in terms of
half-Gaussians for the positive error $\sigma^+_{ZZ}$ and negative error
$\sigma^-_{ZZ}$ each, as done in the case of the \bsmumu\ likelihood above.
$\tau_X=15\%
\times R_{h_{\textrm{sig}}}(X)$ is a very conservative estimate of
the theoretical error based on\cite{Baglio:2010ae}, used for every
channel $X$.  

In addition to constraining $h_{\textrm{sig}}$, there is another
crucial aspect of cases 1 and 2, which is that the second of the two
lightest $CP$-even Higgs bosons must have escaped detection at the LHC (or at LEP if very light),
or be ``hidden.''  In the following we refer to it as
$h_{\textrm{hid}}$. In other words, the production rate of
$h_{\textrm{hid}}$ should be less than what the experiments are
currently sensitive to for all $X$. As a result,
$R_{h_{\textrm{hid}}}(X)$ should also be constrained by experimental
data, so that all points where the rate of $h_{\textrm{hid}}$ is large
enough for it to have been observed are rejected in our analysis. For
this purpose, we construct an ``exclusion'' likelihood. Following the
procedure outlined in\cite{deAustri:2006pe} for exclusion bounds we
first define a step function,

%%%%
\be
\mathcal{L}_{\rm{excl}}^{(\rm{step})}\left(m_{h_{\rm{hid}}},R_{h_{\rm{hid}}}(X),\mu_{95}(X)\right) = \left\{\begin{array}{cc}
1 & ~~~\textrm{for}\ R_{h_{\rm{hid}}}(X) \le \mu_{95}(X)\\
0 & ~~~\textrm{for}\ R_{h_{\rm{hid}}}(X) > \mu_{95}(X),
\end{array} \right.\label{exclstep}
\ee
%%%%
where $\mu_{95}(X)$ is the value of the {\it
  signal strength modifier} $\mu(X) \equiv
\sigma_{h_{\rm{hid}}}(X)/\sigma_{h_{\textrm{SM}}}(X)$ that is excluded at 95\%~C.L. by the LHC searches
for a Higgs with a given mass $m_{H} =
m_{h_{\textrm{hid}}}$, obtained from the latest exclusion plots published by the
CMS Collaboration\cite{CMS_plots}.\footnote{We used the latest exclusion
limits provided based on 5.1/fb of data at
$\sqrt{s}=7\tev$ and 12.2/fb of data at $\sqrt{s}=8\tev$ for
the $ZZ$ channel, on 4.9/fb at $\sqrt{s}=7\tev$ and 12.1/fb at
$\sqrt{s}=8\tev$ for the $WW\rightarrow ll\nu\nu$ channel and on 17/fb
at $\sqrt{s}=7-8\tev$ for the $\tau\tau$ channel. Limits for the
$\gamma\gamma$ channel have not been updated by CMS and are, therefore, still based
on a combination of 5.1/fb at $\sqrt{s}=7\tev$ and 5.3/fb at
$\sqrt{s}=8\tev$ of data.} 
The LEP exclusion limits are taken into account by \nmssmtools\ beforehand. 

Then, in order to include the theoretical error on the true values of the reduced
cross section and the Higgs mass,
$\mathcal{L}_{\rm{excl}}^{(\rm{step})}$ is smeared out further by 
convolving it with Gaussian functions centered around their true
theoretical values $\hat{R}_{h_{\rm{hid}}}(X)$ and
$\hat{m}_{h_{\rm{hid}}}$, respectively, so that the exclusion
likelihood now becomes
%%%%
\begin{multline}
\mathcal{L}_{\rm{excl}}^{(\rm{smear})}\left(m_{h_{\rm{hid}}},R_{h_{\rm{hid}}},\mu\right) 
=\int d\hat{m}_{h_{\rm{hid}}} \int
d\hat{R}_{h_{\rm{hid}}}\,\mathcal{L}_{\rm{excl}}^{(\rm{step})} 
\big(\hat{m}_{h_{\rm{hid}}},\hat{R}_{h_{\rm{hid}}},\mu)\\
\times \exp\left[{-\frac{(\hat{m}_{h_{\rm{hid}}} -
    m_{h_{\rm{hid}}})^2}{2\tau^2}}\right]\,\exp\left[{-\frac{(\hat{R}_{h_{\rm{hid}}}
    - R_{h_{\rm{hid}}})^2}{2\tilde{\tau}^2}}\right]\,,
\label{exclsmear}
\end{multline}
%%%%
where the theoretical errors are taken to be $\tau = 3$\gev\ and
$\tilde{\tau} = 15\%\cdot R_{h_{\rm{hid}}}$\cite{Baglio:2010ae},
respectively. The exclusion likelihood is calculated for $X
=\gamma\gamma$, $ZZ$, $WW$ and $\tau\tau$. Finally, in order for our
exclusion criterion to be consistent with our criterion for signal
observation at $125.8\pm3.1\gev$ (with theory and experimental
errors added in quadrature), we further impose the condition
%%%%
\be
\mathcal{L}_{\rm{excl}}(X) = \left\{\begin{array}{cc}
0 & \textrm{for}\ 122.7\gev \le m_{h_{\rm{hid}}} \le 128.9\gev \\
\mathcal{L}_{\rm{excl}}^{(\rm{smear})}(X) & \textrm{elsewhere}.\
\end{array}\right.
\ee
%%%%

For case~3, the likelihood in Eq.~(\ref{eq:like_sig_conv}) is separately
computed for both $\hone$ and $\htwo$. In this scenario, only the combined
production rate for $\hone$ and $\htwo$ needs to be equal to
$R_{\rm{obs}}(X)$. Hence the observation likelihood is now defined as
%%%%
\be
\label{eq:like_R_2sig}
\like_{\rm{obs}}(X) = \exp\left\{-\left[R_{\rm{obs}}(X) -(R_{\hone}(X)+
      R_{\htwo}(X))\right]^2/2 (\sigma_X^2+\tau_X^2)\right\}. 
\ee
%%%%

The values and uncertainties of our Higgs constraints are given in
Table~\ref{tab:exp_constraints}.

%%%%%%%%%%%%%%%%%%%%%%%%%%%%%%%%%%%%%%%%%%%%%%%%%%%%%%%%%%%
\section{\label{Results}Methodology and Results}

The scanned ranges of the CNMSSM parameters along with the type
of distribution of their prior are listed in Table~\ref{tab:2}. 
Also listed in the table are the input ranges of the nuisance parameters included in the scans. 
The sign of \mueff\ is fixed to $+1$ or $-1$ for a given scan.

The reason for choosing the given range of \lam\ is twofold. First, we have checked that allowing lower values of \lam\ 
hardly increases the number of points allowed by the physicality conditions. Second, allowing very small values of \lam\ 
would have most likely driven the scan towards a purely CMSSM-like scenario, thus preventing us from scrutinizing any 
characteristic features of the CNMSSM, particularly for case~1.

%%%%%%%%%%%%%%%%%%%%%%%%%%%%   T   A   B   L   E   %%%%%%%%%%%%%%%%%%%%%%%%%%%%
% Table showing our prior ranges
\begin{table}[t]\footnotesize
\begin{center}
%\input{./Tables/priors}
%%%%%%%%%%%%%%%%%%%%%%%%%%%%%%%%%%%%%%%%%%%%%%%%%%%%%%%%%%%%%%%%%%%%%%%%%%%%%%%%
\begin{tabular}{|l|l|l|l|}
\hline 
%%%%%%%%%%%%%%%%%%%%%%%%%%%%%%%%%%%%%%%%%%%%%%%%%%%%%%%%%%%%%%%%%%%%%%%%%%%%%%%%
CNMSSM parameter & Description & Prior range & Prior distribution \\
%%%%%%%%%%%%%%%%%%%%%%%%%%%%%%%%%%%%%%%%%%%%%%%%%%%%%%%%%%%%%%%%%%%%%%%%%%%%%%%%
\hline %\midrule
%%%%%%%%%%%%%%%%%%%%%%%%%%%%%%%%%%%%%%%%%%%%%%%%%%%%%%%%%%%%%%%%%%%%%%%%%%%%%%%%
\mzero        	& Universal scalar mass          & 100, 4000 	& Log\\
\mhalf		& Universal gaugino mass         & 100, 2000 	& Log\\
\azero        	& Universal trilinear coupling   & $-7000$, 7000 & Linear\\
\tanb	        & Ratio of Higgs vev's            & 1, 62 	& Linear\\
$\lambda$		& Higgs trilinear coupling        & 0.001, 0.7 & Linear \\
%%%%%%%%%%%%%%%%%%%%%%%%%%%%%%%%%%%%%%%%%%%%%%%%%%%%%%%%%%%%%%%%%%%%%%%%%%%%%%%%%
%\hline %\midrule
%\multicolumn{4}{|l|}{additionally in NUHM} \\
%\hline %\cmidrule(r){1-4}
%%%%%%%%%%%%%%%%%%%%%%%%%%%%%%%%%%%%%%%%%%%%%%%%%%%%%%%%%%%%%%%%%%%%%%%%%%%%%%%%%
%\mhu        	& GUT-scale soft mass of $\hu$          & 100, 4000 	& Log\\
%\mhd		& GUT-scale soft mass of $\hd$          & 100, 4000 	& Log\\
%%%%%%%%%%%%%%%%%%%%%%%%%%%%%%%%%%%%%%%%%%%%%%%%%%%%%%%%%%%%%%%%%%%%%%%%%%%%%%%%
\hline 
%%%%%%%%%%%%%%%%%%%%%%%%%%%%%%%%%%%%%%%%%%%%%%%%%%%%%%%%%%%%%%%%%%%%%%%%%%%%%%%%
Nuisance & Description & Central value $\pm$ std. dev. & Prior distribution \\
%%%%%%%%%%%%%%%%%%%%%%%%%%%%%%%%%%%%%%%%%%%%%%%%%%%%%%%%%%%%%%%%%%%%%%%%%%%%%%%%
\hline 
%%%%%%%%%%%%%%%%%%%%%%%%%%%%%%%%%%%%%%%%%%%%%%%%%%%%%%%%%%%%%%%%%%%%%%%%%%%%%%%%
$M_t$           	& Top quark pole mass 	& $173.5\pm1.0$ 	& Gaussian\\
\mbmbmsbar 	& Bottom quark mass	& $4.18\pm0.03$ 	& Gaussian\\
\alphasmzms	& Strong coupling	& $0.1184\pm0.0007$   	& Gaussian\\
%%%%%%%%%%%%%%%%%%%%%%%%%%%%%%%%%%%%%%%%%%%%%%%%%%%%%%%%%%%%%%%%%%%%%%%%%%%%%%%%
\hline 
\end{tabular}
\caption{Priors for the parameters of the
  model and for the SM nuisance parameters used in our scans. Masses
  and \azero\ are in GeV.}
\label{tab:2}
\end{center}
\end{table}
%%%%%%%%%%%%%%%%%%%%%%%%%%%%%%%%%%%%%%%%%%%%%%%%%%%%%%%%%%%%%%%%%%%%%%%%%%%%%%%%

The analysis was performed using the package BayesFITS which engages
several external, publicly available tools: for sampling it uses
MultiNest\cite{Feroz:2008xx} with 4000 living points, evidence
tolerance factor set to 0.5, and sampling efficiency equal to 0.8.
Mass spectrum, along with \delmbs, is computed with \nmssmtools\
v3.2.1\cite{NMSSMTools} and passed via SUSY Les Houches Accord format to \superiso\
v3.3\cite{superiso} to calculate \brbxsgamma, \brbsmumu, \brbutaunu,
and \deltagmtwomususy.  DM observables, such as the relic density and
direct detection cross sections, are calculated with \micromegas\
2.4.5\cite{micromegas}.

Below we will present the results of our scans as one-dimensional (1D)
or two-dimensional (2D) marginalized posterior pdf maps of parameters
and observables. For example, in evaluating a posterior pdf for a
given parameter, we marginalize over all of the model's other parameters
and the SM nuisance parameters, as described in detail in
Refs.\cite{Fowlie:2012im,Roszkowski:2012uf}.

%%%%%%%%%%%%%%%%%%%%%%%%%%%%%%

Notice that when discussing the results of the global scan for case~1
it will become apparent that this case presents a remarkable CMSSM-like
behavior.  It would therefore be natural to try to compare those
results with our recent CMSSM analysis\cite{Fowlie:2012im}.  In doing
so, one needs to take into account the differences between the
numerical codes and constraints adopted in both studies. We summarize
them here.

1. In the present study we use NMSPEC (included in NMSSMTools) for calculating the supersymmetric
spectrum, while in\cite{Fowlie:2012im} we used \softsusy.  We have
repeatedly cross-checked the spectra obtained in the MSSM limit of the
NMSSM with the ones generated by \softsusy, finding some differences,
especially with respect to loop corrections giving the largest values
of the lightest Higgs
mass.
In some regions of the parameter space the difference between the two
generators can amount to a maximum of $\sim0.5$--1\gev.\footnote{The best agreement between \softsusy\ and 
NMSSMTools is obtained by setting the flag \texttt{precision for the Higgs masses} to zero which, therefore, 
was chosen as the default setting for our calculations.} Given the experimental and
theoretical uncertainties in the Higgs mass, such difference translates
into $\sim0.25$ units of $\chi^2$, which is not significant for the
purpose of the global scan.

2. In this paper we use the value of \brbsmumu\ measured at LHCb\cite{Aaij:2012ct}, which has been incorporated in the 
likelihood as described in Sec.~\ref{sec:bsmlike}. The SM rate rescaled
by the time-dependent asymmetries is now
$\brbsmumu_{SM}=(3.53\pm0.38)\times10^{-9}$\cite{Mahmoudi:2012un}, which is a value more
appropriate for comparison with the experimental rate than the
unscaled, $\sim3.2\times10^{-9}$, one.

3. We have updated the nuisance parameters \mtpole\ and \mbmbmsbar\
%lr *** $m_t$ and $m_b$
following\cite{Beringer:1900zz}; see Table~\ref{tab:2}. The upgrade
in \mtpole\ has significant implications for \mhone. 
%lr the overall scale of loop
The leading one-loop corrections to the Higgs mass squared are given by 
%%%%
\begin{equation}
  \Delta\mhl^2=\frac{3\mtop^4}{4\pi^2v^2}\left[\ln\left(\frac{\msusy^2}{\mtop^2}\right)+\frac{X_t^2}{\msusy^2}\left(1-\frac{X_t^2}{12\msusy^2}\right)\right],
\label{1loopmh}  
\end{equation}
%%%%
where $m_t$ is the running top quark mass,\footnote{Note that the running
  top quark mass is related to the pole mass through the formula
  given in Eq.~(10) of Ref.\cite{Spira:1997dg}.}  \msusy\ is the geometrical average of the physical stop
masses, $\msusy\equiv\sqrt{m_{\tilde{t}_1}m_{\tilde{t}_2}}$, and $X_t=A_t-\mueff\cot\beta$. 
Since $\Delta\mhl^2\propto m_t^4$ it is
now easier to generate Higgs masses in agreement with the experimental
values.  In particular, as we highlighted in\cite{Fowlie:2012im}, a
Higgs mass compatible with the observed excess at 126\gev\ was rather
difficult to achieve over the CMSSM parameter space. That tension has
now become somewhat reduced, and we will show below that the
correct Higgs mass can be obtained in the CMSSM limit of the CNMSSM.
   
% *** \lr{got down to here} ****\\

%We can now move on to discussion the impact of the constraints.

\subsection{Impact of the relic density}

To set the ground for the presentation of our numerical results, we
first comment on the role of the relic density of DM in selecting
favored regions. The relic density is a strong constraint, since it is a
positive measurement (in contrast to a limit) with a rather small
experimental uncertainty (Table~\ref{tab:exp_constraints}). On top of it, it
is well known that in unified SUSY models with neutralino LSP the
corresponding abundance \abundchi\ is typically too large, or in other
words, its annihilation in the early Universe is ``generically'' too
inefficient, so that specific mechanisms for enhancing it are therefore needed.
They, however, are only applicable in specific SUSY configurations. As
a result, in most cases the regions
of high probability in the global posterior will reflect one or more of the
regions of parameter space where \abundchi\ is close to the measured
relic density of DM. The regions that are still allowed by direct SUSY searches are
as follows: 

1. The stau-coannihilation (SC) region\cite{Ellis:1998kh}. As is known, in constrained SUSY models, like the
C(N)MSSM, this is a narrow strip at a sharp angle to the \mhalf\
axis. The values of \azero\ and \tanb\ are also constrained, as only
for $|\azero|$ not exceeding $\sim2$--3\tev\ and \tanb\ not too large
does the mass of the stau become light enough to be comparable with the neutralino mass, 
but not so light as to make it
the LSP. Values of \mhalf\ that are excessively large, on the other
hand, can suppress the annihilation cross section\cite{Nihei:2002sc}.
After other relevant constraints are included, the parameters of
interest are, therefore, $\mzero\lesssim600\gev$,
$\mhalf\lesssim1000\gev$, $|\azero|\lesssim3000\gev$ and, when the
neutralino is close to 100\% bino, $\tanb<30$. A similar effect can
also be obtained for large \azero\ with the stop $\tilde{t}_1$
replacing $\tilde{\tau}_1$\cite{Ellis:2001nx}.

2. The $A$-funnel (AF) region, where neutralinos annihilate through
the resonance with the lightest pseudoscalar\cite{Drees:1992am}. This mechanism can occur over broad ranges of the (\mzero,
\mhalf) plane where the pseudoscalar mass is close to twice the
neutralino LSP mass, and is enhanced by large \tanb\ ($\tanb\gsim 35$) and positive
\azero. 
   
3. The focus point/hyperbolic branch (FP/HB) region\cite{Chan:1997bi,Feng:1999zg}, where the
annihilation cross section can be enhanced by an increased higgsino
component of the neutralino. For this to occur, $\mu$ (or \mueff\ in
the NMSSM) must be of the order of a few hundred GeV, and $\tanb$
cannot be too large, $\tanb\lesssim45$. In the (\mzero, \mhalf) plane
the condition corresponds to the region where $\mzero\gg\mhalf$.

\subsection{Impact of the Higgs mass\label{Sec:HiggsImpact}}

%%%%%%%%%%%%%%%%%%%%%%%%%   F   I   G   U   R   E   %%%%%%%%%%%%%%%%%%%%%%%%%%%%
% 2 by 1: left: plot of 1d pdf of mhl,
% right: chi2 vs mhl
\begin{figure}[ht!]
\centering
\subfloat[]{%
\label{fig:-a}%
\includegraphics[width=0.50\textwidth]{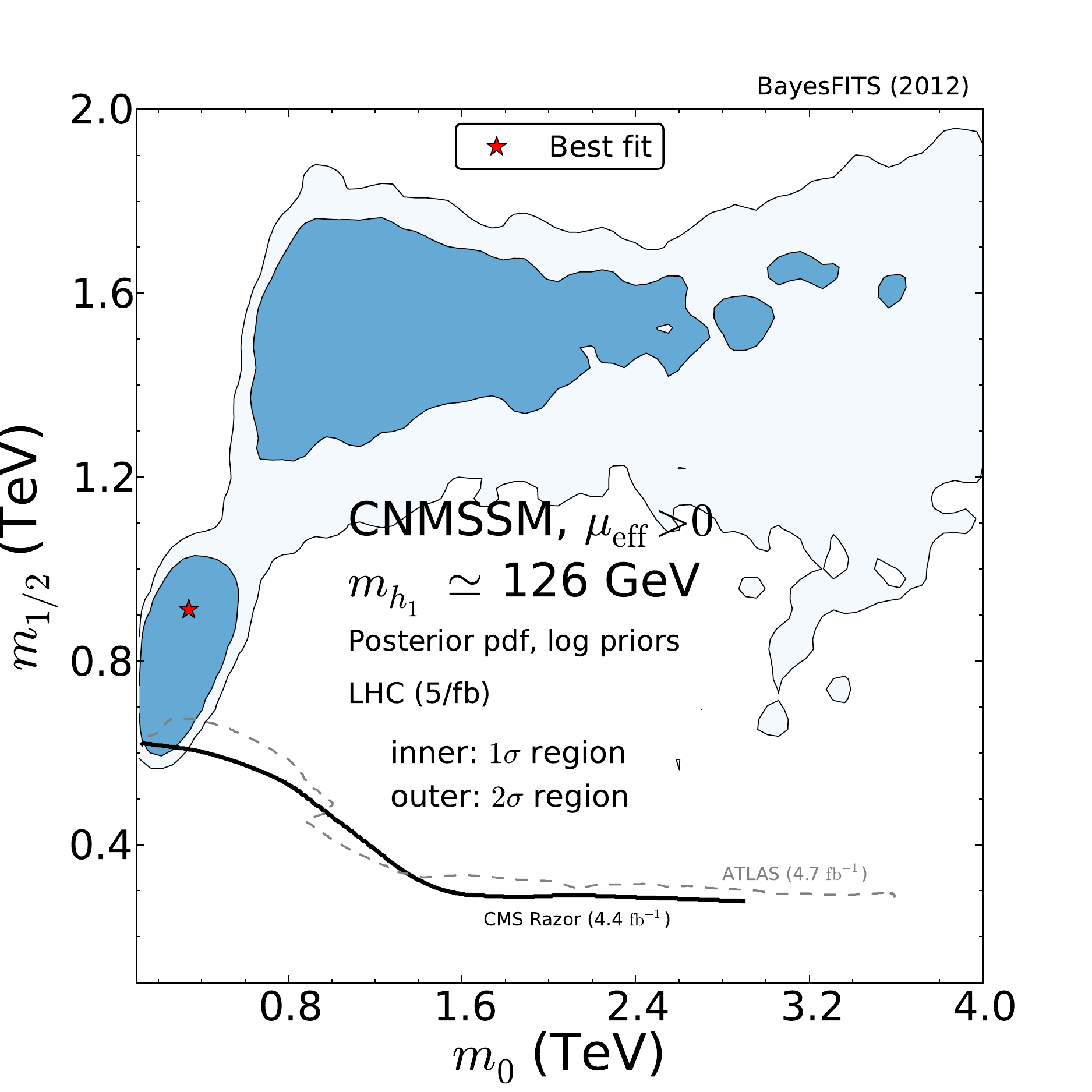}
}%
%\hspace{1pt}%
\subfloat[]{%
\label{fig:-b}%
\includegraphics[width=0.50\textwidth]{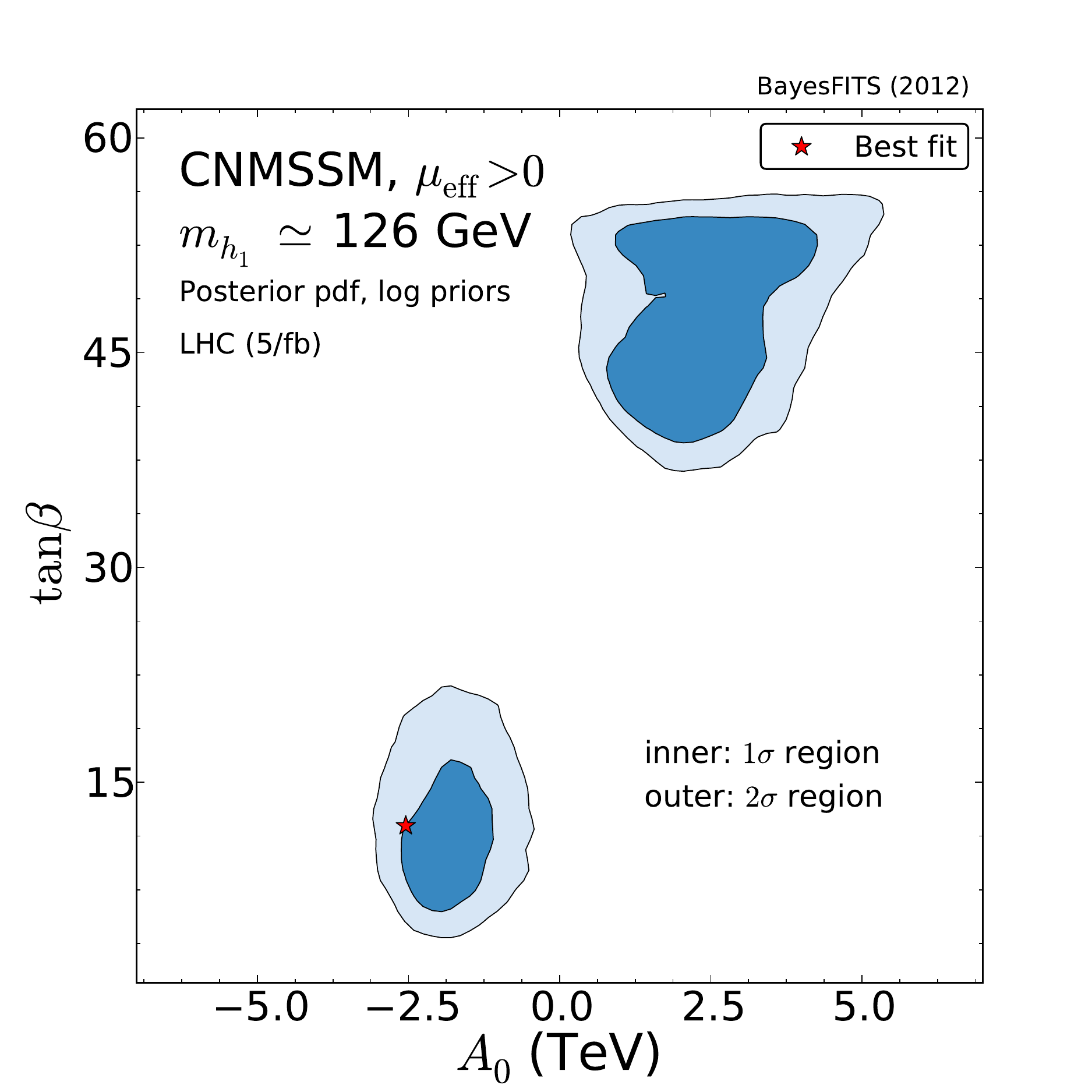}
}%
\caption[]{Marginalized 2D posterior pdf in \subref{fig:-a} the (\mzero, \mhalf)
  plane and \subref{fig:-b} the (\azero, \tanb) plane of the CNMSSM for case~1, constrained by the
  experiments listed in Table~\ref{tab:exp_constraints}. The 68\% credible regions are shown in dark blue, and the 95\% credible regions in light blue.
  The solid black (dashed gray) line shows the CMS razor (ATLAS hadronic) 95\%~C.L. exclusion bound.}
\label{fig:cnmssm_params}
\end{figure} 
%%%%%%%%%%%%%%%%%%%%%%%%%%%%%%%%%%%%%%%%%%%%%%%%%%%%%%%%%%%%%%%%%%%%%%%%%%%%%%%%

The measurement of the Higgs mass has added an
important additional constraint on unified SUSY models. Below we will
be discuss in turn the three cases listed earlier in Sec.~\ref{Higgslike}.

\paragraph{Case~1.} In Fig.~\ref{fig:cnmssm_params}\subref{fig:-a} we
show the marginalized posterior pdf in the (\mzero, \mhalf) plane of
the CNMSSM for case~1, obtained by imposing simultaneously all the
constraints shown in Table~\ref{tab:exp_constraints}.  In these and
the following plots the Bayesian 68\% ($1\sigma$) credible regions are
indicated in dark blue and the 95\% ($2\sigma$) credible regions in
light blue.  Notice that the regions of high probability are located
above the CMS razor limit, which we implemented in the likelihoood as
described in Sec.~\ref{sec:razor}, and which is shown in the plots as
a solid black line.  In case~1 the role of the SM-like Higgs is played
by the lightest $CP$-even scalar (almost purely $H_u^0$-like), while
\htwo\ (almost purely $H_d^0$-like) and $h_3$ (almost purely
singletlike) are usually much heavier and decoupled. This case is
thus expected to present features very similar to the CMSSM.
 
In Fig.~\ref{fig:cnmssm_params}\subref{fig:-a} one can see two main 68\% credibility regions:
the SC region on the lower left side, and the AF region on the top part of the plot. 
As is also the case in the CMSSM,
besides giving the correct relic abundance, the SC region shows also
the better fit to the Higgs mass, $\mhone=124.5\gev$.  This is
because, as we explained in Ref.\cite{Fowlie:2012im}, in the SC region
\azero\ can easily be negative without spoiling the relic abundance
constraint.  Large negative values of \azero\ are necessary to drive
the parameter $A_t$ to even larger negative values at the EW scale,
thus making the stop mixing contribution to the loop corrections of
the Higgs mass maximal. Note as well that the best-fit point is also
located in the SC region.\footnote{We postpone further discussion of 
the best-fit points until Sec.~\ref{bestfit:sec}.} In Fig.~\ref{fig:cnmssm_params}\subref{fig:-b} we show the
marginalized posterior in the (\azero, \tanb) plane. The high
probability ``island'' at negative \azero\ and $\tanb\lesssim25$
corresponds to the SC region.

In the CNMSSM the SC region appears to be more
extended relative to the CMSSM\cite{Fowlie:2012im}, and somewhat larger Higgs masses also
seem preferred, as we will show below. The increased relevance of the SC
region is due to the fact that it is now much easier to obtain values of the Higgs mass closer to the correct one. 
This could be mistakenly thought to
be a specific feature of the CNMSSM extended Higgs sector. We have
checked that this is not the case: the Higgs mass is simply quite
sensitive to the increased central value of the top mass, as we
explained at the beginning of this section.

%%%%%%%%%%%%%%%%%%%%%%%%%   F   I   G   U   R   E   %%%%%%%%%%%%%%%%%%%%%%%%%%%%
% 2 by 1: left: plot of 1d pdf of mhl,
% right: chi2 vs mhl
\begin{figure}[ht!]
\centering
\subfloat[]{%
\label{fig:-a}%
\includegraphics[width=0.50\textwidth]{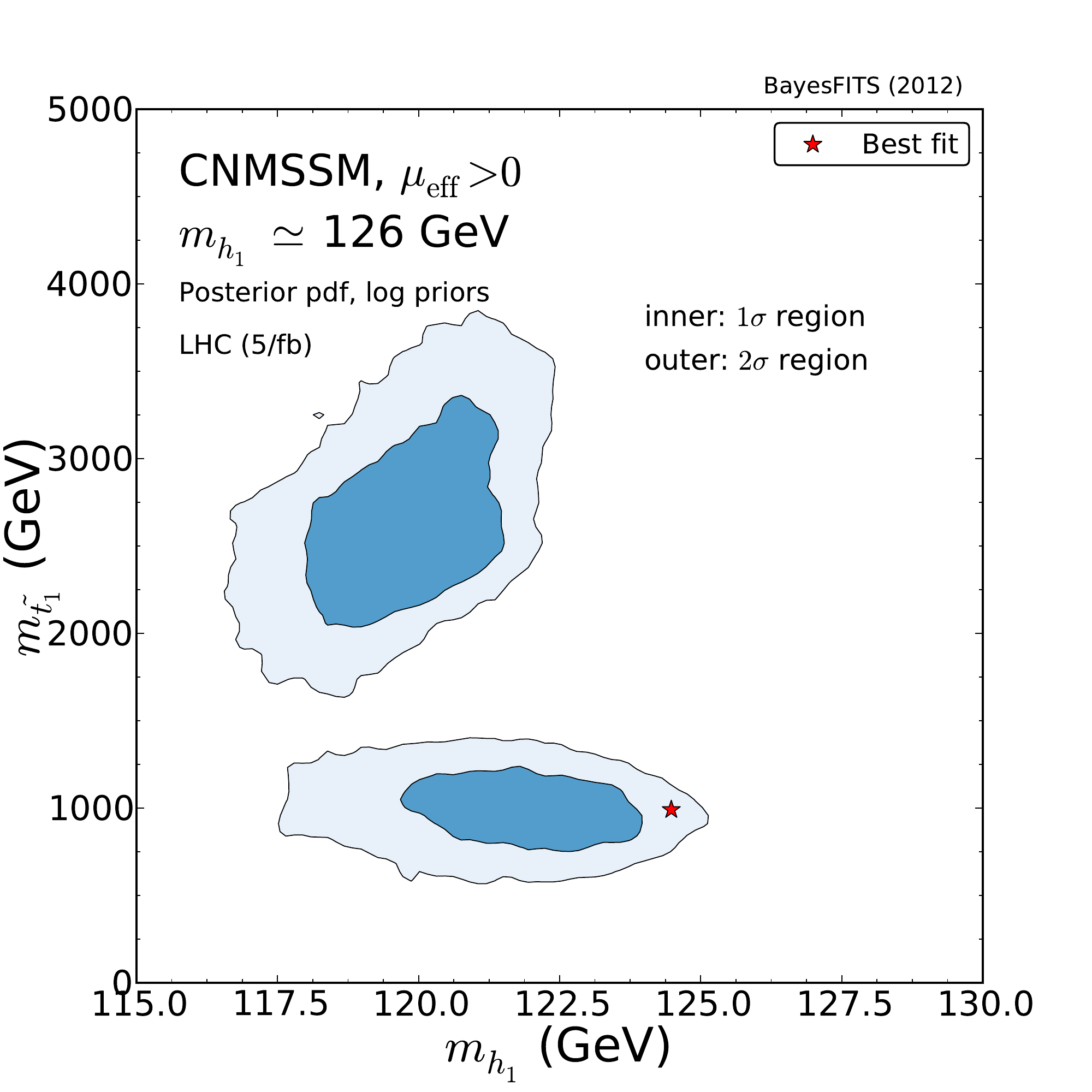}
}%
%\hspace{1pt}%
\subfloat[]{%
\label{fig:-b}%
\includegraphics[width=0.50\textwidth]{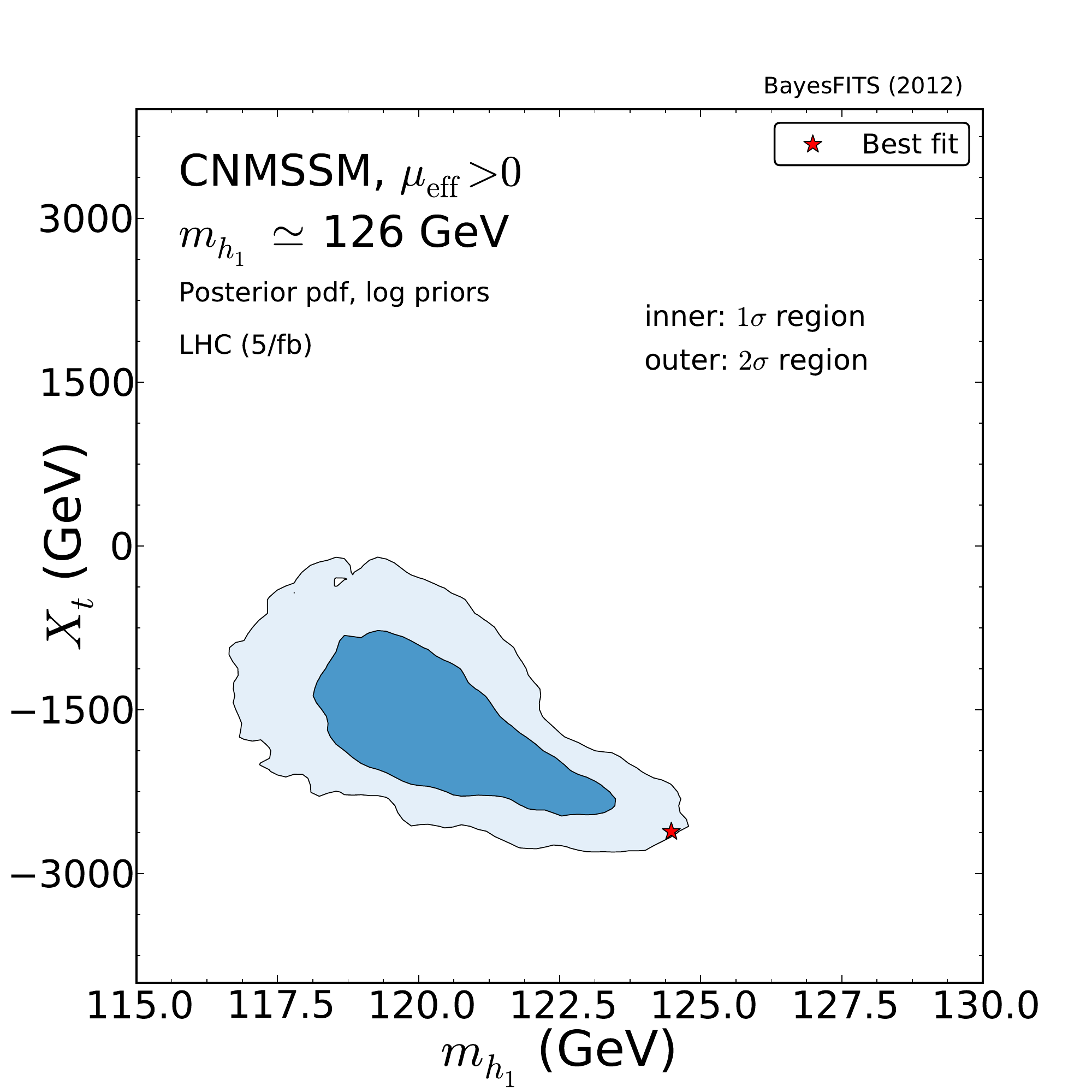}
}%

\caption[]{Marginalized 2D posterior pdf in \subref{fig:-a} the (\mhone, $m_{\tilde{t}_1}$) plane and \subref{fig:-b} the (\mhone, $X_t$)
 plane, for the CNMSSM constrained by the
 experiments listed in Table~\ref{tab:exp_constraints} for case~1. The color code is the same as in Fig.~\ref{fig:cnmssm_params}.}
\label{fig:higgsloop}
\end{figure} 
%%%%%%%%%%%%%%%%%%%%%%%%%%%%%%%%%%%%%%%%%%%%%%%%%%%%%%%%%%%%%%%%%%%%%%%%%%%%%%%%

In Figs.~\ref{fig:higgsloop}\subref{fig:-a} and
\ref{fig:higgsloop}\subref{fig:-b}, we show how the constraints
affect the main observables responsible for the loop corrections to
the Higgs boson mass. In
Fig.~\ref{fig:higgsloop}\subref{fig:-a}, we show the posterior
in the (\mhone, $m_{\tilde{t}_1}$) plane.
Notice that the lightest stop does \textit{not} have to be excessively heavy in
the SC region, $m_{\tilde{t}_1}\simeq1\tev$, since large stop mixing
compensates for smaller
\msusy. The correct Higgs mass is obtained since
$|X_t|\simeq2.5\tev$, as can be seen in
Fig.~\ref{fig:higgsloop}\subref{fig:-b} where we plot the posterior in
the (\mhone, $X_t$) plane. 
On the other hand, one can see that in both figures the best-fit point lies outside of the 68\% credibility regions of the 
marginalized 2D pdf, when the latter is projected to the plane of these observables. 
This is a feature not uncommon in Bayesian analyses where the credibility regions map the ``volume'' 
of scan points satisfying well enough a certain set of constraints, rather than representing isocontours 
of the likelihood function, as is the case in frequentist analyses. 
The fact that the best-fit point is situated outside of the region of highest posterior probability 
for the considered observables simply tells us that, while for this point all the constraints are very well satisfied, 
it is also not very likely to obtain a similarly good value of the observable for similar choices of the input parameters. 
Particularly, in the case of Figs.~\ref{fig:higgsloop}\subref{fig:-a} and
\ref{fig:higgsloop}\subref{fig:-b}, the majority of the points that satisfy all the constraints in the 
SC region present a Higgs mass in the range 120--123\gev, while for the other regions it is even less.
  
%%%%%%%%%%%%%%%%%%%%%%%%%   F   I   G   U   R   E   %%%%%%%%%%%%%%%%%%%%%%%%%%%%
% 2 by 1: left: plot of 1d pdf of mhl,
% right: chi2 vs mhl
\begin{figure}[t]
\centering
\subfloat[]{%
\label{fig:-a}%
\includegraphics[width=0.50\textwidth]{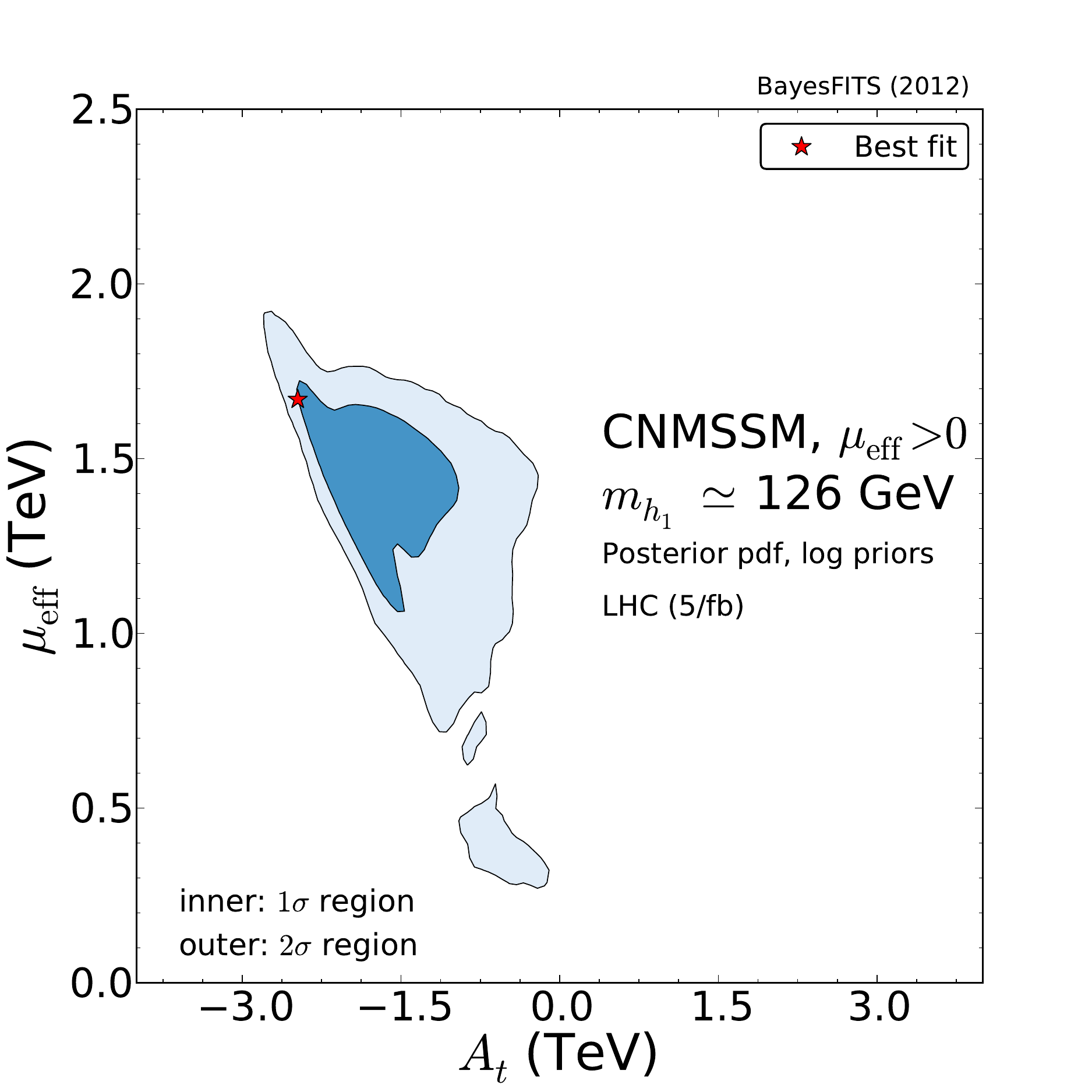}
}%
%\hspace{1pt}%
\subfloat[]{%
\label{fig:-b}%
\includegraphics[width=0.50\textwidth]{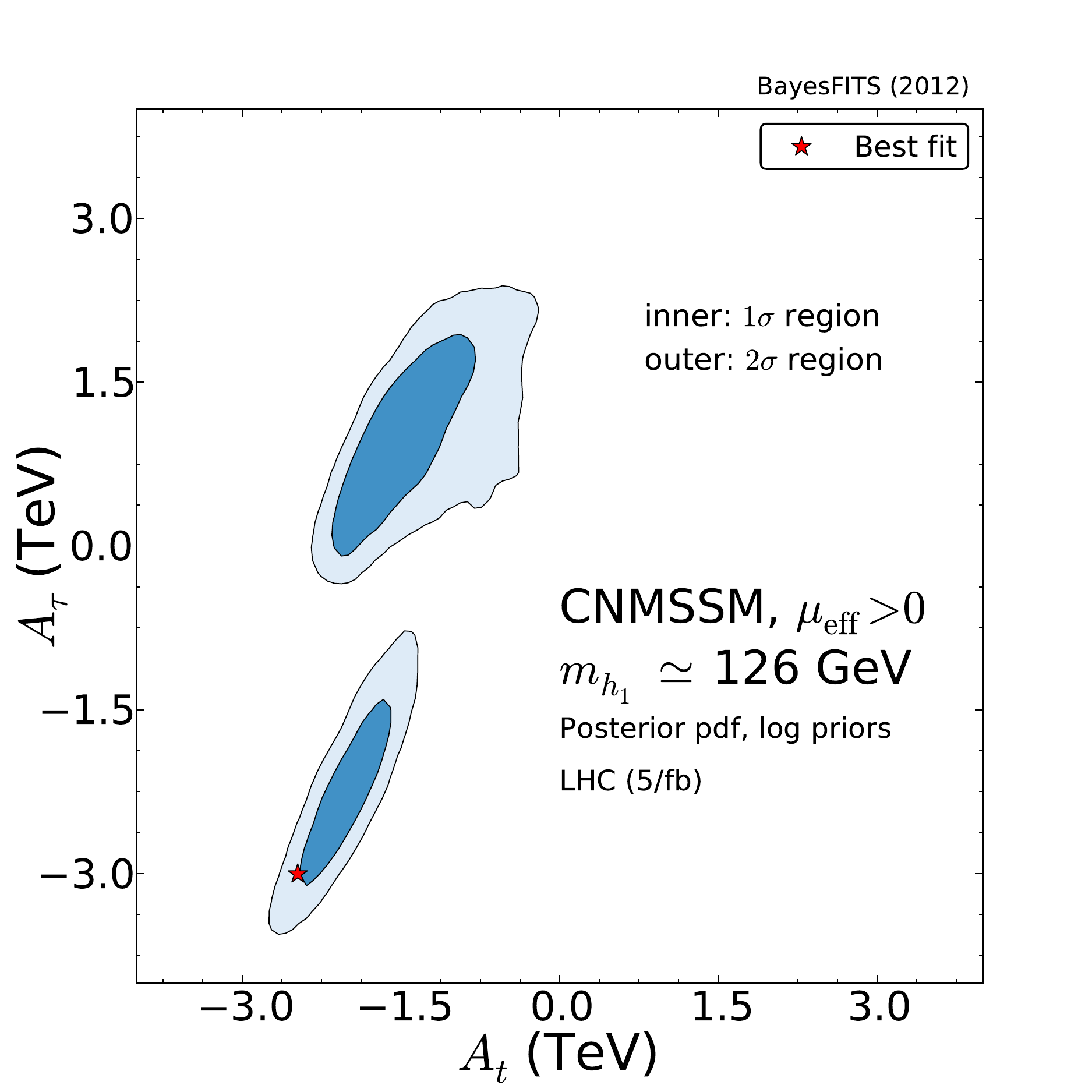}
}%
\caption[]{Marginalized 2D posterior pdf in \subref{fig:-a} the ($A_t$, \mueff) plane and \subref{fig:-b} the ($A_t$, $A_{\tau}$) plane, 
for the CNMSSM constrained by the experiments listed in Table~\ref{tab:exp_constraints} for case~1. 
The color code is the same as in Fig.~\ref{fig:cnmssm_params}.}
\label{fig:cnmssm_at}
\end{figure} 
%%%%%%%%%%%%%%%%%%%%%%%%%%%%%%%%%%%%%%%%%%%%%%%%%%%%%%%%%%%%%%%%%%%%%%%%%%%%%%%%

Figure~\ref{fig:cnmssm_params}\subref{fig:-a} also shows that the
Higgs mass constraint favors the part of the AF region situated at
$\mhalf\gsim1500\gev$ for a broad range of \mzero\ values. Besides, in
Fig.~\ref{fig:cnmssm_params}\subref{fig:-b} one can see that in the
(\azero, \tanb) plane the AF region spans a large range of positive
\azero\ values and \tanb\ is constrained to $\tanb\gsim40$. In
this region one-loop corrections to the Higgs mass are driven up by a
large stop mass ($m_{\tilde{t}_1}\sim2$--4\tev, as shown in
Fig.~\ref{fig:higgsloop}\subref{fig:-a}) rather than large stop mixing
which, given the preferred values of \azero, is minimal (see
Fig.~\ref{fig:higgsloop}\subref{fig:-b}). Large values of \msusy\ push
the Higgs mass close to the experimentally observed value, but not enough to reach it 
(for even larger \msusy, \maone\ becomes too heavy and resonant annihilation of neutralinos is not efficient). This can
be interpreted as a sign of some tension between the relic abundance
and Higgs mass constraints in the AF region, which was investigated
for the CMSSM in our previous paper\cite{Fowlie:2012im}. The tension
persists in the CNMSSM.

Only a limited fraction of the FP/HB region (which appears as a narrow 95\%-credibility ``tail'' at $\mzero\gg\mhalf$) survives,
as was the case for the CMSSM, despite the fact that the
relic density is well satisfied over there. As we pointed out
in\cite{Fowlie:2012im}, some tension with the 126\gev\ Higgs boson mass arises not only in
the AF but also in the FP/HB region where \mueff\ is smaller than
anywhere else. In fact, small values of \mueff\ can be obtained only for relatively small values of $|A_t|$, 
as can be seen in
Fig.~\ref{fig:cnmssm_at}\subref{fig:-a}, where we show the
posterior in the weak scale parameters $A_t$ and \mueff\ (notice that
$A_t\approx X_t$ over all
parameter space). For the chosen range of \mzero, in the FP/HB
region \msusy\ cannot be very large either, so that the correct Higgs
mass cannot be reached. The region of high posterior thus moves up towards
larger \mhalf, where the
neutralino has still a non-negligible higgsino component, but \msusy\
is large enough to give the correct Higgs mass. 

It is worth pointing out that $\azero\simeq 0$ is not realized in the
CNMSSM. This is because the lightest pseudoscalar $\aone$ becomes
nonphysical for such values. The mass of $\aone$ is, for moderate and
large values of \tanb, well approximated by $m_{\aone}^2\approx -3\kap
s \akap$\cite{Miller:2003ay}. \alam\ and \akap\ are unified to \azero\
at the GUT scale, and \akap\ barely runs, since the one-loop
contribution to its $\beta$-function is negligible. As a consequence,
in the CNMSSM \kap\ and \akap\ have always opposite signs and there
are no points in the scan with $\azero=0$ or $\kap=0$.

We present in Fig.~\ref{fig:cnmssm_at}\subref{fig:-b} the marginalized posterior pdf in the ($A_t$, $A_{\tau}$) plane. 
The parameters show a clear linear correlation, which in the SC region (bottom left corner) results in large negative values for both observables, due to the 
fact that the correct Higgs boson mass requires large stop mixing, as discussed above.

%%%%%%%%%%%%%%%%%%%%%%%%%   F   I   G   U   R   E   %%%%%%%%%%%%%%%%%%%%%%%%%%%%
% 2 by 1: left: plot of 1d pdf of mhl,
% right: chi2 vs mhl
\begin{figure}[t]
\centering
%\subfloat[]{%
%\label{fig:-a}%
\includegraphics[width=0.50\textwidth]{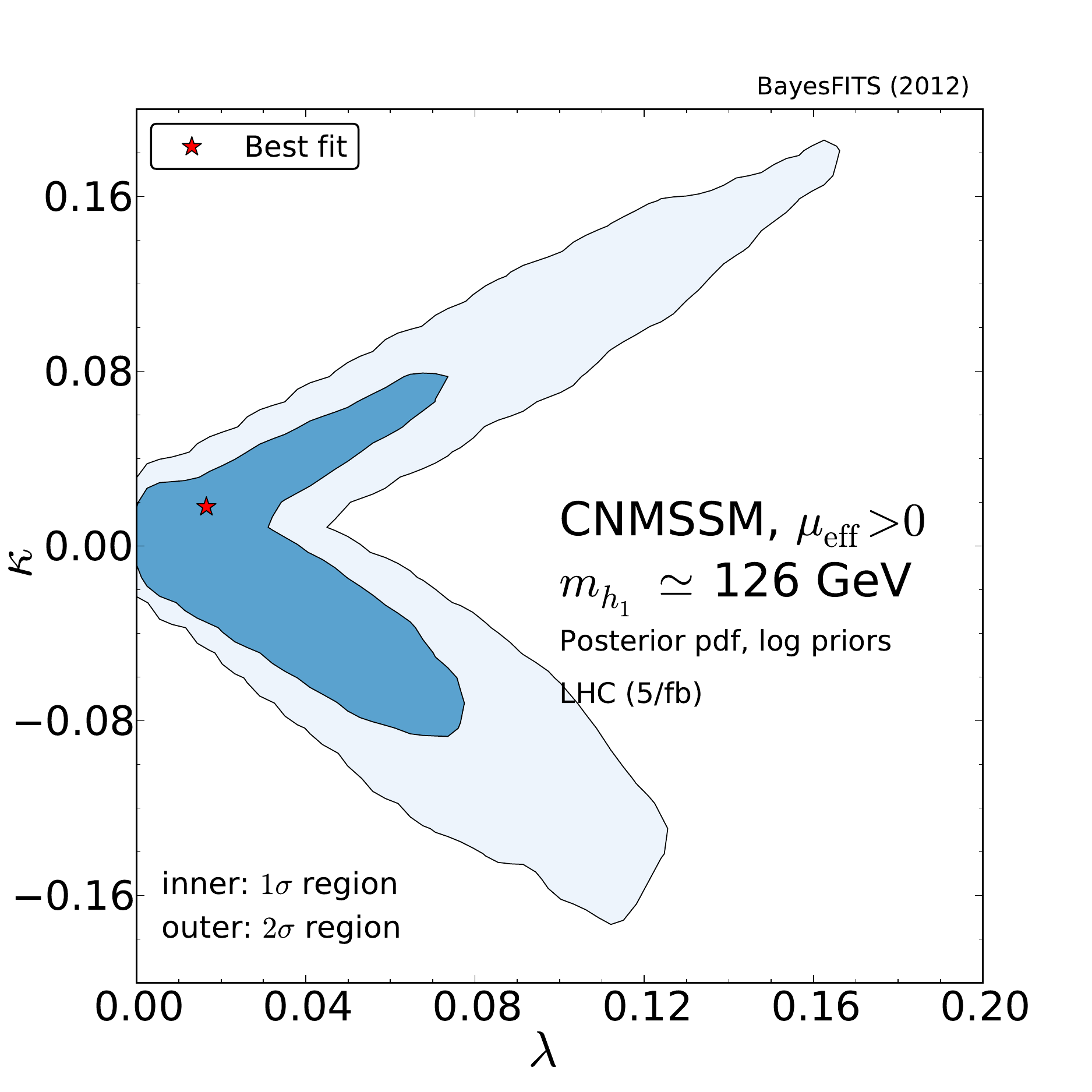}
%}%
\caption[]{Marginalized 2D posterior pdf in the (\lam, \kap)
 plane, for the CNMSSM constrained by the
 experiments listed in Table~\ref{tab:exp_constraints} for case~1. The color code is the same as in Fig.~\ref{fig:cnmssm_params}.}
\label{fig:cnmssm_lamkap}
\end{figure} 
%%%%%%%%%%%%%%%%%%%%%%%%%%%%%%%%%%%%%%%%%%%%%%%%%%%%%%%%%%%%%%%%%%%%%%%%%%%%%%%%

Figure~\ref{fig:cnmssm_lamkap} shows the 2D posterior
in the (\lam, \kap) plane.  One can notice the known correlation
between \kap\ and \lam\cite{LopezFogliani:2009np}, as in general $|\kap|$ cannot
exceed a given value of \lam\ by too much without causing a Landau
pole in RGE running~(see e.g.\cite{Miller:2003ay} for details) and, on the other hand, \lam\ cannot exceed a
given value of $|\kap|$ by too much because the increased mixing between the doublet and singlet 
states lowers \mhone\ much below the observed value, where the likelihood becomes negligible, and eventually below the LEP bounds\cite{Hugonie:2007vd}. 
In the SC region \kap\ is positive since \akap\ is negative, while in the
AF and FP/HB regions these parameters switch signs. Hence, the upper
branch shown in Fig.~\ref{fig:cnmssm_lamkap}
corresponds to the SC region, and the lower one to both the AF and the
FP/HB regions.
Notice also that the upper branch is narrower than the lower one, 
due to the fact that SC occurs for limited ranges of the \mzero\ ($\lesssim500\gev$), \azero\ (large and negative) and \mueff\ ($\sim1$--2\tev) parameters. 
As a consequence, the condition for electroweak symmetry breaking requires \kap\ to be very close to \lam\ in the SC region. 
      
At 95\% credibility, the branches stretching along $|\kap|\sim\lam$ show an absolute upper bound in \lam, which
changes slightly for regions of parameter space characterized by different mechanisms to reduce the relic abundance.
Specifically one can see that, on the one hand, $\lam\lesssim0.16$ for the upper
branch. Again, for the ranges of parameters required by SC, a larger \lam\ lowers \mhone\ below 
the scale preferred by the likelihood function because of the increased singlet-doublet 
mixing\cite{Ellwanger:2009dp}. On the other hand, in the AF region (right-hand part of the lower branch) \lam\ cannot exceed 
0.12 by too much as this would increase \maone, pushing it off the resonance with the lightest neutralino. 
In the FP region (left-hand part of the lower branch) the bound is on negative \kap, and is due to the chosen 
ranges for \mzero\ ($\mzero\le 4000\gev$) and the fact that \mueff\ tends to be small there, as shown in Fig.~\ref{fig:cnmssm_at}\subref{fig:-a}. Simply, 
when \kap\ is significantly below $-0.1$, it becomes difficult to obtain electroweak symmetry breaking.

We want to reemphasize that, when the global
constraints are considered, case~1 presents a very
CMSSM-like character.  The parameter \lam\ is small, and its effect on
\mhone\ is insignificant, given that $\tanb \gsim
4$ over all of the regions of high posterior probability, as shown in
Fig.~\ref{fig:cnmssm_params}\subref{fig:-b}. In the SC region the lightest Higgs mass
can assume values close to 126\gev\ more
easily than what was observed in our work on the CMSSM, but the reason lies in the
updated value of the top mass used for the present analysis.

%%%%%%%%%%%%%%%%%%%%%%%%%   F   I   G   U   R   E   %%%%%%%%%%%%%%%%%%%%%%%%%%%%
% 2 by 1: left: plot of 1d pdf of mhl,
% right: chi2 vs mhl
\begin{figure}[t]
\centering
\subfloat[]{%
\label{fig:-a}%
\includegraphics[width=0.50\textwidth]{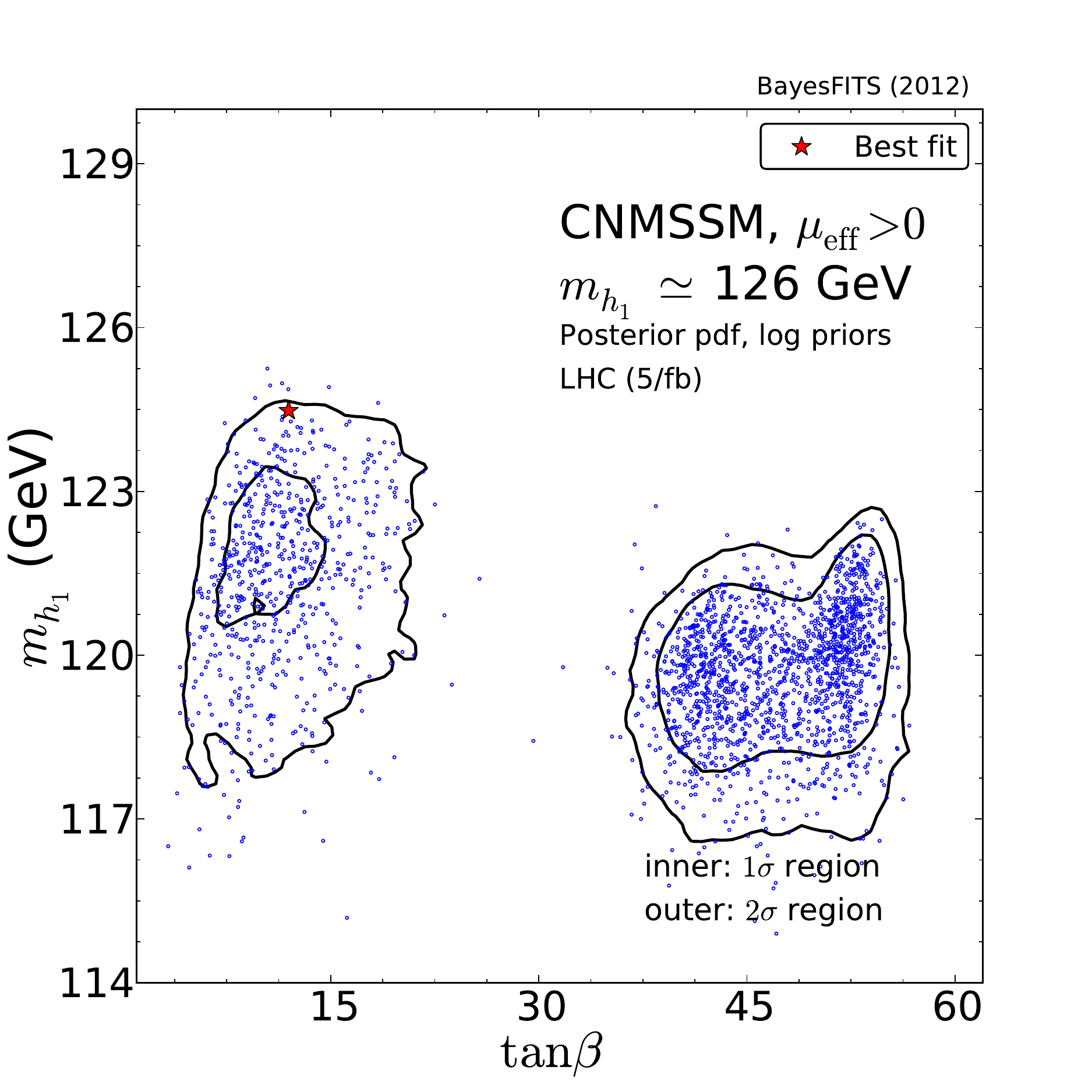}
}%
%\hspace{1pt}%
\subfloat[]{%
\label{fig:-b}%
\includegraphics[width=0.50\textwidth]{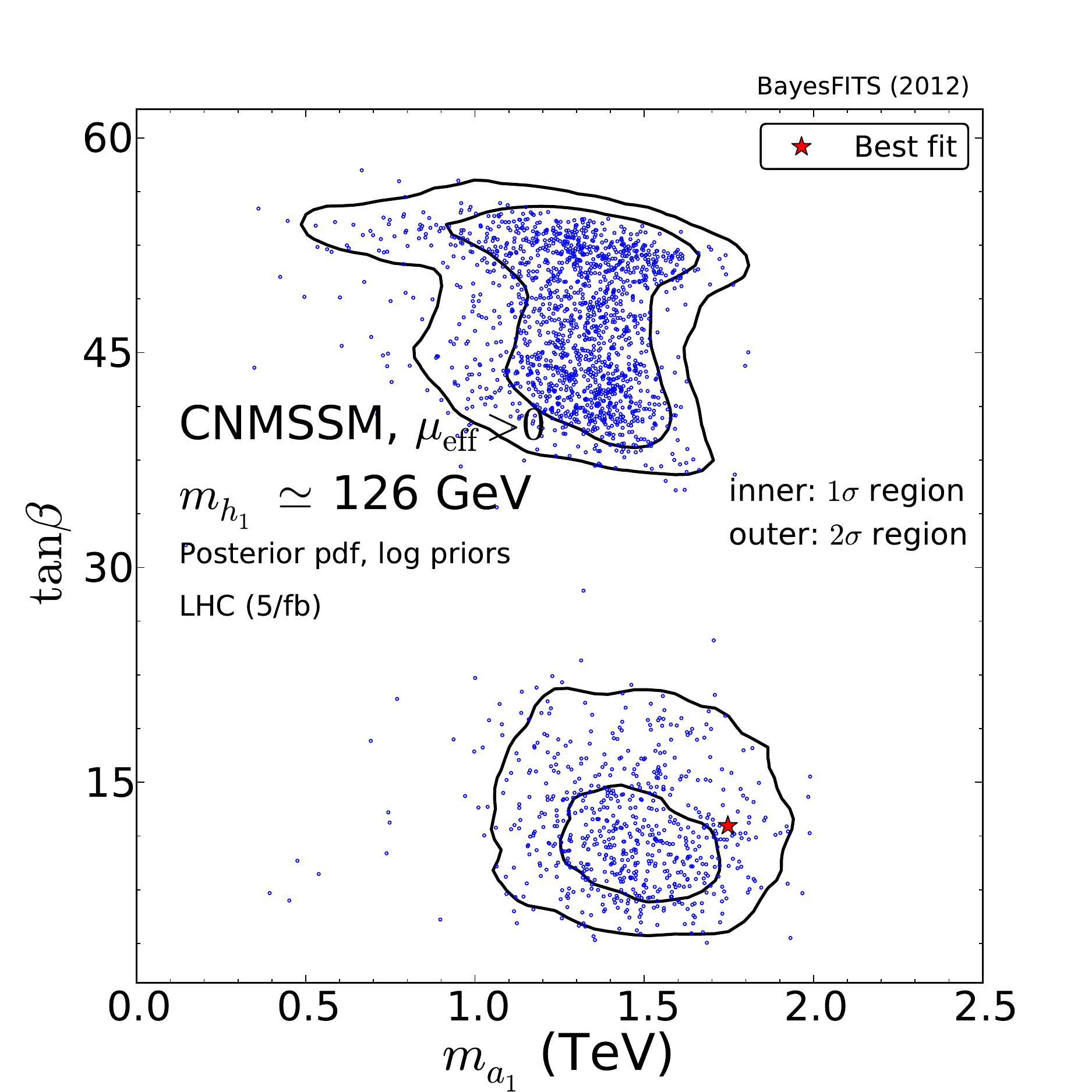}
}%
\caption[]{Marginalized 2D posterior pdf in \subref{fig:-a} the (\tanb, \mhone) plane and \subref{fig:-b} the (\maone, \tanb)
 plane of the CNMSSM constrained by the
 experiments listed in Table~\ref{tab:exp_constraints} for case~1. 
 The outer contours enclose the 95\% credibility regions and the inner contours the 68\% credibility regions. 
 A distribution of samples uniformly selected from our nested sampling chain is superimposed.}
\label{fig:cnmssm_lightHigs}
\end{figure} 
%%%%%%%%%%%%%%%%%%%%%%%%%%%%%%%%%%%%%%%%%%%%%%%%%%%%%%%%%%%%%%%%%%%%%%%%%%%%%%%%

In Fig.~\ref{fig:cnmssm_lightHigs}\subref{fig:-a} we show the
marginalized posterior in the (\tanb, \mhone) plane and in Fig.~\ref{fig:cnmssm_lightHigs}\subref{fig:-b} the posterior in 
the ($m_{\aone}$, \tanb) plane. We also overlap a distribution of samples uniformly selected from our nested sampling chain.
Notice that the density of samples reflects their relative posterior probability.
One can see in Fig.~\ref{fig:cnmssm_lightHigs}\subref{fig:-a} that, in the SC region (left island) the tension between the
correct Higgs mass and the other constraints is much
ameliorated. Figure~\ref{fig:cnmssm_lightHigs}\subref{fig:-b} shows
confirmation of the CMSSM-like nature of this case, as the pattern of
high posterior mirrors the one found in Ref.\cite{Fowlie:2012im}.

%lr \bigskip
\paragraph{Case~2.} The NMSSM allows more freedom in the Higgs sector
than the MSSM, due to the extended number of parameters. Even in its
partially constrained version, there is a possibility of obtaining a
light $\htwo$, which we found to be a mixture of $H_u$ and $S$
fields. A non-negligible singlet component creates the difference in
the Higgs sector between this scenario and case~1. In the rest of
this section we analyze the consistency of an \htwo\ signal with the
observed excess at the LHC.

%%%%%%%%%%%%%%%%%%%%%%%%%   F   I   G   U   R   E   %%%%%%%%%%%%%%%%%%%%%%%%%%%%
% 2 by 1: left: plot of 1d pdf of mhl,
% right: chi2 vs mhl
\begin{figure}[t]
\centering
\subfloat[]{%
\label{fig:-a}%
\includegraphics[width=0.50\textwidth]{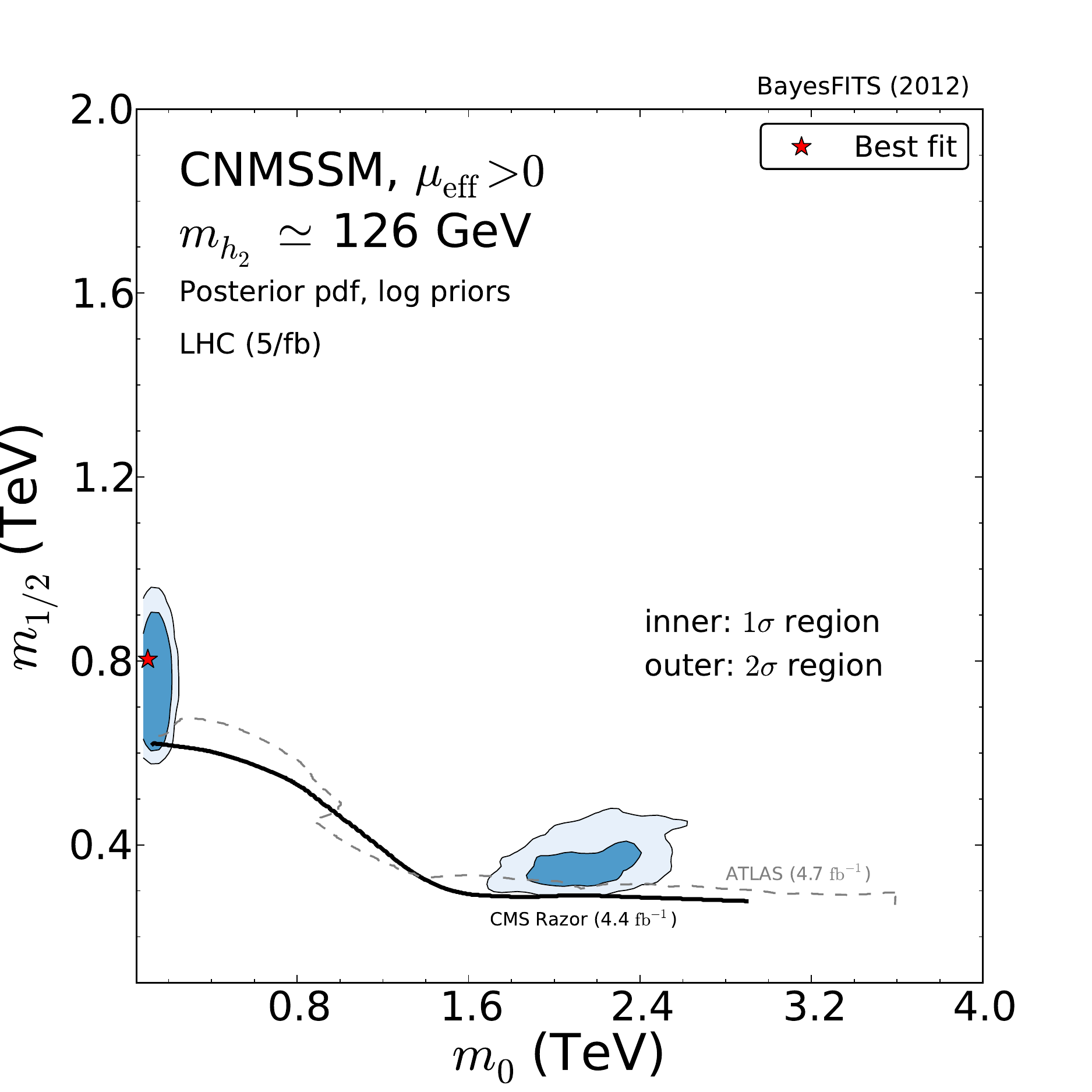}
}%
%\hspace{1pt}%
\subfloat[]{%
\label{fig:-b}%
\includegraphics[width=0.50\textwidth]{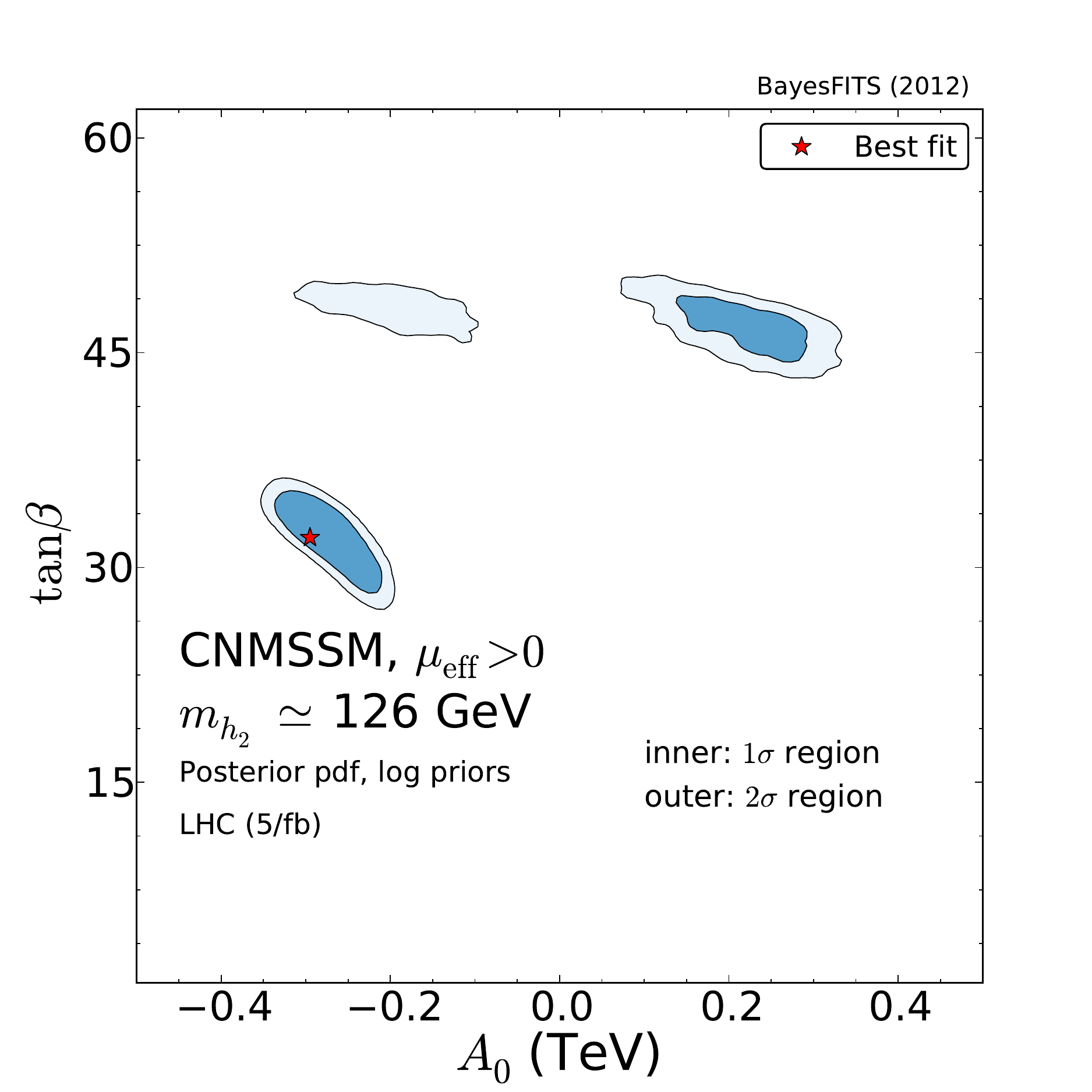}
}%
\caption[]{Marginalized 2D posterior pdf in \subref{fig:-a} the (\mzero, \mhalf) plane and \subref{fig:-b} the (\azero, \tanb)
 plane of the CNMSSM constrained by the
 experiments listed in Table~\ref{tab:exp_constraints} for case~2. The color code is the same as in Fig.~\ref{fig:cnmssm_params}.}
\label{fig:cnmssm_momhalf}
\end{figure} 
%%%%%%%%%%%%%%%%%%%%%%%%%%%%%%%%%%%%%%%%%%%%%%%%%%%%%%%%%%%%%%%%%%%%%%%%%%%%%%%%

In Fig.~\ref{fig:cnmssm_momhalf}\subref{fig:-a} we show the posterior
pdf in the (\mzero, \mhalf) plane for case~2.  The favored parameter
space is now drastically reduced with respect to case~1. Only the SC region, 
where the best-fit point is located, and the FP/HB region survive the requirement of having $\mhtwo\approx 126\gev$.  

In Fig.~\ref{fig:cnmssm_momhalf}\subref{fig:-b} we show the posterior
pdf in the (\azero, \tanb) plane.  In both regions the range preferred
for \azero\ does not extend much from zero, 
$-400\gev<\azero<400\gev$.\footnote{Note that a gap around $\azero\simeq 0$ in
Fig.~\ref{fig:cnmssm_momhalf}\subref{fig:-b} comes from the
physicality condition for $a_1$, as discussed above.}  The reason
is that $|\akap|$ is now limited to values less than 400\gev\ by our
requirement on the mass of $\htwo$.  Specifically, 
under the assumption of a moderate-to-large \tanb\ and as long as the 
parameters \mueff, $\kap s$, and $A_{(\kap,\lam)}$ do not exceed the EW
scale by too much, a good approximation to the masses of the
two lightest $CP$-even Higgs bosons at the tree level was found
in\cite{Miller:2003ay}:
%%%%
\begin{equation}
m_{h_{1,2}}^2\approx\frac{1}{2}\left\{M_Z^2+4(\kap s)^2+\kap s
  \akap\mp\sqrt{\left[M_Z^2-4(\kap s)^2-\kap s
      \akap\right]^2+4\lam^2 v^2\left[2\lam s-\left(\alam+\kap s\right)\sin 2\beta\right]^2}\right\}\,.\label{mh12form}  
\end{equation}
%%%%
In Eq.~(\ref{mh12form}), the second term under the square root is suppressed with respect to the EW scale because \lam\
is small, as we shall see below. One can see that in the regime where $|\kap| s < M_Z$, the mass of $h_2$ is
of order $M_Z$ and \mhone\ scales as $|\kap| s$. Thus, the physicality
condition $\mhone^2\geq0$ translates into the approximate relation
$|\akap|\lesssim 4|\kap| s$. On the other hand, in the regions where $|\kap| s > M_Z$, 
$\mhtwo\sim \kap s$ and $\mhone\sim M_Z$. Values of \mhtwo\ much
greater than 126\gev\ are disfavored by the likelihood function, so
that $|\kap| s$ presents an upper bound, which translates into an
upper bound on $|\akap|$.

Since in most of the parameter space $s$ is very large, and \lam\ and
\kap\ are correlated, the scan also shows upper bounds for \lam\ and
\kap, in a fashion very similar to what is shown in Fig.~\ref{fig:cnmssm_lamkap} for case~1.  
For case~2, in the SC region \lam\ is very small, $\lam\lesssim0.01$, while
in the FP/HB region it can assume slightly larger values, $\lam\lesssim 0.04$.
Obviously, the upper bound on $|\kap| s$ does not depend on any particular position in the 
parameter space, but the bound on \kap\ (or \lam) does, and is affected particularly
by \mueff.  In the SC region $\mueff>600\gev$ while in
the FP/HB region $\mueff\simeq 200\gev$.
%In NMSSMTools \mueff\ is determined by the requirement of EW-symmetry breaking once the other parameters are fixed by the global constraints. 
%Thus, \mueff\ assumes the same values as the parameter $\mu$ in the CMSSM. As a consequence, while \lam\ is a free parameter, the singlet vev $s$ is determined point by point.

Given the strong constraint on $|\azero|$ placed by the mass of \htwo, the only way of obtaining the relic density though coannihilation 
with the lightest stau is if the lightest neutralino is \textit{a nearly pure singlino} and very light. 
This is exactly what is observed in the SC region and, as a consequence, in that region \lam\ and \kap\ are bounded to be much smaller
than in case~1 (the neutralino mass matrix with the
convention used in this paper can be found, e.g.,
in\cite{LopezFogliani:2009np}), \azero\ can only be negative, and \tanb\ assumes
larger values ($\tanb\sim 30-35$, favored also by other constraints,
e.g. \deltagmtwomu) than in the same region of case~1, or of the
CMSSM. 
For smaller values of \tanb, or a positive \azero, the neutralino would be mostly bino and the lightest stau would be the LSP.

%%%%%%%%%%%%%%%%%%%%%%%%%   F   I   G   U   R   E   %%%%%%%%%%%%%%%%%%%%%%%%%%%%
% 2 by 1: left: plot of 1d pdf of mhl,
% right: chi2 vs mhl
\begin{figure}[t]
\centering
\subfloat[]{%
\label{fig:-a}%
\includegraphics[width=0.50\textwidth]{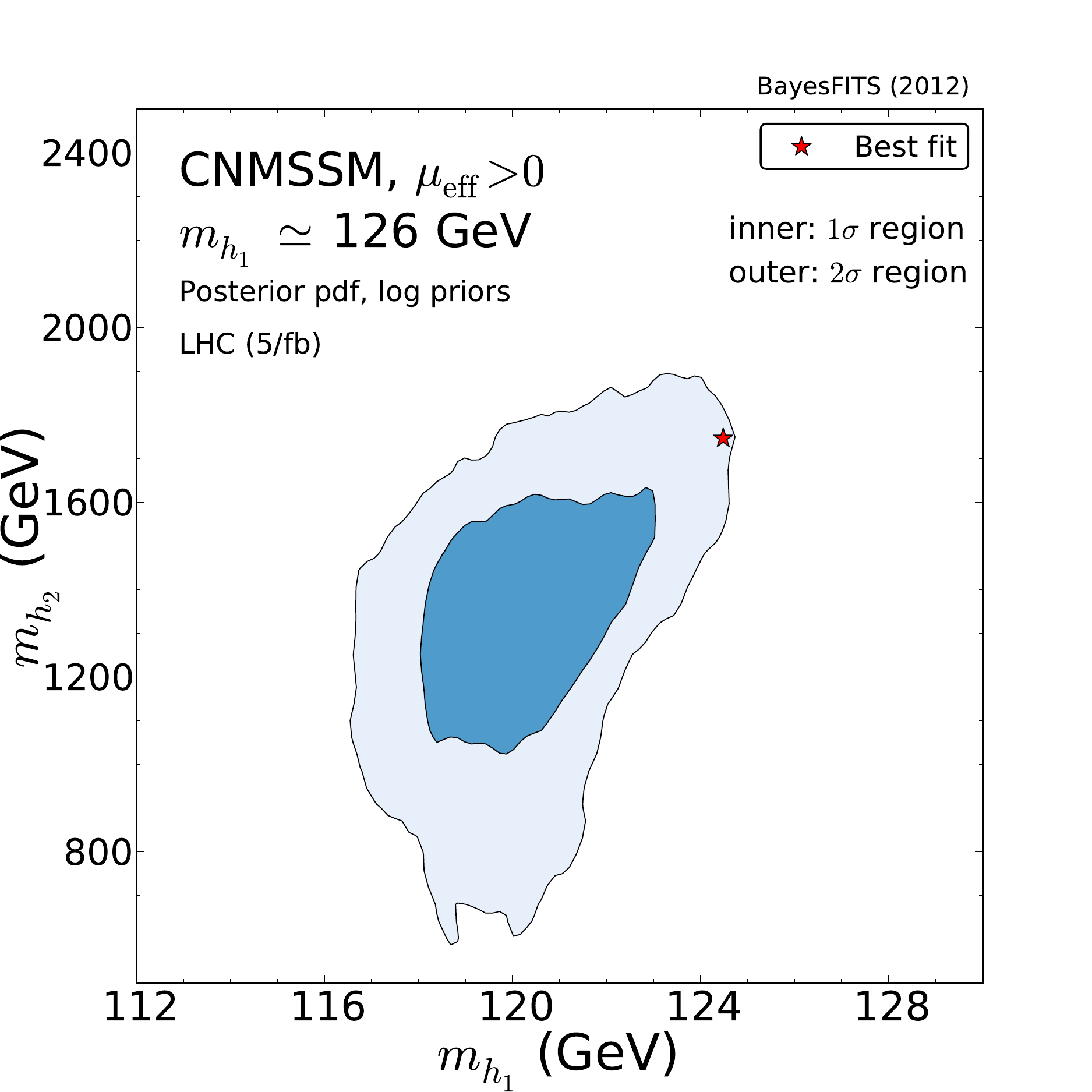}
}%
%hspace{1pt}%
\subfloat[]{%
\label{fig:-b}%
\includegraphics[width=0.50\textwidth]{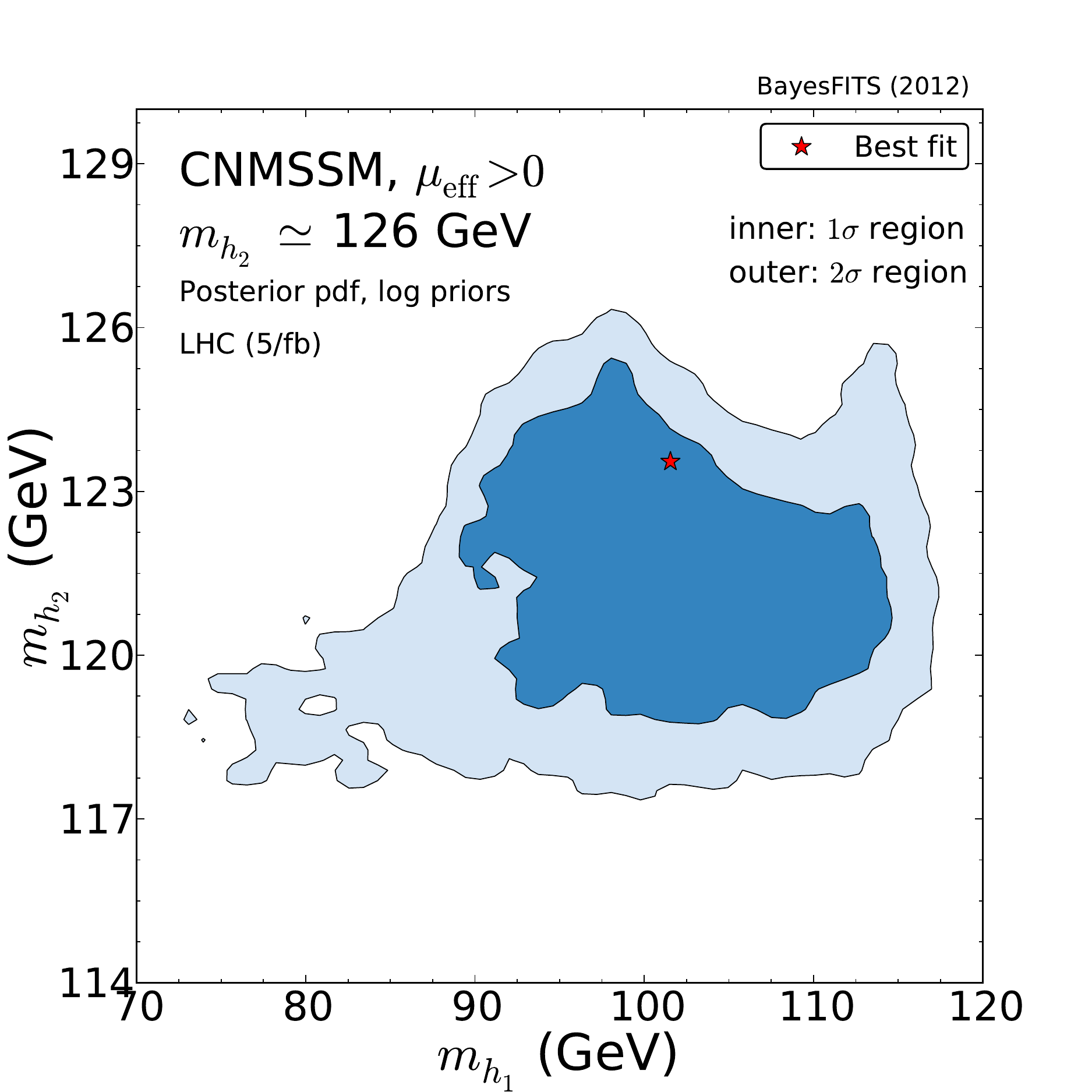}
}%
\caption[]{\subref{fig:-a} Marginalized 2D posterior pdf in the (\mhone, \mhtwo) plane of the CNMSSM constrained by the
 experiments listed in Table~\ref{tab:exp_constraints} for case~1. \subref{fig:-b} Marginalized posterior pdf in the (\mhone, \mhtwo) plane of the CNMSSM for case~2.
 The color code is the same as in Fig.~\ref{fig:cnmssm_params}.}
\label{fig:cnmssm_other_stuff}
\end{figure} 
%%%%%%%%%%%%%%%%%%%%%%%%%%%%%%%%%%%%%%%%%%%%%%%%%%%%%%%%%%%%%%%%%%%%%%%%%%%%%%%%

In Figs.~\ref{fig:cnmssm_other_stuff}\subref{fig:-a} and
\ref{fig:cnmssm_other_stuff}\subref{fig:-b} we show the 2D posterior
in the (\mhone, \mhtwo) plane for case~1 and case~2, respectively. One
can see that in case~1 $\mhtwo\gg\mhone$, while in case~2 the 68\%
credibility region shows a preference for
$90\gev\lesssim\mhone\lesssim115\gev$.

Since in case~2 $\mhtwo\simeq 126\gev$, $m_{\aone}$ assumes values quite
small in the favored regions, $m_{\aone}<500\gev$.  While it
might appear that such a case is excluded by the recent
$H\rightarrow\tau\tau$ searches at the LHC\cite{CMS-PAS-HIG-12-007},
we point out that the limit does not apply to this case, as the
relevant couplings for $\aone\rightarrow\tau\tau$ and
$\aone\rightarrow bb$ are suppressed by the pseudoscalar mixing angle
$\theta_P$, i.e., when $|\cos\theta_P|<1$\cite{Mahmoudi:2010xp}.  We
have checked that $|\cos\theta_P|$ is very close to zero over the
regions of interest.

%lr \bigskip
\paragraph{Case~3.} Case~3 (two degenerate light Higgs bosons) is in fact a
subset of case~2 when it comes to the preferred parameter space, which
is uniquely determined by the requirement of $\mhtwo\simeq 126\gev$ as
was explained above. The 2D posterior pdf's for the input parameters are in this case almost indistinguishable from
the ones showed in Figs.~\ref{fig:cnmssm_momhalf}\subref{fig:-a} and
\ref{fig:cnmssm_momhalf}\subref{fig:-b}, and the best-fit point is again situated in the SC region. We therefore refrain from showing them explicitly.
However, this case presents
also some characteristic features:

$\bullet$ $\hone$ is a mixture of $\hd$ and $\hu$, with a predominance
of the latter, while $\htwo$ is mainly singletlike, with a small
fraction of $\hd$.

%$\bullet$ The importance of the SC region is enhanced with respect to
%Case 2 (68\% credibility region instead of 95\%). It is due to the
%fact that the lightest Higgs with the mass close to 125.3\gev\ is
%easier to obtain in the SC region than in other regions of parameter
%space, as we have already commented while discussing Case 1. The
%best-fit point however remains in the FP/HB region.

$\bullet$ Both \lam\ and \kap\ are now much more limited and very
close to zero. The reason for this is simple: in order to have the two
lightest $CP$-even Higgs bosons almost degenerated in mass, one needs to
minimize the difference $\mhtwo^2-\mhone^2$ from Eq.~(\ref{mh12form}).
This yields negligible values for the parameter \kap\ and, 
consequently, also for \lam. Clearly, as a consequence, the singlino nature of the LSP in the SC region is a feature confirmed for case~3.

\subsection{\label{subsec:Rs} Impact of the cross section rates}

%%%%%%%%%%%%%%%%%%%%%%%%%   F   I   G   U   R   E
%%%%%%%%%%%%%%%%%%%%%%%%%%%%
% 2 by 1: left: plot of 1d pdf of mhl,
% right: chi2 vs mhl
\begin{figure}[t]
\centering
\subfloat[]{%
\label{fig:Rs_a}%
\includegraphics[width=0.50\textwidth]{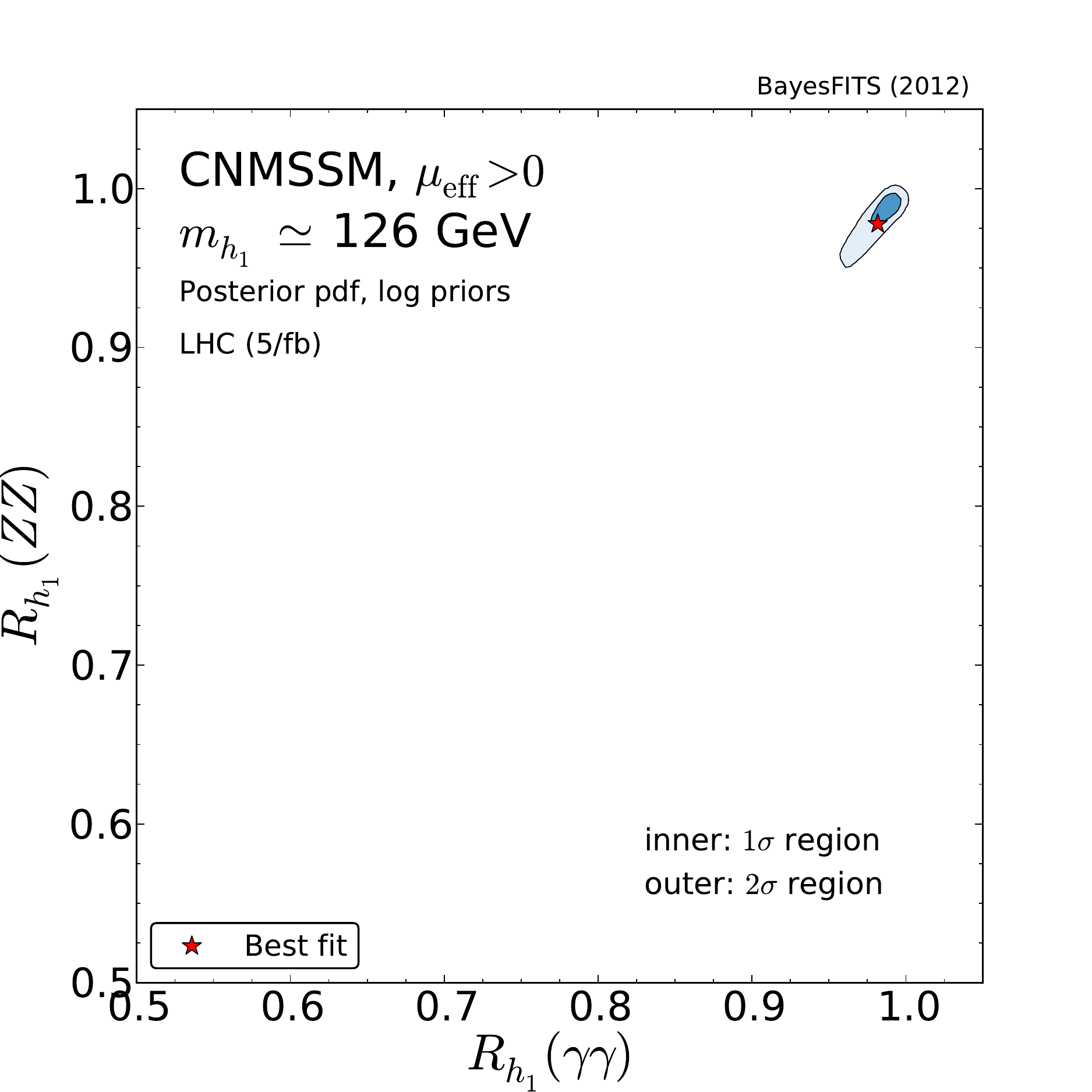}
}%
%\hspace{1pt}%
\subfloat[]{%
\label{fig:Rs_b}%
\includegraphics[width=0.50\textwidth]{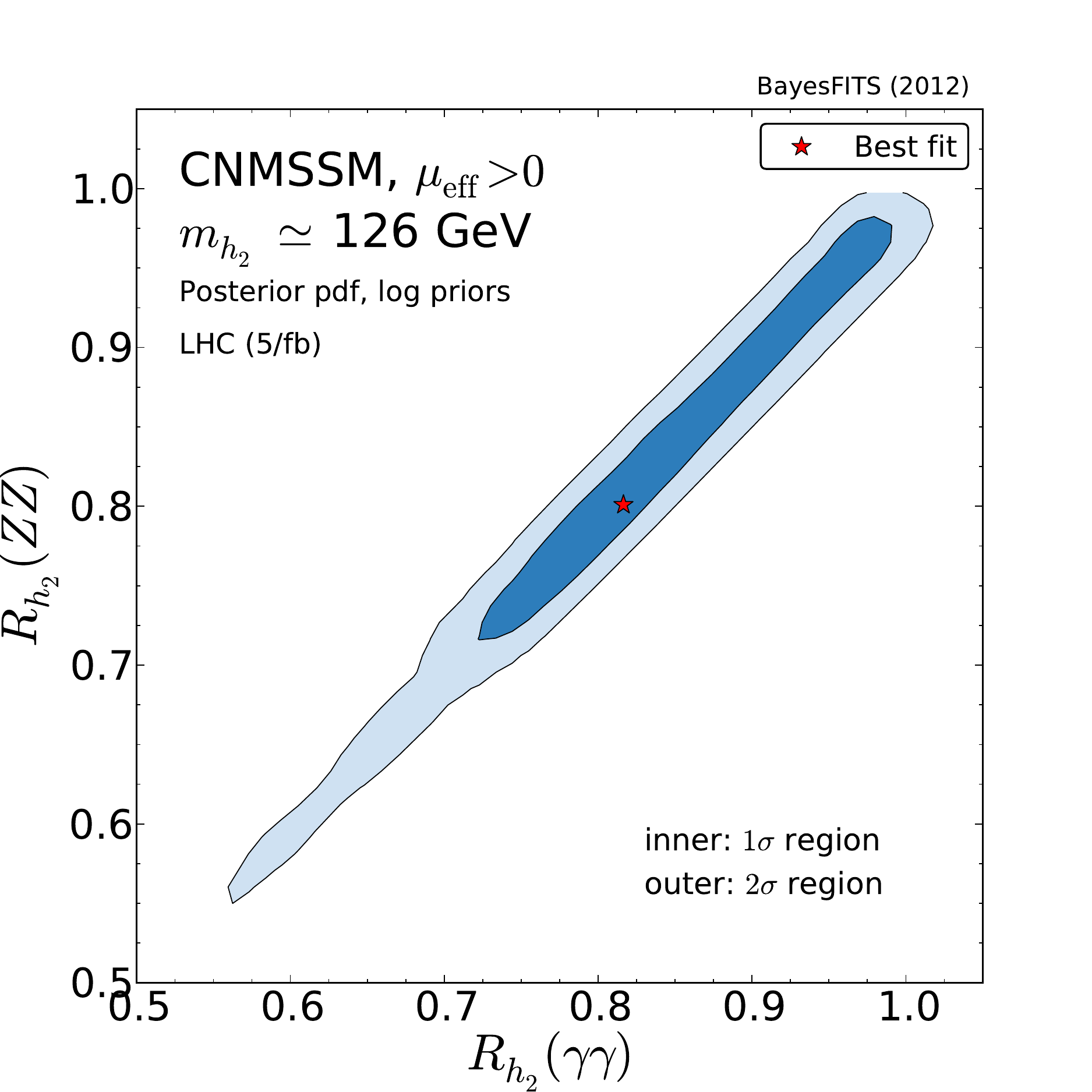}
}%
\caption[]{\subref{fig:-a} Marginalized 2D posterior pdf in the (\ronegg, \ronezz) plane of the CNMSSM constrained by the
 experiments listed in Table~\ref{tab:exp_constraints} for case~1. 
 \subref{fig:-b} Marginalized posterior pdf in the (\rtwogg, \rtwozz) plane of the CNMSSM for case~2.
 The color code is the same as in Fig.~\ref{fig:cnmssm_params}.}
\label{fig:cnmssm_Rs}
\end{figure}
%%%%%%%%%%%%%%%%%%%%%%%%%%%%%%%%%%%%%%%%%%%%%%%%%%%%%%%%%%%%%%%%%%%%%%%%%%%%%%%%

In Fig.~\ref{fig:cnmssm_Rs}\subref{fig:Rs_a} we show the 2D posterior pdf in
the (\ronezz, \ronegg) plane for case~1.
As one can see, the Higgs boson in this case is SM-like.
Actually, it is known\cite{Gunion:2012zd} that the enhancement of the signal
strength in the $\gamma\gamma$ decay channel observed by both CMS and
ATLAS cannot be obtained for the values of \lam\ that are favored by the
global scan.
As a consequence, \ronegg\ cannot be fitted perfectly in the CNMSSM. This
discrepancy effectively adds two units of \chisq\ homogeneously over the
preferred parameter space, but it does not alter the posterior distribution. For the
same reason \ronezz\ cannot be perfectly fitted either, though its
contribution to the total \chisq\ is smaller than 0.5 units of \chisq,
making this observable equally ineffective in constraining the posterior.

In Fig.~\ref{fig:cnmssm_Rs}\subref{fig:Rs_b} we present the posterior
distribution for case~2. Once again, \rtwogg\ can hardly become larger than
1 over the preferred parameter space. The 95\% credible region lies far
from the central value of the observed enhancement and, in fact, even
covers values lower than in case~1. \rtwozz\ presents similar behavior,
although the suppression of the reduced cross section is highly
welcome for this observable, as it places the calculated value closer to the rate observed at CMS. 
Smaller than 1 signal rates indicate less of a SM-like
character for \htwo, which is caused by the suppression of the
SM couplings induced by its increased singlet component.

The posterior distributions presented in
Figs.~\ref{fig:cnmssm_Rs}\subref{fig:Rs_a} and
\ref{fig:cnmssm_Rs}\subref{fig:Rs_b} indicate that, in both case 1 and
case 2 it is in general extremely difficult to obtain the signal
enhancement in the $\gamma\gamma$ channel.  The scan
naturally tends to stay in the regions of parameter space favored by all
constraints.  It is therefore no surprise that among the points scanned for case~1 only two presented 
a $\gamma\gamma$ rate in the range 1.2--2, thanks to the reduced coupling of the signal
Higgs boson to the bottom quarks. Such points present \chisq\ contributions to the relic density of order several 10s, 
and the \chisq\ contribution to \brbsmumu\ is of order 100. In case~2 we found a dozen such points, 
for which the contribution to the relic density is even worse. 
%disfavoured in the global scan by the other constraints, in particular
%by the relic abundance and \bsmumu, and their total \chisq\ is well
%above 100.  For completeness, we list them in
%Table~\ref{tab:Rbenchmarks}.

%However, it is
%not totally impossible. In Table~\ref{tab:Rbenchmarks} we show the points
%for which \rsiggg\ is actually enhanced above the SM value due to the
%reduced coupling of the signal Higgs boson to the b-quarks. Notice, that
%the same mechanism leads also to the enhancement of \rsigzz. All such
%points are strongly disfavoured in the global scan by the other
%constraints, in particular by the relic aboundance and \bsmumu. It is also
%worth to stress that all of them lie in the region of parameter space
%already excluded by the direct LHC SUSY searches.\es{not excluded!}

In case~3 one could expect to obtain an enhancement of \rsiggg\ by
adding the individual rates for both almost degenerate light scalars.
However, the posterior pdf in the ($R_{\hone+\htwo}(\gamma\gamma)$,
$R_{\hone+\htwo}(ZZ)$) plane is remarkably similar to the one shown in
Fig.~\ref{fig:cnmssm_Rs}\subref{fig:Rs_a}, due to the large singlet component of \htwo, and we refrain from showing it again over here. 
In fact, in
case~3 we were not able to find a single point with the enhanced
$\gamma\gamma$ rate.  Since case~3 is a subset of case~2 in terms of
the favored parameter space, and the rates in the $\gamma\gamma$ and
$ZZ$ channel do not show interesting features, we will not consider
it separately from the other cases any further.

\subsection{Prospects for DM direct detection and \brbsmumu}\label{Sec:dm}

In this subsection we will discuss the impact of limits from direct DM
searches on the preferred parameter space of the CNMSSM. This kind of
experiments are complementary to direct LHC SUSY searches, as they are
capable of testing neutralino mass ranges beyond the current and
future reach of the LHC, and therefore could add new pieces of
information to the global picture.

At present the most stringent limit on the spin-independent
cross section \sigsip\ comes from XENON100\cite{Aprile:2012nq}.  
In supersymmetric models it can then be plotted as a function of the neutralino mass in
the form of an exclusion limit in the (\mchi, \sigsip) plane.  

We want to point out that the theory uncertainties are very large (up to a factor of 10) 
and strongly affect the impact of the experimental limit on the parameter space\cite{Roszkowski:2012uf}. 
It was shown that, when smearing out the XENON100 limit with a theoretical uncertainty of order 
10 times the given value of \sigsip, the effect on the posterior is negligible for regions of 
the parameters that appear up to 1 order of magnitude above the experimental line. 
Moreover, the main source of error (the so-called $\Sigma_{\pi N}$ term) arises from different, 
and in fact incompatible, results following from different calculations based on different assumptions and methodologies. 
Therefore, in this
study we decided not to include the upper bound on \sigsip\ from XENON100 in
the likelihood function, but below we will discuss its possible effects on the properties of the model.

%%%%%%%%%%%%%%%%%%%%%%%%%%%%%%%%%%%%%%%%%%%%%%%%%%%%%%%%%%%%%%%%%%%%%%%%%%%%%%%
%%%%%%%%%%%%%%%%%%%%%%%%   F   I   G   U   R   E   %%%%%%%%%%%%%%%%%%%%%%%%%%%%
\begin{figure}[t]
	\centering
    \subfloat[]{%
        \label{fig:h1sigsip-a}
	    \includegraphics[scale=0.4]{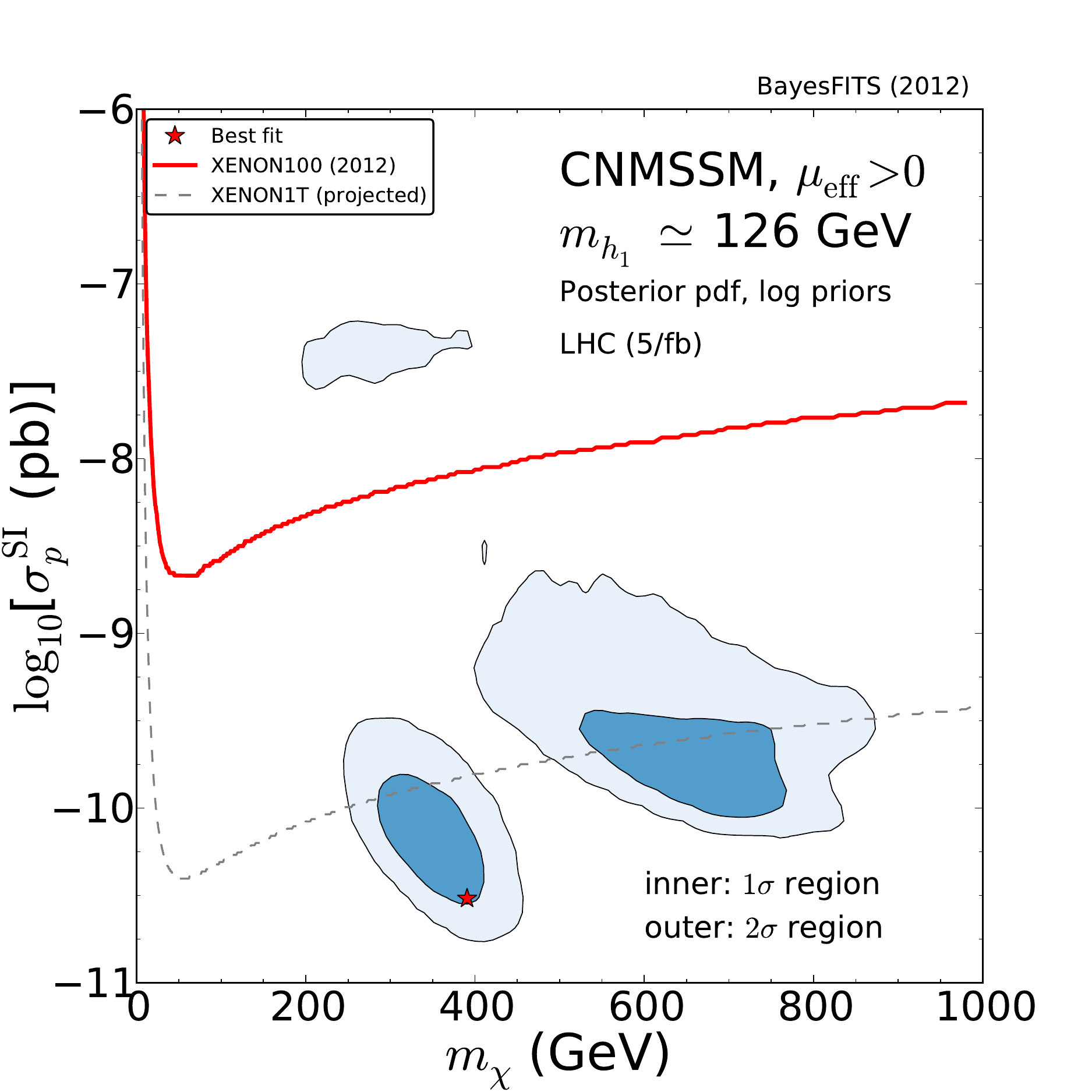}
	}
    \subfloat[]{%
        \label{fig:h1sigsip-b}	    
	    \includegraphics[scale=0.4]{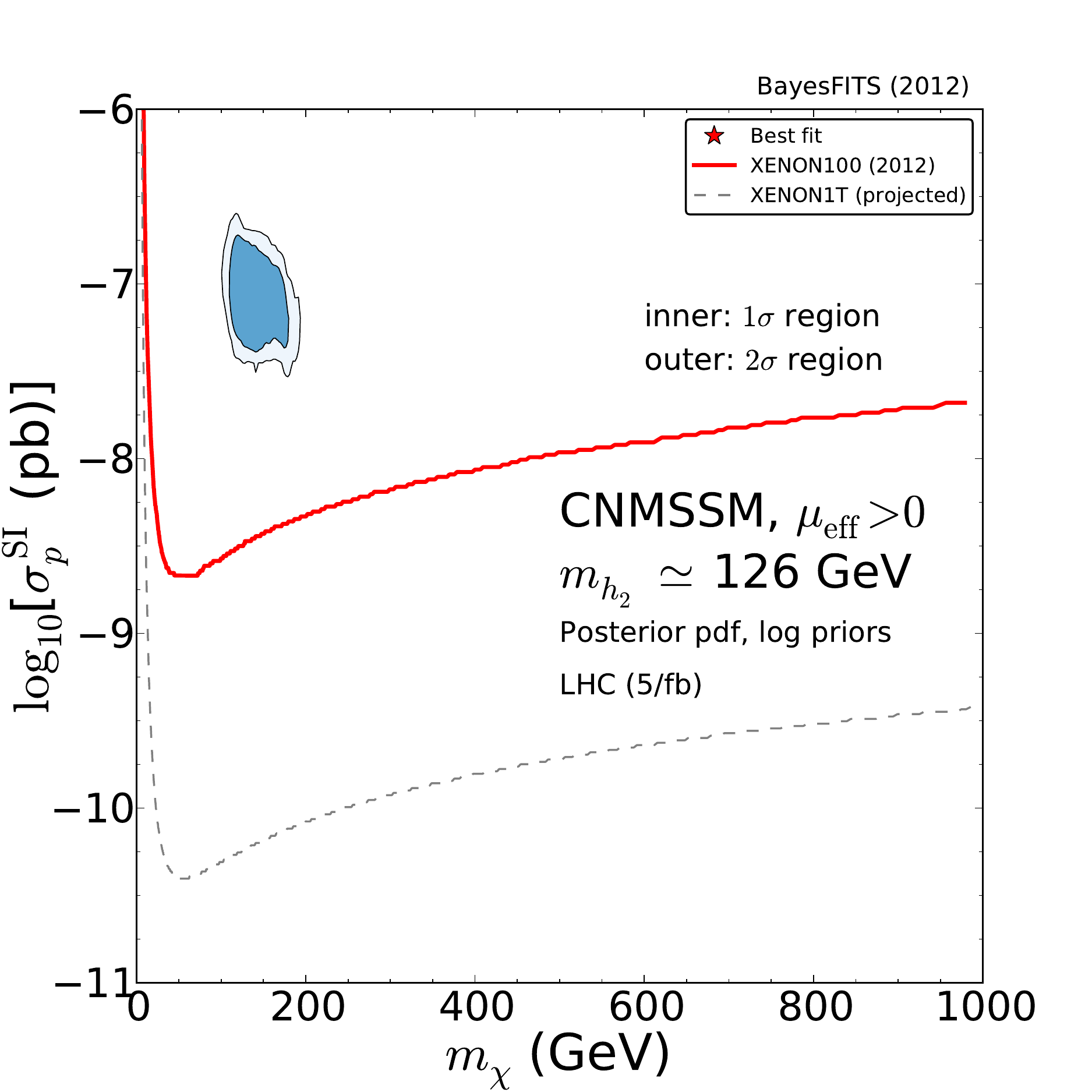}
	}
	\caption[]{Marginalized 2D posterior pdf in the (\mchi, \sigsip) plane of the CNMSSM constrained by the
 experiments listed in Table~\ref{tab:exp_constraints} in \subref{fig:h1sigsip-a} case~1 and \subref{fig:h1sigsip-b} case~2. 
 The solid red line shows the 90\%~C.L. exclusion bound by XENON100 (not included in the likelihood), and the dashed gray line the projected sensitivity for XENON1T.
 The color code is the same as in Fig.~\ref{fig:cnmssm_params}.}
	\label{fig:h1mxsigsip}
\end{figure}
%%%%%%%%%%%%%%%%%%%%%%%%%%%%%%%%%%%%%%%%%%%%%%%%%%%%%%%%%%%%%%%%%%%%%%%%%%%%%%%

In Fig.~\ref{fig:h1mxsigsip}\subref{fig:h1sigsip-a}, we present the
posterior pdf in the (\mchi, \sigsip) plane for the case~1. The solid
red line shows the most recent 90\%~C.L. upper exclusion limit by the
XENON100 Collaboration. One can see that the two high posterior
probability regions (from SC on the left and AF on the right) are
located well below it.  A small 95\% credible region (corresponding to
the FP/HB region with the lightest neutralino being a mixture of bino
and higgsino) would probably be excluded by adding the XENON100 limit to the likelihood, even taking
into account significant theoretical uncertainties. One can also
notice that XENON1T, a future ton-size DM detector (projected
sensitivity represented as the dashed gray line), will be in a
position to test the 68\% credible regions.

The conflict with the XENON100 limit in case~2 (and case~3, which presents similar features) is even stronger. In
Fig.~\ref{fig:h1mxsigsip}\subref{fig:h1sigsip-b} we show the posterior
in the (\mchi, \sigsip) plane. The high probability region above the
experimental line presents the features of the FP/HB region:
$\mueff\simeq 200\gev$, and the lightest neutralino is a mixture of bino
and higgsino. As a consequence, the spin-independent neutralino-proton
cross section is larger and in this case already in strong tension with the XENON100 bound.

%%%%%%%%%%%%%%%%%%%%%%%%   F   I   G   U   R   E   %%%%%%%%%%%%%%%%%%%%%%%%%%%%
\begin{figure}[t]
	\centering
    \subfloat[]{%
        \label{fig:-a}
	    \includegraphics[scale=0.4]{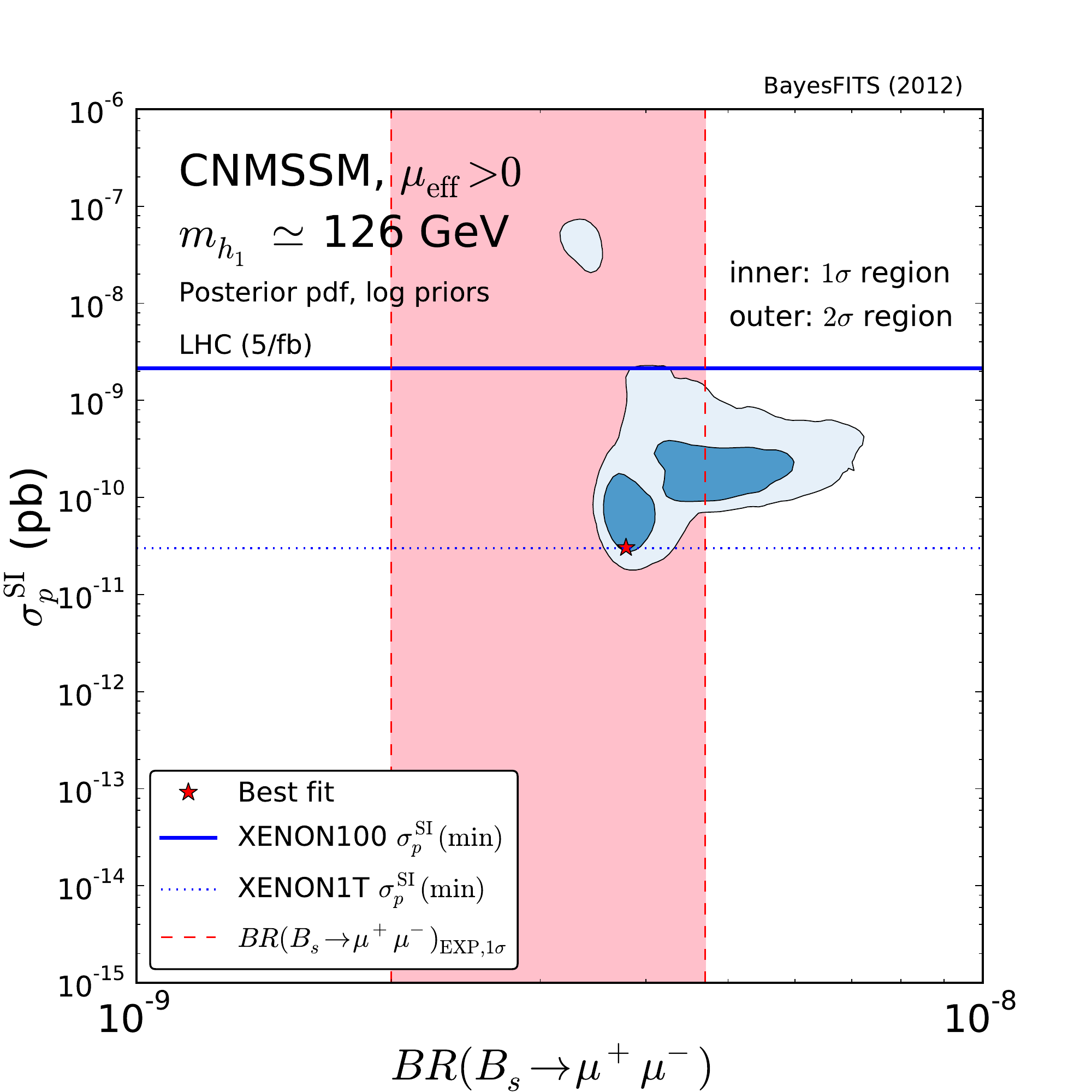}
	}
	\subfloat[]{%
        \label{fig:-a}
	    \includegraphics[scale=0.4]{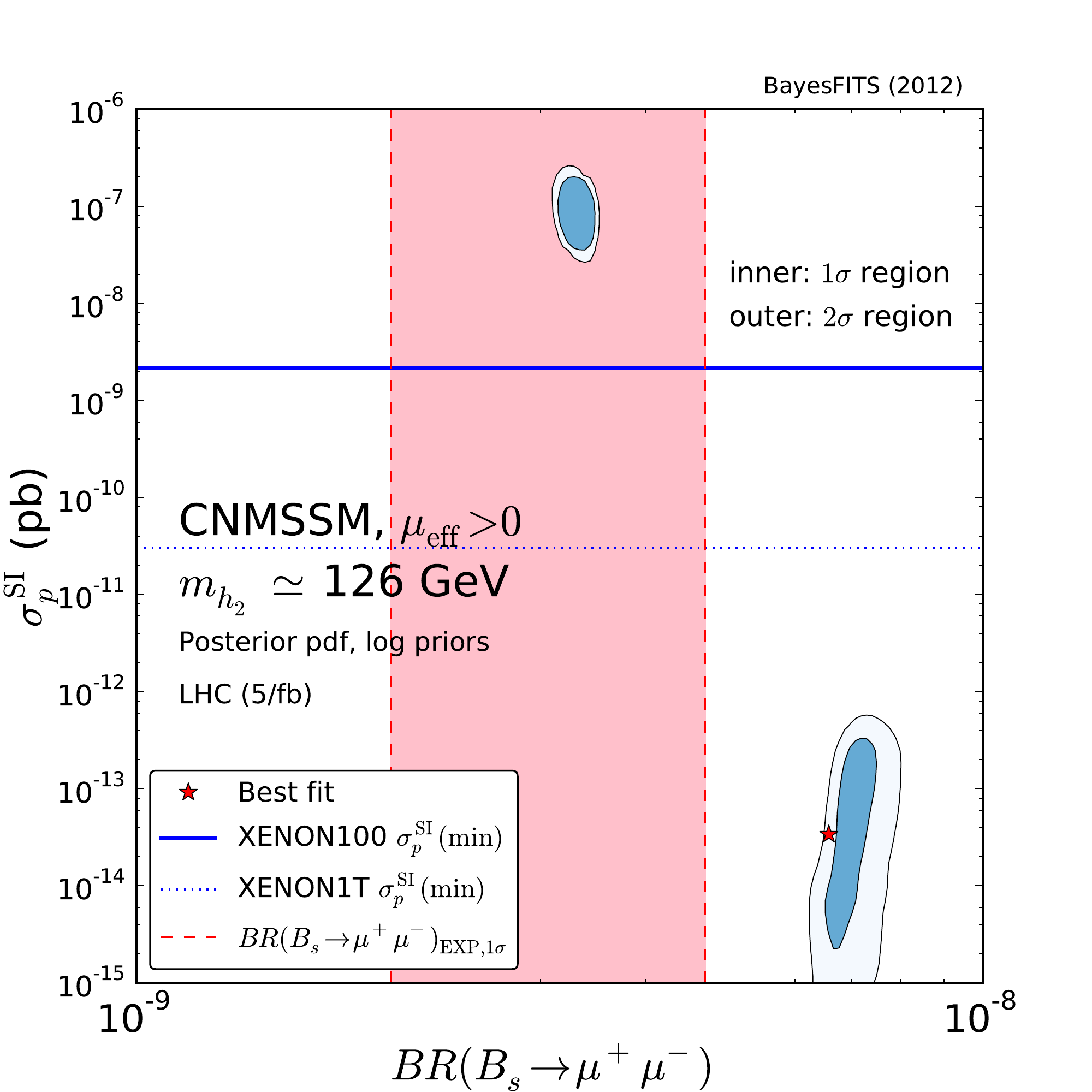}
	}
	\caption[]{Marginalized 2D posterior pdf in the (\brbsmumu, \sigsip) plane of the CNMSSM constrained by the
 experiments listed in Table~\ref{tab:exp_constraints} in \subref{fig:-a} case~1 and \subref{fig:-b} case~2. 
 The color code is the same as in Fig.~\ref{fig:cnmssm_params}. The solid blue horizontal line shows the minimum 90\%~CL
upper bound on \sigsip\ by XENON100 (not included in the likelihood), and the dotted blue horizontal line the corresponding projected sensitivity for XENON1T. The pink vertical band shows the $1\sigma$ experimental uncertainty on the recent measurement of \brbsmumu\cite{Aaij:2012ct} (which is included in the likelihood).}
	\label{fig:MDbsm}
\end{figure}
%%%%%%%%%%%%%%%%%%%%%%%%%%%%%%%%%%%%%%%%%%%%%%%%%%%%%%%%%%%%%%%%%%%%%%%%%%%%%%%

Notice that the SC region is not shown in
Fig.~\ref{fig:h1mxsigsip}\subref{fig:h1sigsip-b} since the neutralino
is a nearly pure singlino there, so that \sigsip\ is several orders of
magnitude below the XENON100 bound (and below the range shown in the
figure). However, we would like to point out that in case 2 (and case
3) strong constraints on the full parameter space can be placed as a
result of the interplay between the limits provided by two completely
different experiments --that test different observables by means of
different experimental techniques --namely LHCb and XENON100.  This
is illustrated in Figs.~\ref{fig:MDbsm}\subref{fig:-a} and
\ref{fig:MDbsm}\subref{fig:-b}, where we show the posterior pdf in the
$\left(\brbsmumu, \sigsip\right)$ plane for case~1 and case~2,
respectively. The solid blue horizontal line shows the minimum 90\%~C.L.
upper bound on \sigsip\ by XENON100 (obtained at $\mchi\simeq 50\gev$), and the dotted blue horizontal line the
corresponding projected sensitivity for XENON1T. The pink vertical band encompasses the $1\sigma$ experimental uncertainty on the recent LHCb measurement of \brbsmumu\cite{Aaij:2012ct}. Figure~\ref{fig:MDbsm}\subref{fig:-b}
shows that, for cases~2 and 3, in the SC region \brbsmumu\ is strongly
enhanced, due to the large values assumed by \tanb\ there, and it
could be excluded by the next updated results from LHCb.

\subsection{Fine-tuning}

In this subsection we will address the issue of fine-tuning. Note that
we will not delve into it, nor will we discuss which values of
fine-tuning are acceptable or not from the point of view of
naturalness. Our aim here is to simply present an estimate of
fine-tuning (provided as an output by NMSSMTools v3.2.1) for the
preferred parameter space of the model, leaving aside the discussion of
the viability of the model itself, which would be a matter of
personal prejudices.
  
The mass of the $Z$ boson (which determines
the EW symmetry breaking scale) can be expressed in terms of the
supersymmetric parameters through the minimization condition of the
Higgs potential, 
%%%%%%%
\be
\frac{M_Z^2}{2}=-\mu^2+\frac{\mhd^2-\mhu^2\tan^2\beta}{(\tan^2\beta-1)}.
\label{mz}
\ee 
%%%%%%%

The fine-tuning problem of the MSSM 
% lr ***was first discussed in
\cite{Barbieri:1987fn}
amounts to the fact that the parameters \mhu, \mhd\ and $\mu$ need to
be simultaneously tuned to a high precision to reproduce the correct
value of $\mz$. In the NMSSM  $\mu$ is replaced by \mueff.

In the framework of the unified theory, at the GUT scale $\mhu=\mhd=\mzero$. They are then evolved down to the EW
scale by means of the RGEs. Therefore, the right-hand side of
Eq.~(\ref{mz}) depends on the parameters of the model at
the GUT scale. 

The measure of fine-tuning associated with the parameter $p_i$ of the
model is defined as\cite{Ellis:1986yg,Barbieri:1987fn} 
%%%%%%%
\be 
\Delta_{p_i}=\left|\frac{\partial\log M_Z^2}{\partial\log p_i^2}\right|.
\ee
%%%%%%%
In the CNMSSM, $p_i=\{\mzero, \mhalf, \azero, \lam,\kap\}$. We do not
present the fine-tuning measure due to the top quark Yukawa coupling,
following the approach adopted in\cite{Feng:1999zg}, and quantifying
the fine-tuning associated only with the supersymmetric parameters. We
define the fine-tuning measure $\Delta$ for a given model point as the
maximal contribution to fine-tuning among all the model's parameters
for that point,
%%%%%%%
\be
\Delta=\textrm{Max}(\Delta_{p_i}).
\ee
%%%%%%%

%%%%%%%%%%%%%%%%%%%%%%%%   F   I   G   U   R   E   %%%%%%%%%%%%%%%%%%%%%%%%%%%%
\begin{figure}[t]
	\centering
    \subfloat[]{%
        \label{fig:-a}
	    \includegraphics[scale=0.4]{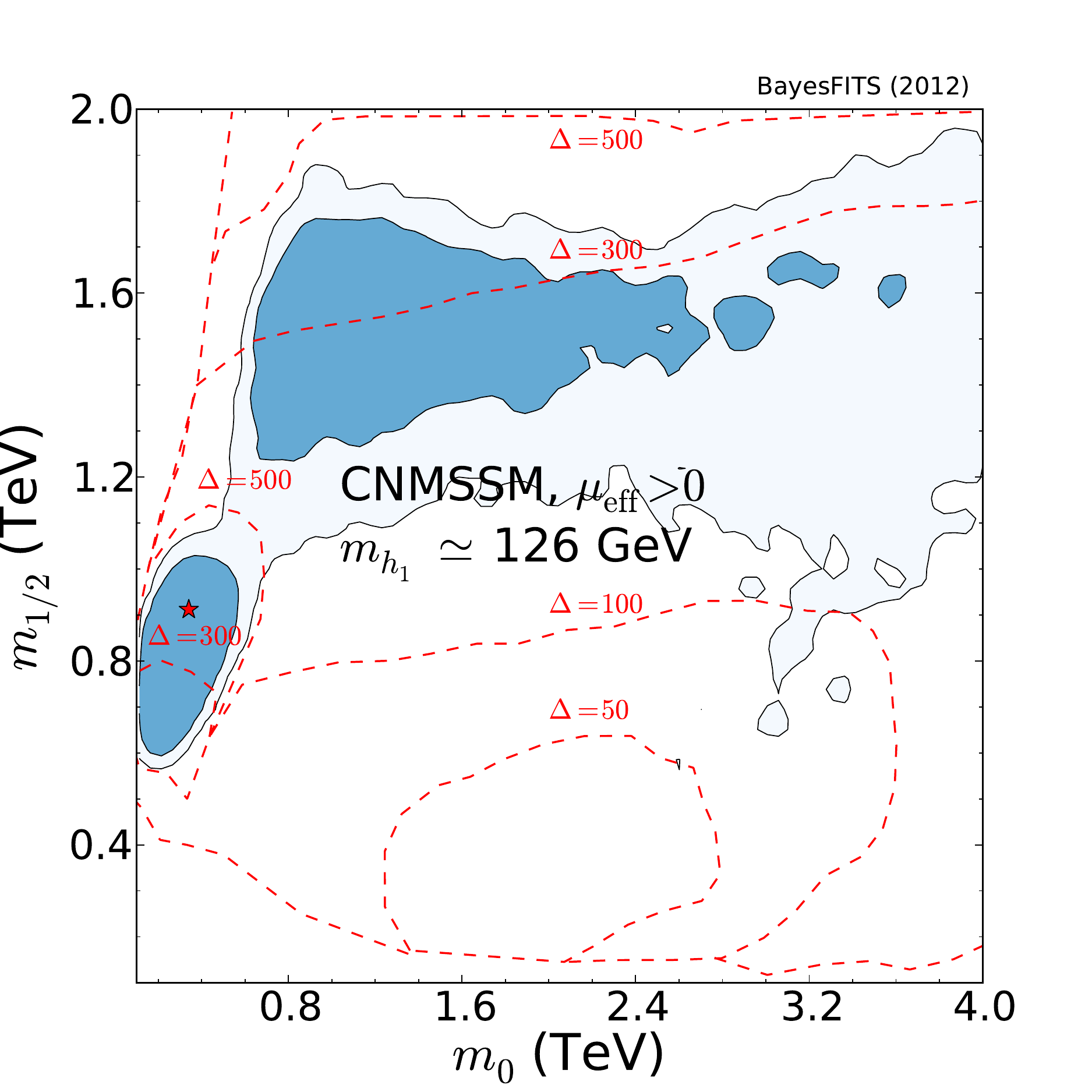}
	}
    \subfloat[]{%
        \label{fig:-b}	    
	    \includegraphics[scale=0.4]{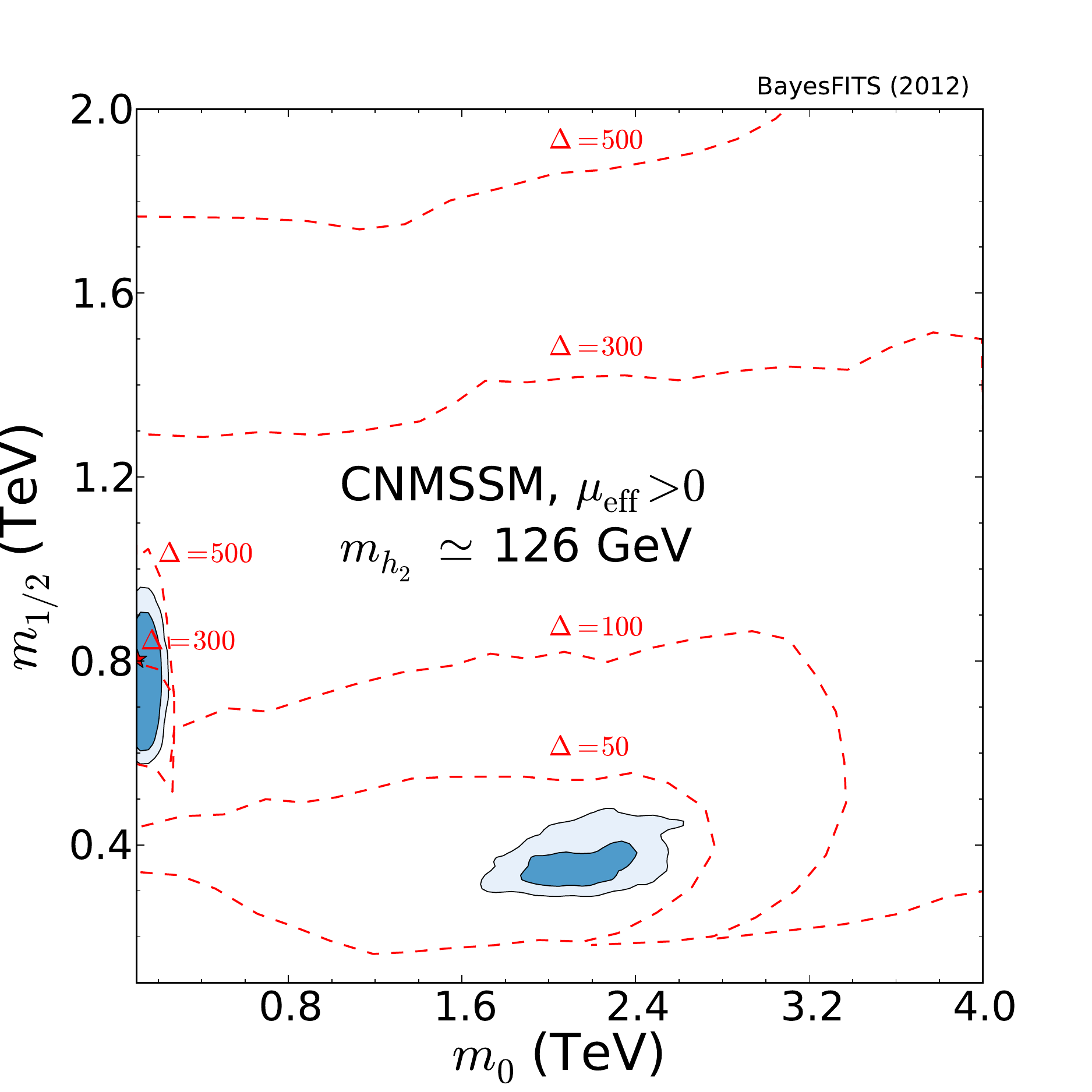}
	}
	\caption[]{Isocontours of the fine-tuning measure $\Delta$ (dashed red lines) superimposed to the 2D posterior 
	pdf for \subref{fig:-a} case~1 and \subref{fig:-b} case~2. The color code is the same as in Fig.~\ref{fig:cnmssm_params}.}
	\label{fig:finetune}
\end{figure}
%%%%%%%%%%%%%%%%%%%%%%%%%%%%%%%%%%%%%%%%%%%%%%%%%%%%%%%%%%%%%%%%%%%%%%%%%%%%%%%

%%%%%%%%%%%%%%%%%%%%%%%%   F   I   G   U   R   E   %%%%%%%%%%%%%%%%%%%%%%%%%%%%
\begin{figure}[b]
        \centering
%    \subfloat[]{%
        \label{fig:-a}
            \includegraphics[scale=0.3]{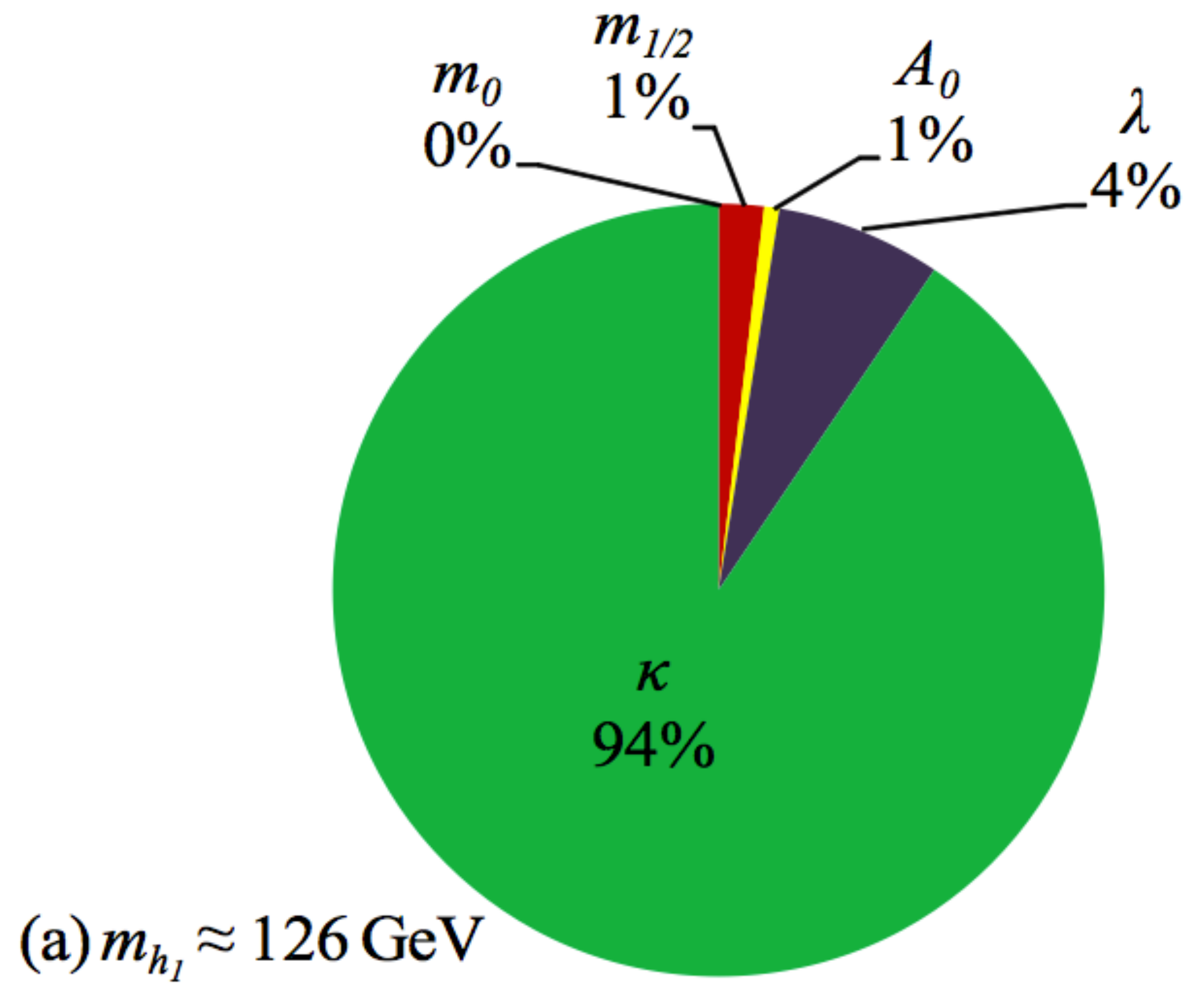}
%       }
%    \subfloat[]{%
\hspace{1.0cm}%
        \label{fig:-b}
            \includegraphics[scale=0.3]{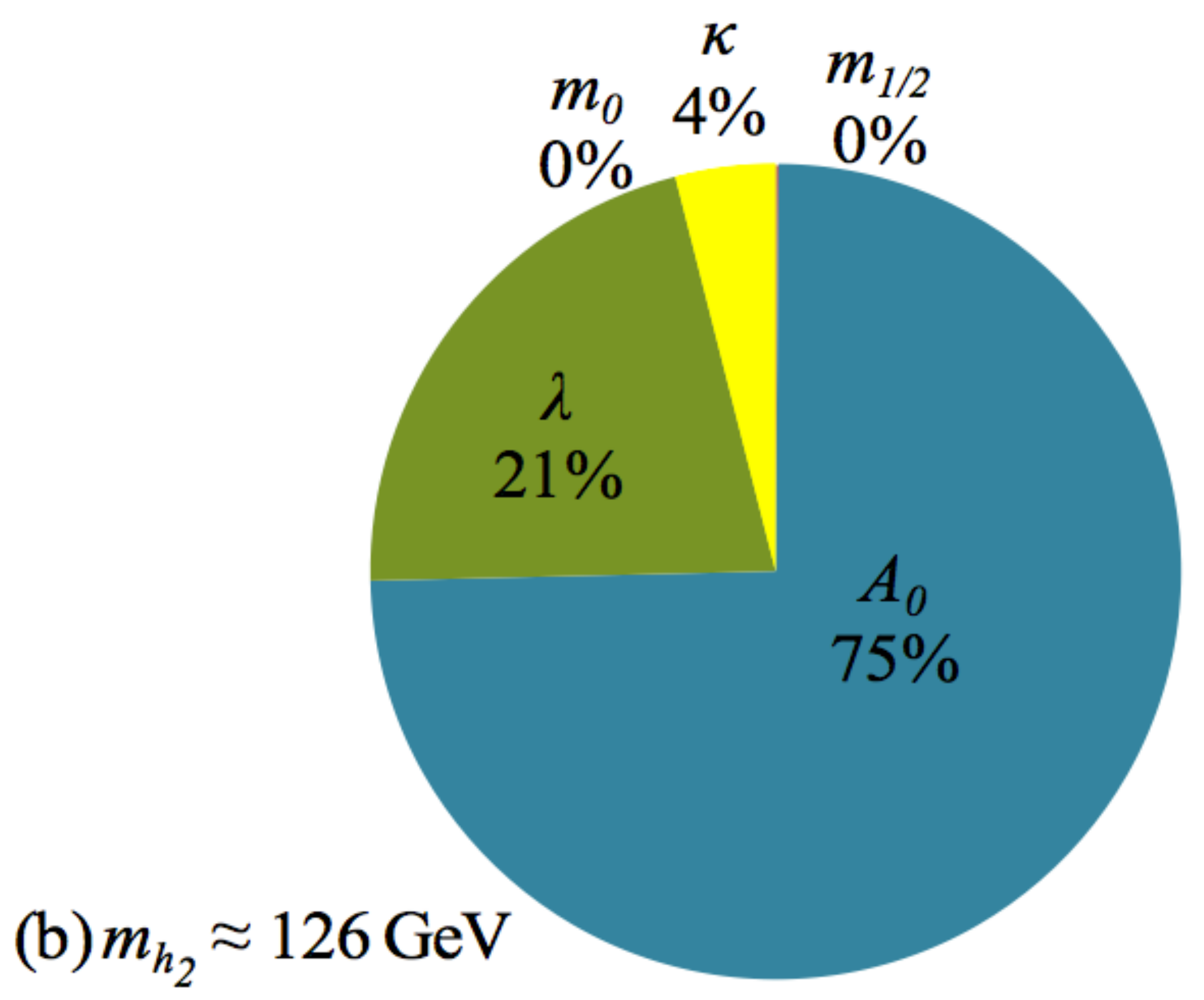}
%       }
        \caption[]{Pie charts showing the fraction of the total number of allowed points that yields maximum 
        fine-tuning by each of the $p_i$'s for \subref{fig:-a} case~1 and \subref{fig:-b} case~2.}
        \label{fig:finetune2}
\end{figure}
%%%%%%%%%%%%%%%%%%%%%%%%%%%%%%%%%%%%%%%%%%%%%%%%%%%%%%%%%%%%%%%%%%%%%%%%%%%%%%%

In Figs.~\ref{fig:finetune}\subref{fig:-a} and
\ref{fig:finetune}\subref{fig:-b} we present the isocontours of the
fine-tuning measure $\Delta$ in the (\mzero, \mhalf) plane for cases~1
and 2, respectively. The isocontours reflect the value of $\Delta$
for the vast majority of the points included.  They are superimposed
on the 2D posterior distributions. Case~1 presents very CMSSM-like
behaviour, where smaller fine-tuning can be achieved only in the FP/HB
region due to the relatively low values of $\mueff$. Note that for the
same reason $\Delta$ is larger in the SC region which is characterized
by larger $\mueff$. On the other hand, case~2 is less fine-tuned
($\Delta<50$ in a vast region of the parameter space preferred at $2\sigma$), which is a reflection of the fact
that the parameter space is already highly constrained by the
requirement of $\mhtwo\simeq 126\gev$. In
Figs.~\ref{fig:finetune2}\subref{fig:-a} and
\ref{fig:finetune2}\subref{fig:-b} we show for what percentage of the
total number of allowed points each of the $p_i$'s yields maximal
fine-tuning. For example, \kap\ gives maximal fine-tuning for 94\% of
the points in case~1, while \azero\ is the main contributor for the
vast majority of the points in case~2.

\subsection{\label{Sec:nog2} Relaxing \gmtwo\ and the case of  negative \mueff}

%%%%%%%%%%%%%%%%%%%%%%%%%   F   I   G   U   R   E   %%%%%%%%%%%%%%%%%%%%%%%%%%%%
% 2 by 1: left: plot of 1d pdf of mhl,
% right: chi2 vs mhl
\begin{figure}[t]
\centering
\subfloat[]{%
\label{fig:-a}%
\includegraphics[width=0.50\textwidth]{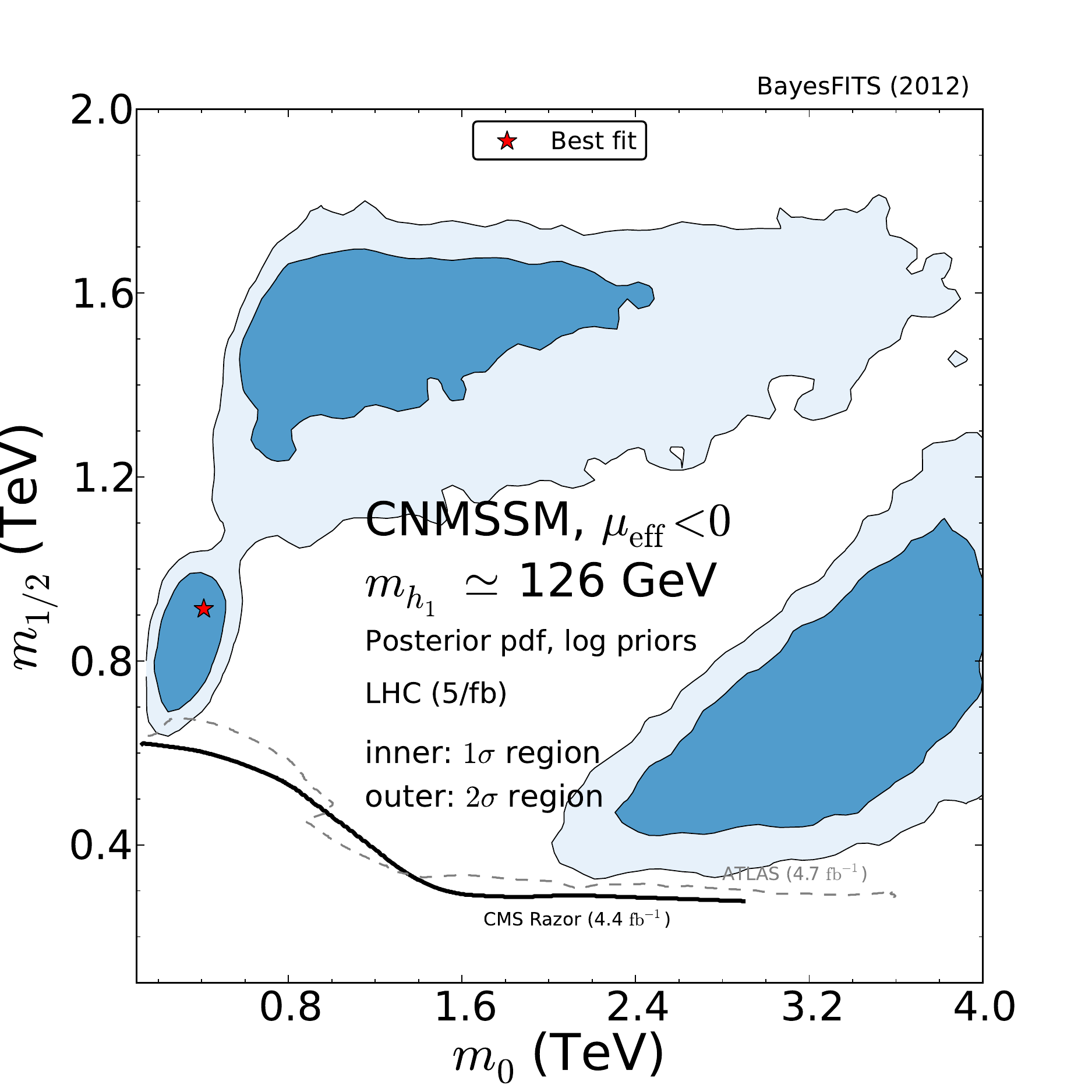}
}%
%\hspace{1pt}%
\subfloat[]{%
\label{fig:-b}%
\includegraphics[width=0.50\textwidth]{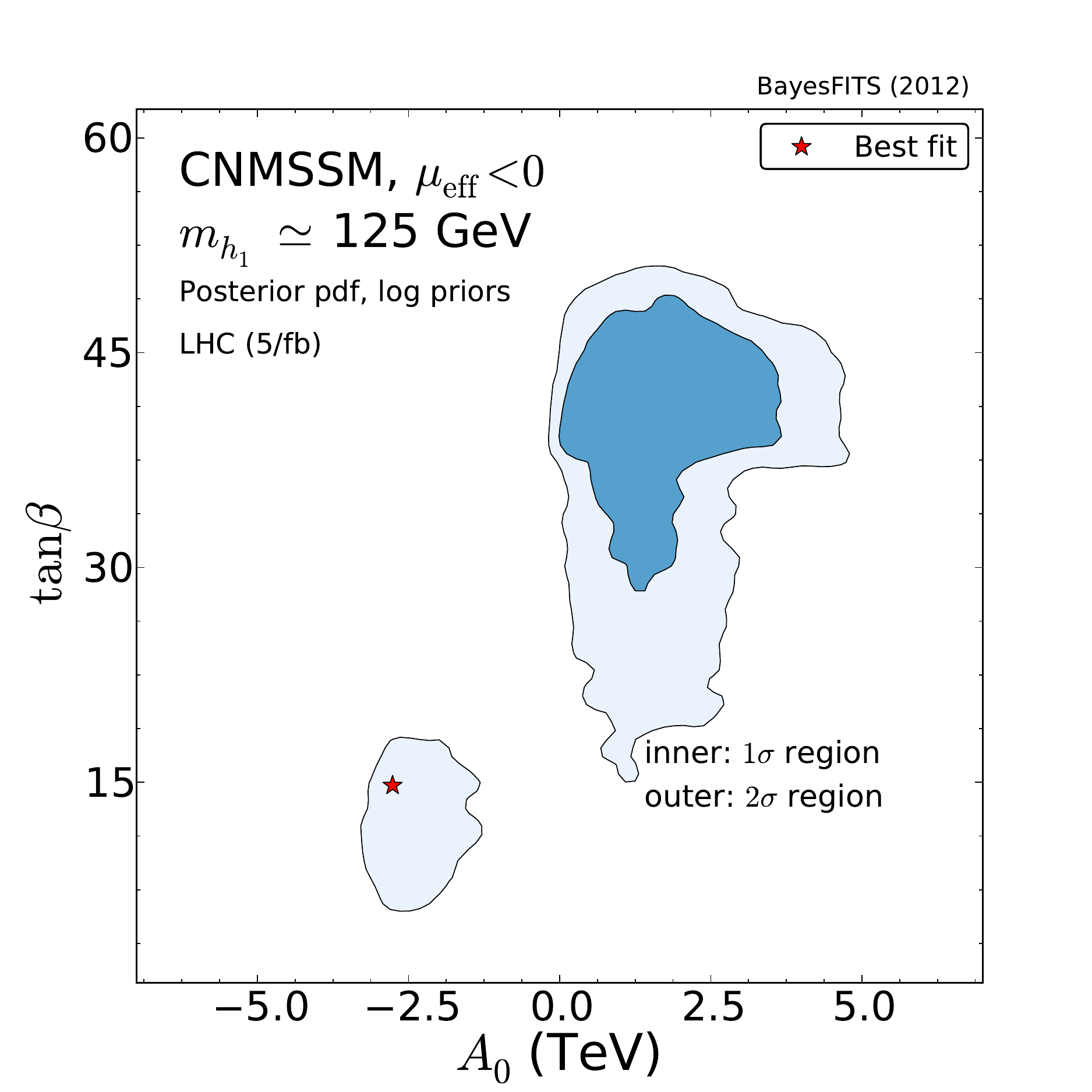}
}%
\caption[]{Marginalized 2D posterior pdf in \subref{fig:-a} the (\mzero, \mhalf)
  plane and \subref{fig:-b} the (\azero, \tanb) plane of the CNMSSM constrained by the
 experiments listed in Table~\ref{tab:exp_constraints}, for case~1 without \gmtwo\ and with $\mueff<0$. The color code is the same as in Fig.~\ref{fig:cnmssm_params}.}
\label{fig:cnmssm_negmu}
\end{figure} 
%%%%%%%%%%%%%%%%%%%%%%%%%%%%%%%%%%%%%%%%%%%%%%%%%%%%%%%%%%%%%%%%%%%%%%%%%%%%%%%

In our recent study of the CMSSM\cite{Fowlie:2012im} we considered the
effect of relaxing the \deltagmtwomu\ constraint.  The reason was
based on the observation that the poor fit of the CMSSM was the result
of basically that single constraint, which simply could not be
reproduced after including especially direct superpartner mass limits
from the LHC. This appears to be a general feature of all SUSY models
where slepton and squark masses are assumed to be comparable,
including simple unified models like the CMSSM or CNMSSM.  We simply
state it as a conclusion reached by many studies.  While we do not
feel to be in a position to comment on the reliability of theoretical
calculations of the SM \gmtwo\ which are strongly affected by
nonperturbative effects related to low-energy strong interactions,
especially the hadronic light-by-light contribution (for more details
see, e.g., the Introduction of\cite{Fowlie:2012im} and references
therein), we believe that it thus makes sense to consider global scans
where the constraint is removed.  We showed in\cite{Fowlie:2012im}
that for the CMSSM the better fit was obtained with $\mu<0$ thanks to
the constraints from $b$-physics, which present a much better \chisq\
for this choice of parameters.

Relaxing the \gmtwo\ constraint in the CNMSSM while keeping \mueff\
positive has no apparent impact on the posterior distributions both
for the parameters of the model as for the measured observables. Such
behavior was to be expected for case~1, as it was already observed for the CMSSM
in\cite{Fowlie:2012im}. We checked that this is true also for case~2 and case~3, 
where relaxing the \gmtwo\ constraint while keeping \mueff\ positive has little effect. 
Specifically, it reduces the statistical relevance of the SC region and slightly increases the size of 
the posterior in the FP/HB region where, as we discussed in Sec.~\ref{Sec:dm}, it is disfavored by the XENON100 bound.

We also confirm that, for case~1, the better overall fit is
obtained with $\mueff<0$ for basically the same reasons as in the
CMSSM. In Fig.~\ref{fig:cnmssm_negmu}\subref{fig:-a} we show the 2D posterior
pdf in the (\mzero, \mhalf) plane for case~1, where we ignored the
constraint from \deltagmtwomu\ and set $\textrm{sgn}(\mueff)=-1$.
Differently from the CMSSM, where one could observe a clear
predominance of the AF region, in the CNMSSM the SC, AF and FP/HB
regions of parameter space now seem to be equally probable. This is
not likely to be an intrinsic difference between the CMSSM and the
CNMSSM; it is more probably due to the fact that the constraints are
implemented with different numerical tools in the two models, as
explained exhaustively at the beginning of Sec.~\ref{Results}.  We
note here that the global likelihood becomes flat over the regions of
parameter space preferred by the relic density, and very little
information can be extracted from the posterior.
   
In Fig.~\ref{fig:cnmssm_negmu}\subref{fig:-b} we show the same
posterior pdf in the (\azero, \tanb) plane for case~1. As was the case
for the CMSSM, one can see that the distribution of \tanb\ tends to
favor slightly smaller values than in the positive \mueff\ case,
particularly in the AF region.

\begin{figure}[t]
\centering
\subfloat[]{%
\label{fig:-a}%
\includegraphics[width=0.50\textwidth]{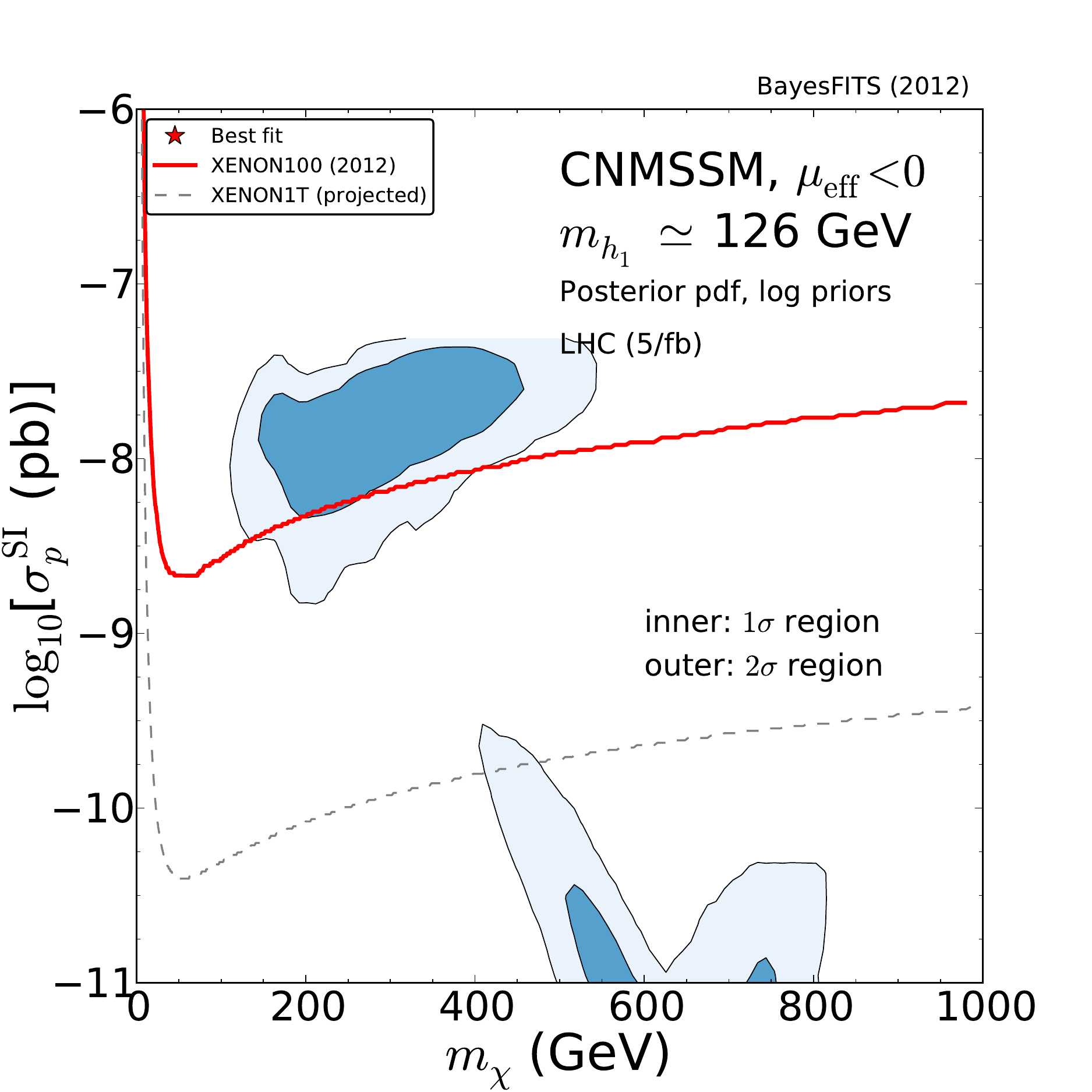}
}%
%\hspace{1pt}%
\subfloat[]{%
\label{fig:-b}%
\includegraphics[width=0.50\textwidth]{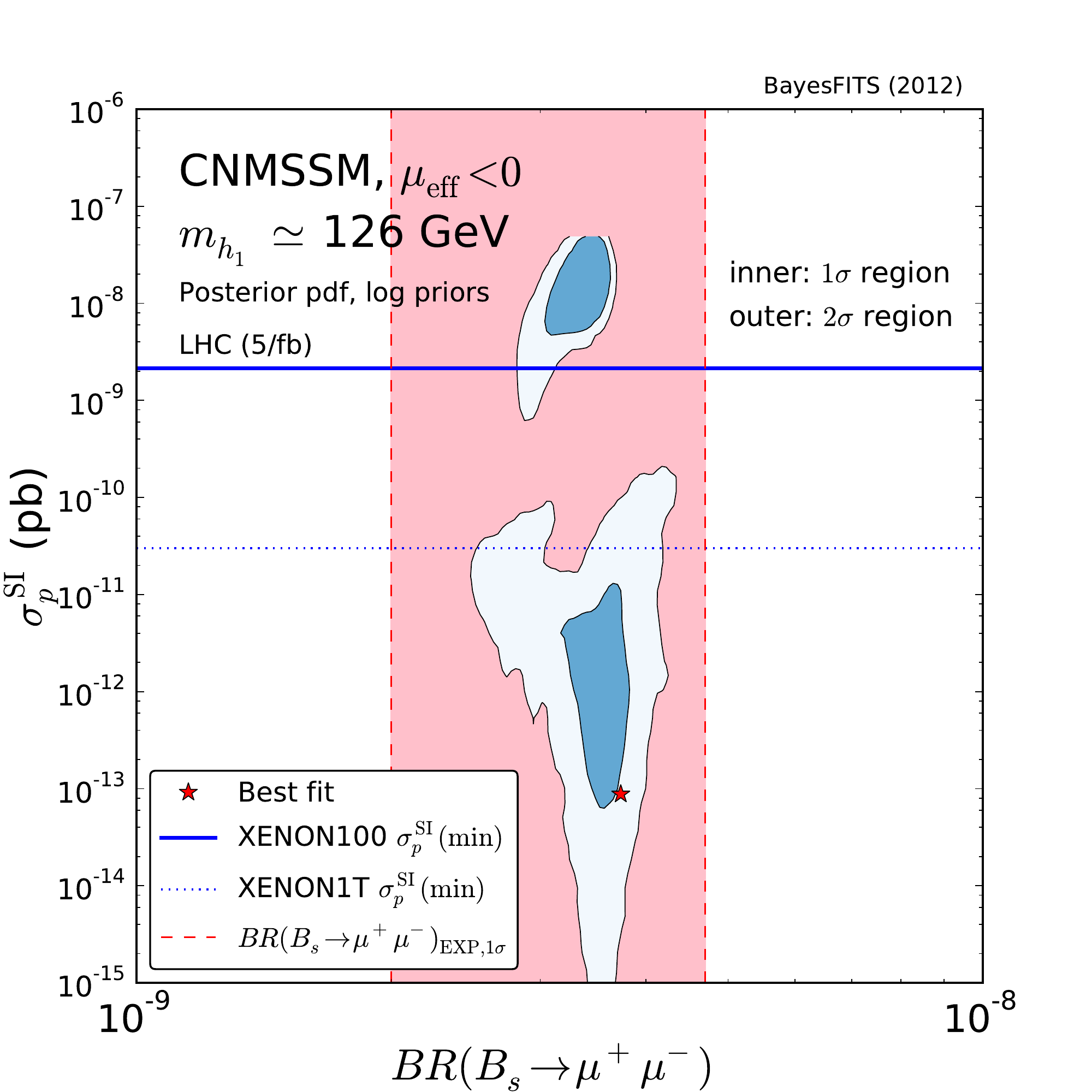}
}%
\caption[]{Marginalized 2D posterior pdf in \subref{fig:-a} the (\mchi, \sigsip)
  plane and \subref{fig:-b} the (\brbsmumu, \sigsip) plane of the CNMSSM constrained by the
 experiments listed in Table~\ref{tab:exp_constraints}, for case~1 without \gmtwo\ and with $\mueff<0$. The color code is the same as in Figs.~\ref{fig:h1mxsigsip} and \ref{fig:MDbsm}.}
\label{fig:othernegmu}
\end{figure} 
%%%%%%%%%%%%%%%%%%%%%%%%%%%%%%%%%%%%%%%%%%%%%%%%%%%%%%%%%%%%%%%%%%%%%%%%%%%%%%%%

In Fig.~\ref{fig:othernegmu}\subref{fig:-a} we show the 2D posterior
pdf in the (\mchi, \sigsip) plane for case~1. Since the posterior in
the FP/HB region is much more extended with respect to the positive
\mueff\ case, a large region of parameter space lies above the
XENON100 bound, and has the potential of being tested with modest 
improvements in sensitivity. On the other hand, as was the case for
%lr *** the CMSSM, the AF region ($\mchi>400\gev$ and $\sigsip<10^{-9}\pb$)
the CMSSM, the AF region ($\mchi>400\gev$ and $\sigsip<10^{-10}\pb$) is
not likely to be further constrained by the new spin-independent cross
section measurements planned for the next years, including XENON1T.

%%%%%%%%%%%%%%%%%%%%%%%%%   F   I   G   U   R   E   %%%%%%%%%%%%%%%%%%%%%%%%%%%%
% 2 by 1: left: plot of 1d pdf of mhl,
% right: chi2 vs mhl
\begin{figure}[b]
\centering
\begin{tabular}{c}
\subfloat[]{%
\label{fig:-a}%
\includegraphics[width=0.50\textwidth]{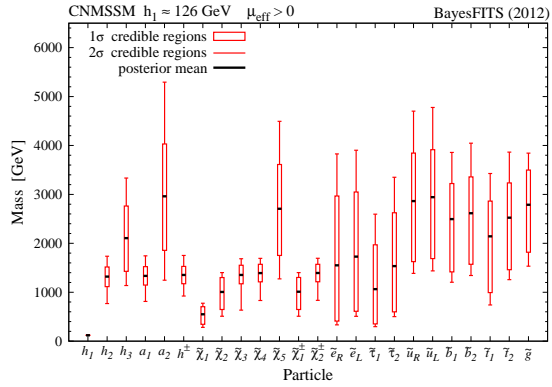}
}%
\\
%\vspace{1pt}%
\subfloat[]{%
\label{fig:-b}%
\includegraphics[width=0.50\textwidth]{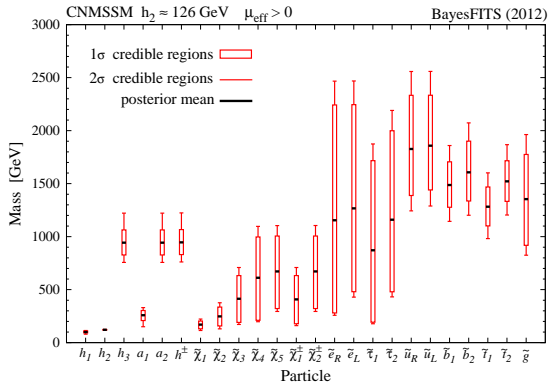}
}%

\subfloat[]{%
\label{fig:-c}%
\includegraphics[width=0.50\textwidth]{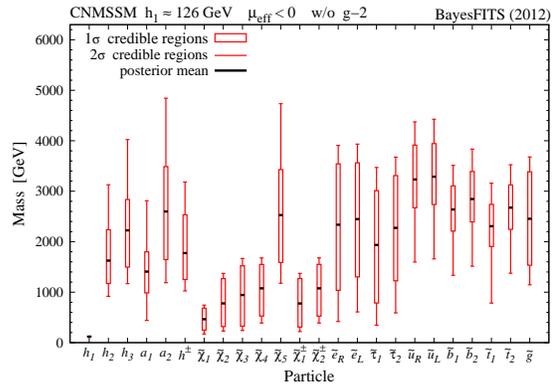}
}%

\end{tabular}
\caption[]{SUSY mass spectra of the CNMSSM with $\mueff>0$ for
  \subref{fig:-a} case~1 and \subref{fig:-b} case~2. \subref{fig:-c} SUSY spectrum with $\mueff<0$ and no \gmtwo\ in case~1.
  The narrow lines indicate the 95\% credibility ranges and the thick
  bars the 68\% credibility regions.}
\label{fig:mass_spectrum}
\end{figure} 
%%%%%%%%%%%%%%%%%%%%%%%%%%%%%%%%%%%%%%%%%%%%%%%%%%%%%%%%%%%%%%%%%%%%%%%%%%%%%%%

We shall analyze the \chisq\ contributions from the individual
constraints in the next subsection. Here we only repeat that, when the \gmtwo\
constraint is relaxed, the better \chisq\ can be obtained for
$\mueff<0$ thanks to the improved fit to \brbsmumu\ and
\brbxsgamma. To illustrate this feature we show in
Fig.~\ref{fig:othernegmu}\subref{fig:-b} the 2D posterior pdf in the
(\brbsmumu, \sigsip) plane for case~1. When $\mueff<0$, a cancellation between 
the pseudoscalar and axial vector form factors\cite{Bobeth:2001sq} takes place, with the result of improving the fit to \brbsmumu\ over all of the 
parameter space, and particularly in the AF region, where the calculated values are pushed below the SM value.

For case~2 and case~3 it was not possible to perform the analysis of the effects of $\mueff<0$, 
as the simultaneous interplay of different constraints strongly disfavors the entire parameter space of the model and no region of good fit appears. 
This can be understood by remembering that $\mhtwo\simeq 126\gev$ decisively favors small values of $|\akap|$ ($\approx|\azero|\approx|\alam|$ in the CNMSSM). 
On the other hand, physicality requires $\maone^2>0$, which for $\mueff<0$ means $B_{\textrm{eff}}\equiv\alam+\kap s<0$\cite{Ellwanger:2009dp}, or $\alam<-\kap s$.
Both constraints cannot be simultaneously satisfied for $\azero<0$ (\akap, \alam\ and \kap\ negative) 
so that the scan tends to prefer small positive values of \azero. 
This generates a conflict with the relic density constraint: 
in the SC region $\azero<0$ is required for singlino LSP, whereas in the FP/HB region, 
when $\mueff<0$ and $\azero>0$ the higgsino component of the neutralino is reduced and the relic density becomes too large. 

\subsection{Mass spectrum and the best-fit points}\label{bestfit:sec}

%%%%%%%%%%%%%%%%%%%%%%%%%   F   I   G   U   R   E   %%%%%%%%%%%%%%%%%%%%%%%%%%%%
% 2 by 1: left: plot of 1d pdf of mhl,
% right: chi2 vs mhl
\begin{figure}[t]
\centering
\begin{tabular}{c}
%\subfloat[]{%
%\label{fig:-a}%
%\includegraphics[width=0.65\textwidth]{figs_v5/spectrum_new_h1_negmu.pdf}
%}%
%\\
%\subfloat[]{%
%\label{fig:-b}%
\includegraphics[width=0.70\textwidth]{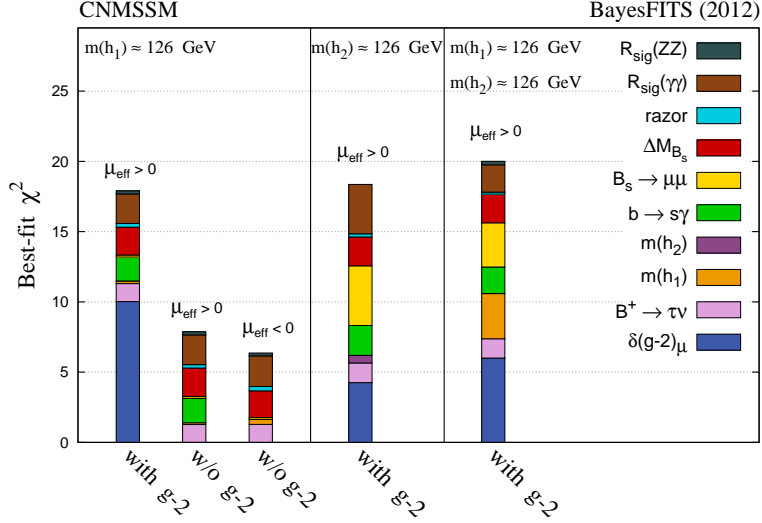}
%}%
\end{tabular}
\caption[]{Individual \chisq\ contributions to the best-fit points of the scans considered.}
\label{fig:chi2cont}
\end{figure} 
%%%%%%%%%%%%%%%%%%%%%%%%%%%%%%%%%%%%%%%%%%%%%%%%%%%%%%%%%%%%%%%%%%%%%%%%%%%%%%%

In Figs.~\ref{fig:mass_spectrum}\subref{fig:-a} and
\ref{fig:mass_spectrum}\subref{fig:-b}, we show the 1D marginalized
posterior pdf's for the SUSY mass spectrum in case~1 and case~2,
respectively. The narrow lines indicate the 95\% credibility regions,
and the thick bars the 68\% credibility regions.

Figure~\ref{fig:mass_spectrum}\subref{fig:-a} shows the CMSSM-like
character of case~1, since $h_3$, \atwo\ and $\chi_5$ are all heavy
and effectively decoupled from the low scale spectrum.  For case~2,
Fig.~\ref{fig:mass_spectrum}\subref{fig:-b} shows that requiring
$\mhtwo\simeq 126\gev$ forces the particles of the Higgs sector to be
quite light, while the sfermions remain heavy, although they tend
to be lighter than in case~1, since \mhone\ is lighter and it requires
smaller loop corrections. The latter affect the scale of the stop and,
through the assumption of unification, the other sfermions.

In Fig.~\ref{fig:mass_spectrum}\subref{fig:-c} we show the mass spectrum
for case~1 where we neglect \deltagmtwomu\ and set $\mueff<0$.
Again we confirm the CMSSM-like nature.

In Fig.~\ref{fig:chi2cont} we show the individual
\chisq\ contributions of the applied constraints to the best-fit
point, for the cases considered in this study. As could be expected,
in case 1 the main contribution comes from \deltagmtwomu, once more confirming the CMSSM-like character of this scenario. 
In case 2 the contribution from \deltagmtwomu\ 
is comparable to the one from \brbsmumu, due to the singlino character of
the neutralino which favors higher \tanb, as explained in Sec.~\ref{Sec:dm}.
When $\mueff>0$, relaxing the \gmtwo\ constraint does not change
significantly the contributions to \chisqmin\ due to the other
constraints.  This shows that the \chisq\ due to \gmtwo\ is fairly
homogeneous throughout the parameter space not yet excluded by
direct LHC SUSY searches.  However, for negative \mueff\ the total
\chisq\ improves by around 1--2 units, thanks to the improved fit to
%\brbsmumu, as explained in Sec.~\ref{Sec:nog2}, and also to
\brbxsgamma, for which the contribution from the chargino-stop loop
changes sign and tends to enhance the branching ratio, towards the
experimental value\cite{Roszkowski:2007fd}. As a consequence, the
overall fit to the experimental measurement improves.

One can finally notice that the contributions to \chisqmin\ due to
\rsiggg\ are sizeable in all three cases, as was explained in
Sec.~\ref{subsec:Rs}. 

% The contribution due to
%\rtwogg\ is also significant, as noted in \refsec{ImpactRs} **\sm{add
%  this label to the subsection 'Impact of the Rs'}**. For $\mu<0$ without \gmtwo\ the contribution due to \bsgamma\ disappears implying a perfect fit to the data. On the other hand, the contribution due to $m_{h_1}$ is slightly enhanced, since \kap\ is negative for this point and the minimization equations require small positive $A_\lambda$ $\sim A_0$, thus not allowing $h_1$ to  acquire a large enough mass. The combined contribution due to $R^{h_2}(X)$ does not show since for Case 1 $h_2$ is almost always too heavy for the exclusion limits to be applicable. 
%
%For Case 2, the contribution due to \brbsgamma\ is even further
%enhanced along with that from $R^{h_2}(\gamma\gamma)$ compared to the
%Case 1. The reason for the latter being that an $h_2$ with mass
%$\simeq 125$\gev\ is highly singlet dominated as noted in
%\refsec{ImpactHmass} **\SM{Add this label to subsection 'Impact of the Higgs mass'} and hence its decay cross-sections are even further away from being SM-like, resulting in a poorer fit to the data. Also, the combined contribution due to $R^(h_1)(X)$ is sizable, since $m_{h_1}$ is always smaller than $m_{h_2} \simeq$~125\gev\, and is thus bounded more strongly by the LHC exclusion limits implemented. 

In Table~\ref{tab:bestfit_params} we show  the parameters of the best-fit points obtained in our scans.

%%%%%%%%%%%%%%%%%%%%%%%%%%%% T A B L E %%%%%%%%%%%%%%%%%%%%%%%%%%%%
% Table showing parameters for best-fit points from various scans
\begin{table}[t]
\centering
%%%%%%%%%%%%%%%%%%%%%%%%%%%%%%%%%%%%%%%%%%%%%%%%%%%%%%%%%%%%%%%%%%%%%%%%%%%%%%%%
\begin{tabular}{|c|c|c|c|c|c|c|c|c|}
\hline %\toprule
%%%%%%%%%%%%%%%%%%%%%%%%%%%%%%%%%%%%%%%%%%%%%%%%%%%%%%%%%%%%%%%%%%%%%%%%%%%%%%%%
& \mzero & \mhalf & \azero & \tanb & \lam & $m_{h_{\textrm{sig}}}$ & $\Delta$ & $\chi^2_{min}$ \\
\hline %\midrule
%%%%%%%%%%%%%%%%%%%%%%%%%%%%%%%%%%%%%%%%%%%%%%%%%%%%%%%%%%%%%%%%%%%%%%%%%%%%%%%%
\multicolumn{9}{|c|}{With \gmtwo}\\
\hline
Case~1, $\mueff>0$ & 340.9 & 911.9 & $-2546$ & 12.0 & 0.017 & 124.5 & 427 & 17.91 \\ \hline
Case~2, $\mueff>0$ & 102.8 & 803.6 & $-295.0$ & 32.1 & 0.006 & 123.5 & 396 & 18.45 \\ \hline
Case~3, $\mueff>0$ & 124.7 & 958.8 & $-356.7$ & 33.7 & 0.001 & 120.3, 126.2 & 480 & 20.08 \\ \hline
\multicolumn{9}{|c|}{No \gmtwo}\\
\hline
Case~1, $\mueff<0$ & 410.2 & 913.3 & $-2764$ & 14.8 & 0.017 & 123.9 & 455 & 6.40 \\ \hline
%%%%%%%%%%%%%%%%%%%%%%%%%%%%%%%%%%%%%%%%%%%%%%%%%%%%%%%%%%%%%%%%%%%%%%%%%%%%%%%%
%\hline %\bottomrule
\end{tabular}
%%%%%%%%%%%%%%%%%%%%
\caption{CNMSSM input parameters, Higgs masses and fine-tuning measure for the
best-fit points of our scans. Masses and $A_0$ are in \gev.
}
\label{tab:bestfit_params}
\end{table}
%%%%%%%%%%%%%%%%%%%%%%%%%%%%%%%%%%%%%%%%%%%%%%%%%%%%%%%%%%%%%%%

%%%%%%%%%%%%%%%%%%%%%%%%%%%%%%
\section{\label{Summary}Summary and conclusions}

In this paper we have presented the first global analysis of the
CNMSSM which included the measurement of the mass and decay
cross sections of the Higgs-like resonance observed at the LHC, the first evidence of nonzero \brbsmumu\ at LHCb, 
SUSY mass limits from direct searches, as well as the updated value of the
top mass. We discussed in detail the
possibility of either one of the lightest $CP$-even Higgs bosons of the
model playing the role of the SM-like Higgs (case~1 and case~2 in the
text) and that of the observed excess being due to a combination of
two of them, both lying within the range $\sim 123$--129\gev, after inclusion of a 3\gev\ theoretical error on the 
Higgs mass calculation (case~3 in the text).

In contrast to the case with $m_S^2=m_0^2$, explored for example in\cite{Arbey:2011ab},
we found that with our choice of parameters the CNMSSM allows for a 
SM-like $h_1$ as heavy as 125\gev, 
especially in the stau-coannihilation region where the scan obtains a good compromise between
the relic density and the $h_1$ mass constraint, owing to large
negative $A_0$ values. The calculated value of \mhone\ is larger than
what was previously found for the CMSSM\cite{Fowlie:2012im} thanks to the
increased value of the top pole mass. The overall fit is somewhat
spoiled by the requirement of having \ronegg\ consistent with the
value reported by the CMS Collaboration, since it is virtually
impossible, given our parameter choice and ranges, 
to obtain an enhancement in the cross section rate in
case~1. A similar conclusion can be drawn for case~2 where, in turn,
$h_2$ is required to have a mass around 126\gev, which can also be
achieved but at the cost of a poor fit to other observables. Case~3,
with almost degenerate $h_1$ and $h_2$, was found not to be as
interesting as anticipated, since the combined $R_{h_1 +
  h_2}(\gamma\gamma)$ almost never exceeds 1, besides the fact that
the high credibility posterior regions mimic those of case~2.

As mentioned in Sec.~\ref{Model}, we have adopted in this paper 
the choice $\akap=\azero$ at the GUT scale.
We checked with a few preliminary scans that this choice does not affect the 
shape and position of the posterior pdf's in case~1. This make sense since, as we explained in Sec.~\ref{Results}, 
to satisfy all the constraints the model tends to its CMSSM limit, 
and the singlet field effectively decouples from the theory. 
Similarly, relaxing the unification condition on \akap\ would have little impact on the 
(\mzero, \mhalf) 2D posterior in case~2 and case~3.
In both cases the requirement of a very light $h_1$ constrains \mzero\ and \azero\
substantially through \msusy\ and the stop mixing parameter, independently of the value of \akap\ at the GUT scale.  
As a consequence, the relic density can only be satisfied in the regions of the
(\mzero, \mhalf) plane shown in Fig.~\ref{fig:cnmssm_momhalf}\subref{fig:-a}.  
Some differences, on the other hand, can be expected in the distribution of \tanb\ in case~2 and
case~3. Particularly, in the stau-coannihilation region, where the LSP is singlinolike,
the region of high probability is likely to extend to values of \tanb\ lower than the ones 
favored in this study. 
Consequently, the impact of the \brbsmumu\ constraint is likely to be reduced.   
     
On the other hand, even by disunifying \akap, we would not expect changes 
relative to our analysis of the di-boson Higgs decay rates.
In a recent paper\cite{Ellwanger:2012ke} it was shown that 
a $\gamma\gamma$ enhancement consistent with the observation is easily obtained
in case~2. But we remind the reader that the study in question analyzed a model   
substantially less constrained than the one explored in here. In particular 
we have checked that, when allowing \kap\ and \mueff\ to be free at the SUSY
scale (as was done in\cite{Ellwanger:2012ke}), the size of the favored parameter space in case~2
can increase significantly.

Finally, in this paper we also provided estimates of fine-tuning due to the various input
parameters of the model in the form of isocontours in the (\mzero,
\mhalf) plane for cases~1 and 2.
We noted that the maximum fine-tuning for most of the parameter space
comes from two different sources in the two cases.

We assessed the effects of abandoning the \gmtwo\ constraint
since it cannot be reproduced in the CNMSSM, and more generally SUSY
models with slepton-squark unification. In this case 
the overall fit actually improves considerably in case~1 for $\mueff<0$, due to a
better agreement of the model's predictions for $b$-physics observables
with experimental data, similarly to the CMSSM. Case~2 and case~3, on the other hand, are 
strongly disfavored for $\mueff<0$ by the simultaneous impact of several constraints.

\bigskip
%%%%%%%%%%%%%%%%%%%%%%%%%%%%%%%%%%%%%%%%%%%%%%%%%%%%%%%%%%%%%%%%%%%%%%%%%%%%%%%%
\begin{center}
\textbf{ACKNOWLEDGMENTS}
\end{center}

S.M. would like to thank Cyril Hugonie for useful e-mail exchange regarding model implementation in \nmssmtools. 
L.R. thanks Gino Isidori and Matteo Palutan for helpful comments about \brbsmumu.

  This work has been funded in part by the Welcome Programme
  of the Foundation for Polish Science. 
  K.K. is supported by the EU and MSHE Grant No. POIG.02.03.00-00-013/09.
  L.R. is also supported in part by the Polish National Science Centre Grant No. N202 167440, an STFC
  consortium grant of Lancaster, Manchester and Sheffield Universities
  and by the EC 6th Framework Programme MRTN-CT-2006-035505. The use of the CIS computer cluster at NCBJ is gratefully acknowledged. 
%%%%%%%%%%%%%%%%%%%%%%%%%%%%%%%%%%%%%%%%%%%%%%%%%%%%%%%%%%%%%%%%%%%%%%%%%%%%%%%%

\bibliographystyle{utphysmcite}	% (uses file "plain.bst")

\bibliography{myref}

\end{document}